\pdfoutput=1
\documentclass[fleqn,usenatbib,titlepage]{mnras}
\usepackage{newtxtext,newtxmath}
\usepackage{enumitem}
\usepackage{mathptmx}
\usepackage[T1]{fontenc}
\usepackage{ae,aecompl}
\usepackage{longtable}
\usepackage{booktabs}
\usepackage{graphicx}    % Including figure files
\usepackage{amsmath}    % Advanced maths commands
\usepackage{amssymb}    % Extra maths symbols
\usepackage{pdflscape}
\usepackage{etoolbox}
\usepackage{CJKutf8}
\pretocmd{\abstractname}{\newpage}{}{}

\title[\textit{XMM-Newton} point-source catalog for the XMM-LSS field]{
The XMM-SERVS survey: new {\it XMM-Newton} point-source catalog for the XMM-LSS field
}

\author[Chien-Ting Chen et al.]
{C.-T.J. Chen\begin{CJK*}{UTF8}{bsmi}(陳建廷)\end{CJK*}$^{1,2}$\thanks{E-mail: ctchen@psu.edu},
W.N. Brandt$^{1,2,3}$,
B. Luo$^{4,5}$,
P. Ranalli$^{6}$,
G. Yang$^{1,2}$,
\newauthor
D.M. Alexander$^{7}$,
F.E. Bauer$^{8,9,10}$,
D.D. Kelson$^{11}$, 
M. Lacy$^{12}$,
K. Nyland$^{12}$, 
\newauthor
P. Tozzi$^{13}$,
F. Vito$^{1,2}$,
M. Cirasuolo$^{14}$,
R. Gilli$^{15}$,
M.J. Jarvis$^{16,17}$,
B.D. Lehmer$^{18}$,
\newauthor
M. Paolillo$^{19}$,
D.P. Schneider$^{1,2}$,
O. Shemmer$^{20}$,
I. Smail$^{7}$,
M. Sun$^{21,22}$
\newauthor
M. Tanaka$^{23}$,
M. Vaccari$^{17,24}$,
C. Vignali$^{25,15}$,
Y.Q. Xue$^{21,22}$,
M. Banerji$^{26}$,
\newauthor
K.E. Chow$^{27}$,
B. H\"au\ss ler$^{28}$, 
R.P. Norris$^{29,27}$,
J.D. Silverman$^{30}$, and
J.R. Trump$^{31}$\\
$^{1}$Department of Astronomy \& Astrophysics, 525 Davey Lab, 
The Pennsylvania State University, University Park, PA 16802, USA\\
$^{2}$Institute for Gravitation and the Cosmos, The Pennsylvania State University, University Park, PA 16802, USA\\
$^{3}$Department of Physics, The Pennsylvania State University, University Park, PA 16802, USA\\
$^{4}$School of Astronomy and Space Science, Nanjing University, Nanjing 210093, China\\
$^{5}$Key Laboratory of Modern Astronomy and Astrophysics (Nanjing University), Ministry of Education, Nanjing, Jiangsu 210093, China
}

\date{Accepted for publication in MNRAS}

\pubyear{2018}

\newcommand{\xmm}{\hbox{\textit{XMM-Newton}}}
\newcommand{\xray}{\hbox{X-ray}}
\newcommand{\chandra}{\hbox{\textit{Chandra}}}

\newcommand{\ngpz}{449}
\newcommand{\nallpz}{529,913}
\newcommand{\nallgpz}{390,900}
\newcommand{\nallpzsz}{42985}
\begin{document}
\label{firstpage}
\pagerange{\pageref{firstpage}--\pageref{lastpage}}
\maketitle

% Abstract of the paper
\begin{abstract}
We present an \hbox{X-ray} point-source catalog from the XMM-Large Scale Structure survey region (XMM-LSS), one of the XMM-Spitzer Extragalactic Representative Volume Survey (XMM-SERVS) fields. We target the XMM-LSS region with $1.3$~Ms of new {\it XMM-Newton} AO-15 observations, transforming the archival \hbox{X-ray} coverage in this region into a 5.3~deg$^2$ contiguous field with uniform \hbox{X-ray} coverage totaling $2.7$~Ms of flare-filtered exposure, with a $46$~ks median PN exposure time. We provide an \hbox{X-ray} catalog of 5242 sources detected in the soft (0.5--2~keV), hard (2--10~keV), and/or full (\hbox{0.5--10~keV}) bands 
with a 1\% expected spurious fraction determined from simulations.
A total of 2381 new \hbox{X-ray} sources are detected compared to previous source catalogs in the same area. 
Our survey has flux limits of 
$1.7\times10^{-15}$,
$1.3\times10^{-14}$, and
$6.5\times10^{-15}$~erg~cm$^{-2}$~s$^{-1}$ over 90\% of its area 
in the soft, hard, and full bands, respectively, which is comparable to those of the XMM-COSMOS survey. 
We identify multiwavelength counterpart candidates for 99.9\% of the \hbox{X-ray} 
sources, of which 93\% are considered as reliable based on their matching 
likelihood ratios. 
The reliabilities of these high-likelihood-ratio counterparts 
are further confirmed to be $\approx 97\%$ reliable based on deep \chandra\ 
coverage over $\approx 5\%$ of the XMM-LSS region. Results of multiwavelength 
identifications are also included in the source catalog, along with basic 
optical-to-infrared photometry and spectroscopic redshifts from publicly 
available surveys. We compute photometric redshifts for \hbox{X-ray} sources in 4.5~deg$^2$ of our field 
where forced-aperture multi-band photometry is available; $>70$\% of the \hbox{X-ray} sources in this subfield have either spectroscopic or high-quality photometric redshifts. 
\end{abstract}

\maketitle
\footnotetext{
\\
$^{6}$Lund Observatory, Box 43, 22100 Lund, Sweden\\
$^{7}$Centre for Extragalactic Astronomy, Department of Physics, Durham University, South Road, Durham, DH1 3LE, UK\\
$^{8}$Instituto de Astrof{\'{\i}}sica and Centro de Astroingenier{\'{\i}}a, Facultad de F{\'{i}}sica, Pontificia Universidad Cat{\'{o}}lica de Chile, Casilla 306, Santiago 22, Chile\\
$^{9}$Millennium Institute of Astrophysics (MAS), Chile\\
$^{10}$Space Science Institute, 4750 Walnut Street, Suite 205, Boulder, Colorado 80301, USA\\
$^{11}$The Observatories, The Carnegie Institution for Science, 813 Santa Barbara St., Pasadena, CA 91101\\
$^{12}$National Radio Astronomy Observatory, 520 Edgemont Road, Charlottesville, VA 22903, USA\\
$^{13}$INAF, Osservatorio Astrofisico di Arcetri, Largo E. Fermi 5, I-50125, Firenze, Italy\\
$^{14}$European Southern Observatory, Karl-Schwarzschild-Str. 2, 85748 Garching b. M{\"u}nchen, Germany\\
$^{15}$INAF -- Osservatorio Astronomico di Bologna, Via Gobetti 93/3, 40129 Bologna, Italy\\
$^{16}$Oxford Astrophysics, Denys Wilkinson Building, University of Oxford, Keble Road, Oxford OX1 3RH, UK\\
$^{17}$Department of Physics, University of the Western Cape, Bellville 7535, South Africa\\
$^{18}$Department of Physics, University of Arkansas, 226 Physics Building, 825 West Dickson Street, Fayetteville, AR 72701, USA\\
$^{19}$Dip.di Fisica Ettore Pancini, Università di Napoli Federico II, via Cintia, 80126, Napoli, Italy\\
$^{20}$Department of Physics, University of North Texas, Denton, TX 76203, USA\\
$^{21}$CAS Key Laboratory for Research in Galaxies and Cosmology, Department of Astronomy, University of Science and Technology of China, Hefei 230026, China\\
$^{22}$School of Astronomy and Space Science, University of Science and Technology of China, Hefei 230026, China\\
$^{23}$National Astronomical Observatory of Japan, 2-21-1 Osawa, Mitaka, Tokyo 181-8588, Japan\\
$^{24}$INAF -- Istituto di Radioastronomia, via Gobetti 101, 40129 Bologna, Italy\\
$^{25}$Dipartimento di Fisica e Astronomia, Universit\`a degli Studi di Bologna, Via Gobetti 93/2, 40129 Bologna, Italy\\
$^{26}$Institute of Astronomy, University of Cambridge, Madingley Road, Cambridge CB3 0HA, United Kingdom\\
$^{27}$CSIRO Astronomy and Space Science, PO Pox 76, Epping, NSW, 1710, Australia\\
$^{28}$European Southern Observatory, Alonso de Cordova 3107, Vitacura, Santiago, Chile\\
$^{29}$Western Sydney University, Locked Bag 1797, Penrith South, NSW 1797, Australia\\
$^{30}$Kavli Institute for the Physics and Mathematics of the Universe, The University of Tokyo, Kashiwa, Japan 277-8583 (Kavli IPMU, WPI) \\
$^{31}$Department of Physics, University of Connecticut, 2152 Hillside Road, Storrs, CT 06269, USA   
}

\section{Introduction}

\begin{table*}
\footnotesize
\caption{\label{tab:servsmw}
Current and scheduled 1--10 deg$^2$ multiwavelength coverage of the XMM-SERVS fields. 
References: 
[a] \citet{fran15};
[b] \citet{catmeerkat};
[c] \citet{cathermes};
[d] \citet{catswire}; 
[e] \citet{catservs}. Note that SERVS has recently 
been expanded to cover the full LSST deep drilling fields ({\it Spitzer} Program ID 11086).
[f] \citet{catvideo};
[g] \url{http://www.ast.cam.ac.uk/~mbanerji/VEILS/veils_index.html};
[h] \url{http://euclid2017.london/slides/Monday/Session3/SurveyStatus-Scaramella.pdf};
[i] \citet{dieh14};
[j] \citet{cathscpdr1};
[k] \citet{catpanstarrs1};
[l] \citet{catvoice};
[m] \url{http://www.lsst.org/News/enews/deep-drilling-201202.html};
[n] \citet{catspeczcsi,catspeczcsi2};
[o] \citet{catspeczprimus};
[p] \url{https://devilsurvey.org/wp/};
[q] \url{http://www.roe.ac.uk/~ciras/MOONS/VLT-MOONS.html};
[r] \citet{catpfs};
[s] \url{http://www.galex.caltech.edu/researcher/techdoc-ch2.html}.
[t] \url{http://personal.psu.edu/wnb3/xmmservs/xmmservs.html}.
}
\begin{tabular}{lll}
\noalign{\smallskip}\hline\noalign{\smallskip}
{Band}                   &
{Survey Name} &
{Coverage (XMM-LSS, W-CDF-S, ELAIS-S1)}; Notes          \\
\noalign{\smallskip}\hline \hline\noalign{\smallskip}

Radio    & Australia Telescope Large Area Survey ({\bf ATLAS})$^{\rm a}$ & --, 3.7, 2.7 deg$^2$; 15~$\mu$Jy rms depth at 1.4~GHz      \\
& {\bf MIGHTEE} Survey (Starting Soon)$^{\rm b}$ & 4.5, 3, 4.5 deg$^2$; 1~$\mu$Jy rms depth at 1.4~GHz\\
\noalign{\smallskip}\hline\noalign{\smallskip}

FIR& {\it Herschel} Multi-tiered Extragal.\ Surv.\ ({\bf HerMES})$^{\rm c}$ & 0.6--18 deg$^2$;  5--60~mJy depth at 100--500~$\mu$m\\
\noalign{\smallskip}\hline\noalign{\smallskip}

MIR &
{\it Spitzer} Wide-area IR Extragal.\ Survey ({\bf SWIRE})$^{\rm d}$ & 9.4, 8.2, 7.0 deg$^2$;  0.04--30~mJy depth at 3.6--160~$\mu$m\\
\noalign{\smallskip}\hline\noalign{\smallskip}

NIR  &  {\it Spitzer} Extragal.\ Rep.\ Vol.\ Survey ({\bf SERVS})$^{\rm e}$ &
4.5, 3, 4.5 deg$^2$; 2~$\mu$Jy depth at 3.6 and 4.5~$\mu$m\\
& VISTA Deep Extragal.\ Obs.\ Survey ({\bf VIDEO})$^{\rm f}$ & 4.5, 3, 4.5 deg$^2$; $ZYJHK_s$ to $m_{\rm AB}\approx23.8$--25.7\\
& VISTA Extragal.\ Infr.\ Legacy Survey ({\bf VEILS})$^{\rm g}$ & 3, 3, 3 deg$^2$; $JK_s$ to $m_{\rm AB}\approx24.5$--25.5\\
& \textbf{\textit{Euclid}} Deep Field$^{\rm h}$ & --, 10, --~deg$^2$; $YJH$ to $m_{\rm AB}\approx26$, VIS to $m_{\rm AB}\approx26.5$\\
\noalign{\smallskip}\hline\noalign{\smallskip}

Optical
& Dark Energy Survey ({\bf DES})$^{\rm i}$ & 9, 6, 9~deg$^2$; Multi-epoch
$griz$, $m_{\rm AB}\approx27$ co-added\\
Photometry& Hyper Suprime-Cam ({\bf HSC}) Deep Survey$^{\rm j}$ & 5.3, --, --~deg$^2$; $grizy$ to $m_{\rm AB}\approx25.3$--27.5\\
& Pan-STARRS1 Medium-Deep Survey ({\bf PS1MD})$^{\rm k}$ & 8, --, 8 deg$^2$; Multi-epoch $grizy$, $m_{\rm AB}\approx26$ co-added\\
& VST Opt. Imaging of CDF-S and ES1 ({\bf VOICE})$^{\rm l}$ & --, 4.5, 3~deg$^2$; Multi-epoch $ugri$, $m_{\rm AB}\approx26$ co-added\\
& SWIRE optical imaging$^{\rm d}$ & 8, 7, 6 deg$^2$; $u'g'r'i'z'$ to $m_{\rm AB}\approx24$--26 \\
& {\bf LSST} deep-drilling field (Planned)$^{\rm m}$ & 10, 10, 10 deg$^2$;
$ugrizy$, $\gtrsim 10\,000$ visits per field\\
\noalign{\smallskip}\hline\noalign{\smallskip}
Optical/NIR & Carnegie-{\it Spitzer}-IMACS Survey ({\bf CSI})$^{\rm n}$ & 6.9, 4.8, 3.6~deg$^2$; $140\,000$ redshifts, 3.6~$\mu$m selected\\
Spectroscopy &PRIsm MUlti-object Survey ({\bf PRIMUS})$^{\rm o}$ & 2.9, 2.0, 0.9~deg$^2$; 77\,000 redshifts
to $i_{\rm AB}\approx23.5$\\
& AAT Deep Extragal.\ Legacy Survey ({\bf DEVILS})$^{\rm p}$ & 3.0, 1.5, -- deg$^2$; 43\,500 redshifts
to $Y=21.2$\\
& VLT {\bf MOONS} Survey (Scheduled)$^{\rm q}$ & 4.5, 3, 4.5 deg$^2$; $210\,000$ redshifts to $H_{\rm AB}\approx23.5$\\
& Subaru {\bf PFS} survey (Planned)$^{\rm r}$ & 5.3, --, --~deg$^2$; $J \approx 23.4$ for HSC deep fields. \\
\noalign{\smallskip}\hline\noalign{\smallskip}

UV & {\it GALEX} Deep Imaging Survey$^{\rm s}$ &  8, 7, 7~deg$^2$;  Depth $m_{\rm AB}\approx25$\\
\noalign{\smallskip}\hline\noalign{\smallskip}
X-ray & \textbf{XMM-SERVS}$^{\rm t}$ &  5.3, 4.5, 3~deg$^2$; 4.7~Ms {\it XMM-Newton} time, $\approx50$~ks depth\\
\noalign{\smallskip}\hline\noalign{\smallskip}

\end{tabular}
%{\tiny }
\end{table*}
%add PFS, J ≃ 23.4 in HSC-DEEP

Due to the penetrating nature of \hbox{X-ray} emission and its 
ubiquity from accreting supermassive black holes (SMBHs), extragalactic 
\hbox{X-ray} surveys have provided an effective census of active 
galactic nuclei (AGNs), including obscured systems, in the distant
universe. Over at least the past three decades, the overall design 
of cosmic \hbox{X-ray} surveys has followed a ``wedding cake'' strategy. 
At the extremes of this strategy, some surveys have ultra-deep \hbox{X-ray} 
coverage and a narrow ``pencil-beam'' survey area ($\lesssim 1$~deg$^2$), 
while others have shallow \hbox{X-ray} coverage over a wide survey 
area (\hbox{$\approx$ 10--10$^4$}~deg$^2$). The wealth 
of data from cosmic \hbox{X-ray} surveys (and their co-located 
multiwavelength surveys) have provided a primary source of information 
in shaping understanding of how SMBHs grow through cosmic time, where 
deep surveys generally sample high-redshift, moderately luminous AGNs, 
and wide-field surveys generally probe the high-luminosity, rare 
objects that are missed by surveys covering smaller volumes. 
However, narrow-field surveys lack the contiguous volume to 
encompass a wide range of cosmic large-scale structures, and wide-field 
surveys generally lack the \hbox{X-ray} sensitivity to track the bulk of the 
AGN population through the era of massive galaxy 
assembly (see \citealt{xrayreview15} for a recent review).

Among extragalactic \hbox{X-ray} surveys, the medium-deep COSMOS 
survey over $\approx 2$~deg$^2$ has the necessary sensitivity-area 
combination to begin to track how a large fraction of distant SMBH 
growth relates to cosmic large-scale structures \citep[e.g., ][]{hasi07xmmcos,civa16}.
However, even COSMOS cannot sample the full range of cosmic 
environments. The largest structures found in cold dark matter 
simulations are already as large as the angular extent of COSMOS 
at $z\approx 1$ (80--100 Mpc in comoving size, which 
covers 2--3 deg$^2$; e.g., see \citealt{klyp16}). Clustering analyses 
also demonstrate that COSMOS-sized fields are still subject to 
significant cosmic variance \citep[e.g.,][]{mene09,torr10,skib14}. 

Therefore, to study SMBH growth across the full range of 
cosmic environments and minimize cosmic variance, it is necessary 
to obtain multiple medium-deep \hbox{X-ray} surveys in distinct 
sky regions \citep[e.g.,][]{driv10,most11} 
with multiwavelength data comparable to those of COSMOS. 
In this work, we present a catalog of 5242 {\it XMM-Newton} sources detected 
over 5.3~deg$^2$ in one of the well-studied {\it Spitzer} Extragalactic 
Representative Volume Survey \citep[SERVS, ][]{catservs} fields, 
the XMM-Large Scale Structure (XMM-LSS) region. This is the first 
field of the broader XMM-SERVS survey which aims to expand the parameter 
space of \hbox{X-ray} surveys with three \hbox{$>3$~deg$^{2}$} 
surveys reaching XMM-COSMOS-like depths, including XMM-LSS, 
Wide {\it Chandra} Deep Field-South (W-CDF-S), 
and ELAIS-S1.\footnote{{\it XMM-Newton} observations of W-CDF-S and ELAIS-S1 have been allocated via the AO-17 {\it XMM-Newton} Multi-year Heritage Program.} These three extragalactic fields have been chosen based 
on their excellent multiwavelength coverage and superior legacy value. 
We list the current and scheduled multiwavelength coverage of XMM-SERVS 
in Table~\ref{tab:servsmw}.

The \hbox{X-ray} source catalog presented here has been generated using a 
total of 1.3~Ms of \xmm\ AO-15 observations in the XMM-LSS
field (specifically the region covered by SERVS), plus all 
archival \xmm\ data in this same region. Our AO-15 observations 
target the central part of XMM-LSS adjacent to (and partly including) the 
Subaru {\it XMM-Newton} Deep Survey \citep[SXDS,][]{catsxdf}, 
transforming the complex archival \xmm\ coverage in this region 
into a contiguous 5.3~deg$^2$ field with relatively uniform \hbox{X-ray} coverage.
The median clean exposure time with the PN instrument is $\approx 46$~ks, 
reaching survey depths comparable to those of XMM-COSMOS \citep[e.g.,][]{capp09xcos} 
and SXDS. We also present multiwavelength counterparts, basic photometric 
properties, and spectroscopic redshifts obtained from the literature. 
Photometric redshifts are derived over a 4.5~deg$^2$ region
using the forced-photometry catalog of Nyland et al. 2018 (in preparation).
The excellent multiwavelength coverage in the XMM-SERVS XMM-LSS 
field will provide the necessary data for studying the general 
galaxy population and tracing large-scale structures. The 
combination of these multiwavelength data and the new \hbox{X-ray} 
source catalog (along with similar data for COSMOS and the other XMM-SERVS fields) will enable 
potent studies of SMBH growth across the full range of cosmic environments, from voids to 
massive clusters, while minimizing cosmic variance effects. 
The \xmm\ source catalog and several associated data products are being 
made publicly available along with this paper.\footnote{\url{http://personal.psu.edu/wnb3/xmmservs/xmmservs.html}.}

This paper is organized as follows: in \S\ref{sec:datasum} 
we present the details of the new and archival observations, 
and the procedures for data reduction. 
In \S\ref{sec:mainx} we describe the \hbox{X-ray} source-searching 
strategies and the details of the production of the \hbox{X-ray} point-source catalog. 
We also outline the reliability assessment of the \hbox{X-ray} catalog 
using simulated \hbox{X-ray} observations. 
The survey sensitivity and the number counts are also presented here. 
In \S\ref{sec:mw}, we describe the multiwavelength counterpart identification methods and 
reliability assessments. 
In \S\ref{sec:redshifts}, we describe the 
spectroscopic and photometric redshifts of the X-ray sources. 
The basic multiwavelength properties and the source classifications are presented in 
\S\ref{sec:class}. 
A summary is given in \S\ref{sec:sum}. 
The source catalog, including the properties of the multiwavelength counterparts identified with likelihood-ratio matching methods, and the descriptions of columns are included in Appendix \ref{sec:catcols}. 
Multiwavelength matching results using the Bayesian matching code {\sc NWAY} are included in Appendix \ref{sec:nwaycols}. 
In addition to the X-ray sources, we also present the photometric redshifts 
for the galaxies in our survey region in Appendix \ref{sec:pzapp}.
Throughout the paper, we assume a $\Lambda$CDM cosmology with $H_0 = 70$ km s$^{-1}$ Mpc$^{-1}$, $\Omega_m = 0.3$, and $\Omega_\Lambda = 0.7$. 
We adopt a Galactic column density $N_{\rm H} = 3.57\times10^{20}$ cm$^{-2}$
along the line of sight to the center of the source-detection region
at RA$=35.6625^{\circ}$, DEC$=-4.795^{\circ}$ \citep[e.g.,][]{refgalactic}.\footnote{Derived using the {\sc colden} task included in the {\sc CIAO} software package.} 
AB magnitudes are used unless noted otherwise.  

\begin{figure}
    \includegraphics[width=\columnwidth]{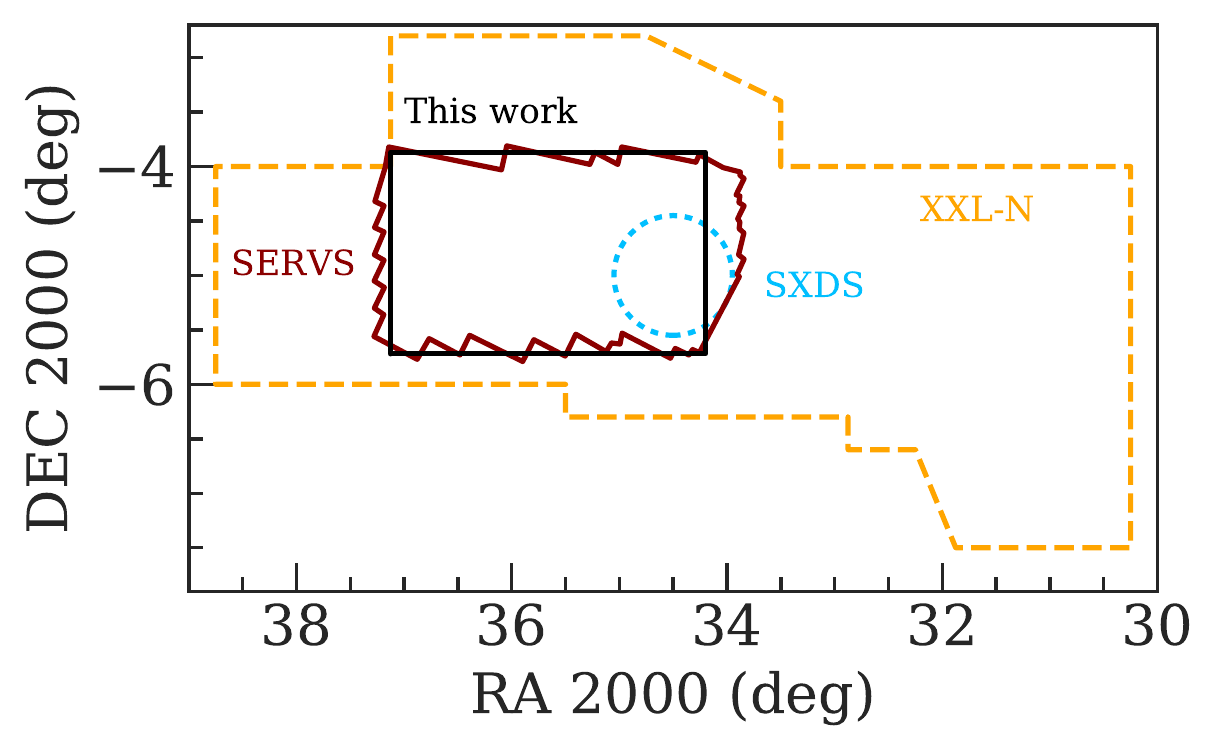}
    \vspace{-0.5cm}
    \caption{
    Illustration of the survey regions of XMM-XXL-North (XXL-N, \citealt{catxxl1}, orange dashed line), the Subaru {\it XMM-Newton} Deep Survey (SXDS, \citealt{catsxdf}, blue dotted circle), and the XMM-SERVS survey of XMM-LSS
    presented in this work (black box). The {\it Spitzer} SERVS coverage of XMM-LSS is also shown as the red polygon.
    }
    \label{fig:maps}
\end{figure}

\section{{\it XMM-Newton} Observations in the XMM-LSS region and data reduction}\label{sec:datasum}
\begin{table*}
\scriptsize
    \caption{\label{tab:xmmdata}
The {\it XMM-Newton} data used to create the source catalog include 
155 pointings with a total of 2.7~Ms of flare-filtered exposure time, of which 1.1~Ms is from 
the new AO-15 observations.$^a$ Columns from left to right: target field, {\it XMM-Newton} revolution, 
{\it XMM-Newton} ObsID, observation starting date/time, Right Ascension and Declination of the pointing center 
(J2000, degrees), cleaned exposure time for PN, MOS1, and MOS2 in each pointing. This table is available in its entirety online.}
    \begin{tabular}{ccccccccc} % four columns, alignment for each
        \hline
        Field                    & Revolution & ObsID        & Date & R.A. & Decl. & GTI (PN) & GTI (MOS1) & GTI (MOS2)\\
                                    &            &              & (UT) &      &       & (ks)     & (ks)     & (ks)     \\
        \hline
AO-15 & 3054 & 0780450101 & 2016-08-13T01:34:06 & 35.81072 & $-5.15989$ & 20.91 & 23.61 & 23.61\\
XMM-LSS & 1205 & 0404965101 & 2006-07-09T08:08:08 & 35.80953 & $-5.48532$& 3.44 & 10.36 & 9.91\\
XMDS & 287 & 0111110401 & 2001-07-03T14:01:54 & 35.97582 & $-5.15253$ & 21.40 & 27.20 & 27.40\\
SXDS & 118 & 0112370101 & 2000-07-31T21:57:54 & 34.47819 & $-4.98115$ & 39.13 & 42.70 & 42.83\\
XMM-XXL-North & 2137 & 0677580101 & 2011-08-10T01:53:35 & 37.16867 & $-4.49993$ & 4.94 & 5.93 & 5.52\\
XMM-XXL-North & 2137 & 0677580101 & 2011-08-10T01:53:35 & 37.33404 & $-4.49993$ & 2.01 & 6.47 & 6.67\\
XLSSJ022404.0--041328    & 0928 & 0210490101 & 2005-01-01T19:08:30 & 36.03267 & $-4.20230$ & 80.28 & 87.98 & 87.98\\
\hline
\end{tabular}\\   
$^a$: {\tiny MOS only (MOS1 and MOS2 have the same exposure time). For PN, the total flare-filtered time is 2.3~Ms, 
of which 0.9~Ms is from the new AO-15 observations.}     
\end{table*}

\subsection{{\it XMM-Newton} and {\it Chandra} data in the XMM-LSS region}
The XMM-LSS field has been targeted by a number of \xmm\ surveys of different
sensitivities (e.g., see Fig.~3 of \citealt{xrayreview15} and Fig. 1 of \citealt{xue17review}). 
The original XMM-LSS survey was an $\approx 11$ deg$^2$ field typically covered by \xmm\ 
observations of $\approx 10$~ks exposure time per pointing \citep[][]{catxmmlss06,catxxl1}. %, each separated by 20$^\prime$ 
Within the 11 deg$^2$ field, $\approx 4$ deg$^2$ were observed by the 
{\it XMM-Newton} Medium Deep Survey \citep[XMDS, $20-25$~ks exposure depth,][]{catxmds}.
In addition, the Subaru {\it XMM-Newton} Deep Survey \citep[SXDS,][]{catsxdf}, adjacent to the
XMDS field, covers a 1.14 deg$^2$ area and reaches a nominal $\approx 50$~ks exposure per pointing
\citep{catsxdf}. 
Moreover, the XMM-LSS field recently became a part of the 25 deg$^2$ 
XMM-XXL-North field \citep{catxxl1}, which has similar \xmm\ coverage as the original XMM-LSS survey (i.e., $\approx 10$~ks depth).

\begin{figure*}
    %\hspace*{0.1in}
    \includegraphics[width=\textwidth]{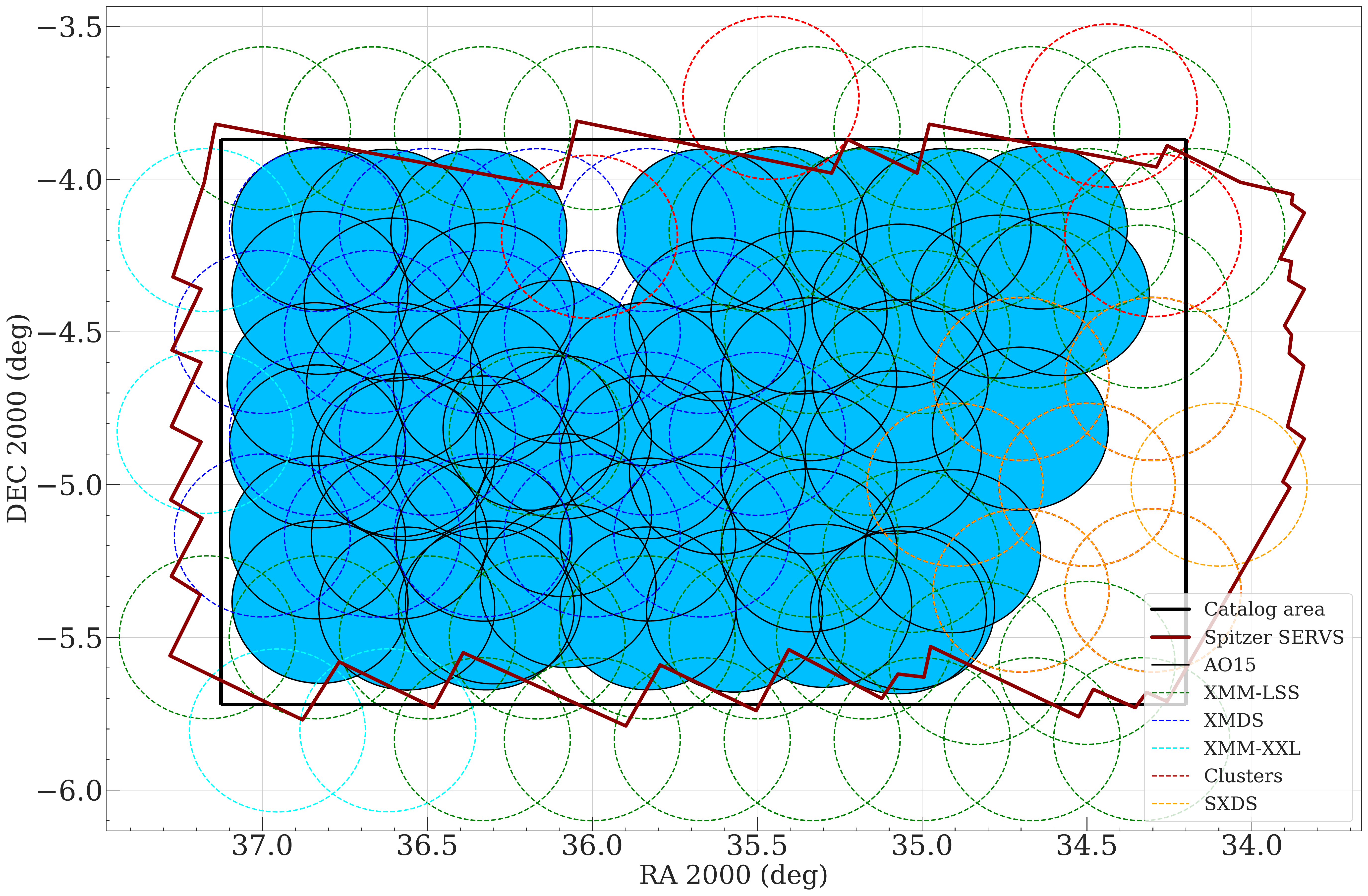}
    \vspace{-0.5cm}
    \caption{Locations of the {\it XMM-Newton} observations used in this work. 
    The AO-15 observations are marked as the blue-filled circles with solid boundaries. 
    The archival observations are marked as dashed circles. 
    Circles with green, orange, blue, and cyan colors are for XMM-LSS, SXDS, XMDS, and XMM-XXL observations, respectively. 
    The RA/DEC range of our catalog selection area is indicated by the black rectangle, 
    and the {\it Spitzer} SERVS footprint is marked as the dark-red polygon.
    Our AO-15 observations do not cover the entirety of the SERVS region,
    because the existing data from SXDS (bottom-right corner, orange circles) and from deep \hbox{X-ray} cluster observations 
    (top-middle and top-right, the red circles) reached the desired depth. 
    }
    \label{fig:cell}
\end{figure*}

In addition to the \xmm\ data, the XMM-LSS region has extensive
multiwavelength coverage (see Table~\ref{tab:servsmw} for a summary, also see \citealt{cathelp}). 
In particular, the central $\approx 5$ deg$^2$ area of the XMM-XXL-North field 
(i.e., the combination of the XMDS and SXDS fields, 
see Fig.~\ref{fig:maps} for an illustration of the relative positions of different surveys.)  
was selected to be one of the SERVS fields. 
This sky region is covered uniformly by 
multiple photometric and spectroscopic surveys (see Sec.~\ref{sec:mw} for more details),
and it is one of the deep drilling fields of the Dark Energy Survey \citep{dieh14} 
and the upcoming Large Synoptic Survey Telescope (LSST) surveys (see Table~\ref{tab:servsmw}).
However, compared to the relatively uniform multiwavelength data,
archival \xmm\ observations covering this sky region 
span a wide range of exposure time (see Table~\ref{tab:xmmdata}).
In order to advance studies of accreting SMBHs and their environments,
deep \hbox{X-ray} observations with similar areal coverage are  
required in addition to the rich multiwavelength data in this field.
To this end, we obtained \xmm\ AO-15 observations taken between July 2016 and February 2017 with a total of 1.3~Ms exposure time. The relative sky coverage of our survey region, XMM-XXL-North, and SXDS are displayed in Fig~\ref{fig:maps}. 
Our AO-15 data include 67 {\it XMM-Newton} observations. 
All of these 67 observations were carried out with a THIN filter for the EPIC cameras.
The choice of the THIN filter maximizes the signal-to-noise ratio. Since the XMM-LSS field is far from the Galactic plane and thus the number of bright stars is small, the optical loading effects are negligible for almost all detected \hbox{X-ray} sources. Even for the brightest star in XMM-LSS, HD 14417, the optical loading effects are only limited to a few pixels at its position.
In addition to the new data, 
we made use of all the overlapping archival \xmm\ observations
to create a uniform, sensitive {\it XMM-Newton} survey contiguously covering most of the SERVS data in the XMM-LSS region.
After excluding observations that were completely lost due to flaring background (see \S\ref{subsec:dataprep}), 
the archival data used here include 51 observations culled from the $10$~ks XMM-LSS survey,
18 observations from XMDS with $20-25$~ks exposures, four mosaic-mode observations\footnote{Each mosaic-mode observation is comprised of a number of 10~ks exposures, see \url{https://xmm-tools.cosmos.esa.int/external/xmm_user_support/documentation/uhb/mosaic.html}.} obtained as part of the XMM-XXL survey \citep{catxxl1}, 
four archival {\it XMM-Newton} observations targeting galaxy clusters identified in the XMM-XXL-North and XMM-LSS surveys 
($\approx 30-100$~ks), 
and the ten $50$~ks observations from SXDS.
We present the details of each observation in Table~\ref{tab:xmmdata},
and show the positions of each \xmm\ observation used in this work in Fig.~\ref{fig:cell}.

Our AO-15 observations were separated into two epochs to minimize the
effects of background flaring. We first observed the XMM-LSS sky region in the SERVS footprint 
with $\approx 1$~Ms of {\it XMM-Newton} exposure time during July--August 2016. 
These first observations were screened for flaring backgrounds (\S\ref{subsec:dataprep}); we then 
re-observed the background-contaminated sky regions using the remaining 0.3~Ms. 
We also observed the SXDS region in which one of the SXDS observations
carried out in 2002 was severely affected by background flares. 
In this work, we present an \hbox{X-ray} source catalog obtained from a 5.3~deg$^2$ sky-region with $34.2^{\circ} \leq \alpha_{\rm J2000} \leq 37.125^{\circ}$ and 
$-5.72^{\circ} \leq \delta_{\rm J2000} \leq -3.87^{\circ}$\footnote{This is equivalent to 
the Galactic coordinates $170.25184^{\circ} < l < 172.07153^{\circ}$, $-60.49169^{\circ} < b < -57.17011^{\circ}$.
} 
(black rectangle in Fig~\ref{fig:maps} and Fig~\ref{fig:cell}). 
The sky region is primarily selected by the footprint of our AO-15 observations, with additional SXDS data within the SERVS footprint in the south-west corner. 
A total of $3.0$~Ms of raw {\it XMM-Newton} observations are used for generating the \hbox{X-ray} source catalog.

In addition to the \xmm\ data, there are also a number of {\it Chandra} observations 
in our source-search region, including 18 observations of 10--90~ks exposure depth following up \hbox{X-ray} galaxy clusters identified in the XMM-LSS and XMM-XXL surveys 
(PIs: 
Andreon, S.;
Jones, L.;
Mantz, A.;
Maughan, B.;
Murray, S.; 
Pierre, M.); these observations occupy a wide RA/DEC range in our catalog region.
In \S\ref{subsec:lrmatching} and \S\ref{subsec:matchingcheck}, we make use of the {\it Chandra} sources 
in these observations culled from the \chandra\ Source Catalog 2.0 \citep[CSC 2.0;][]{catcsc}.\footnote{We use the CSC Preliminary Detections List \url{http://cxc.harvard.edu/csc2/pd2/}.}
There are a total of 328 \hbox{{\it Chandra}} sources 
from CSC 2.0 in our survey region. Note that the source-flux information is not yet available for the CSC 2.0 Preliminary Detections List. Of these 328 {\it Chandra} sources, 201 of them are in CSC 1.1 \citep{catcsc}.
Their 0.5--7 keV band fluxes range from $3\times10^{-16}-1.7 \times 10^{-13}$ erg~cm$^{-2}$~s$^{-1}$, with a median value of $9.7\times 10^{-15}$ erg~cm$^{-2}$~s$^{-1}$.
We use these {\it Chandra} sources as a means to improve and assess the multiwavelength counterpart identification reliabilities, since {\it Chandra} has better angular resolution and astrometric accuracy than those of {\it XMM-Newton}.

\subsection{Data preparation and background-flare filtering}\label{subsec:dataprep}

\begin{figure*}
    \includegraphics[width=0.53\textwidth]{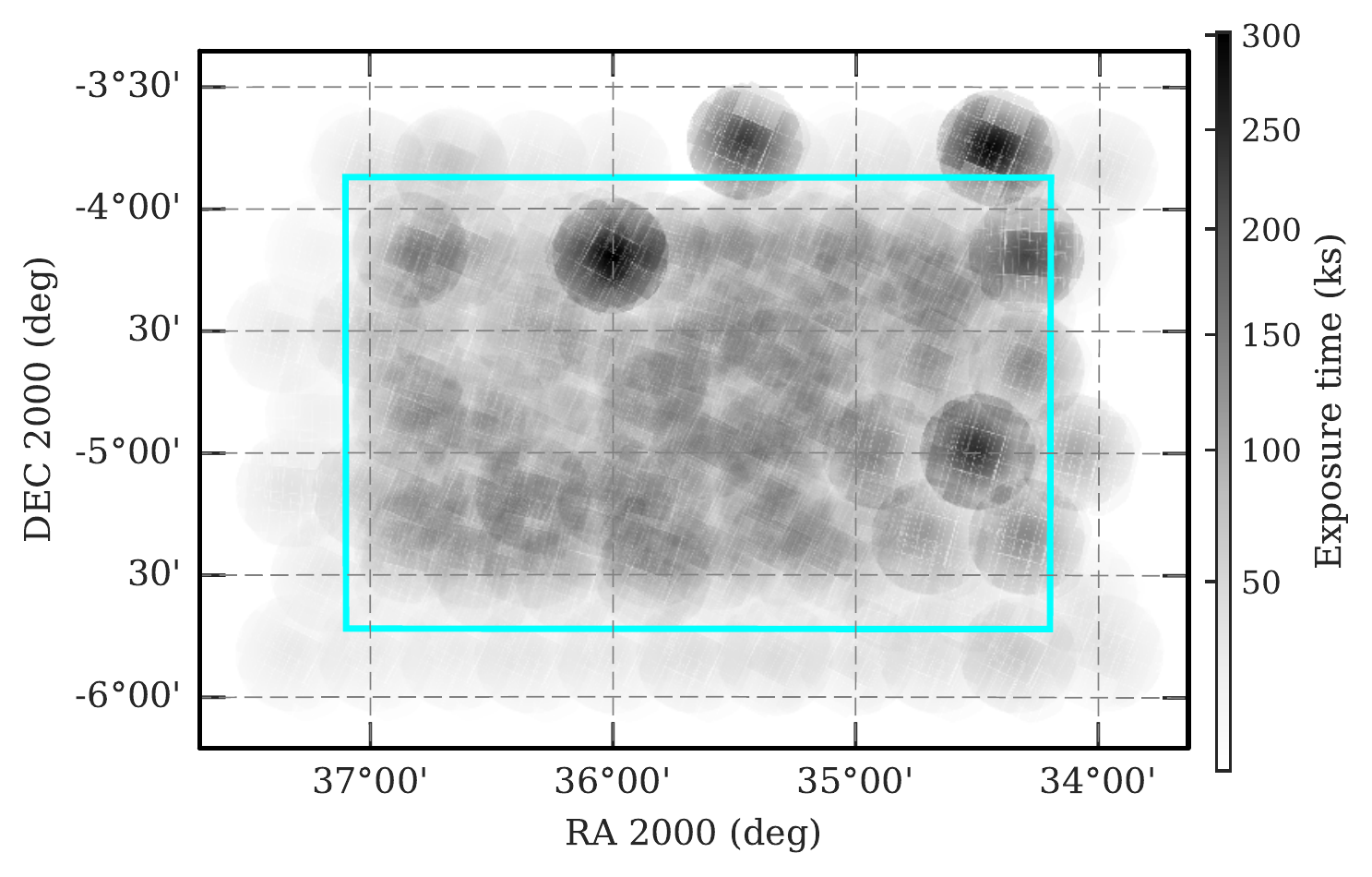}
    \includegraphics[width=0.44\textwidth]{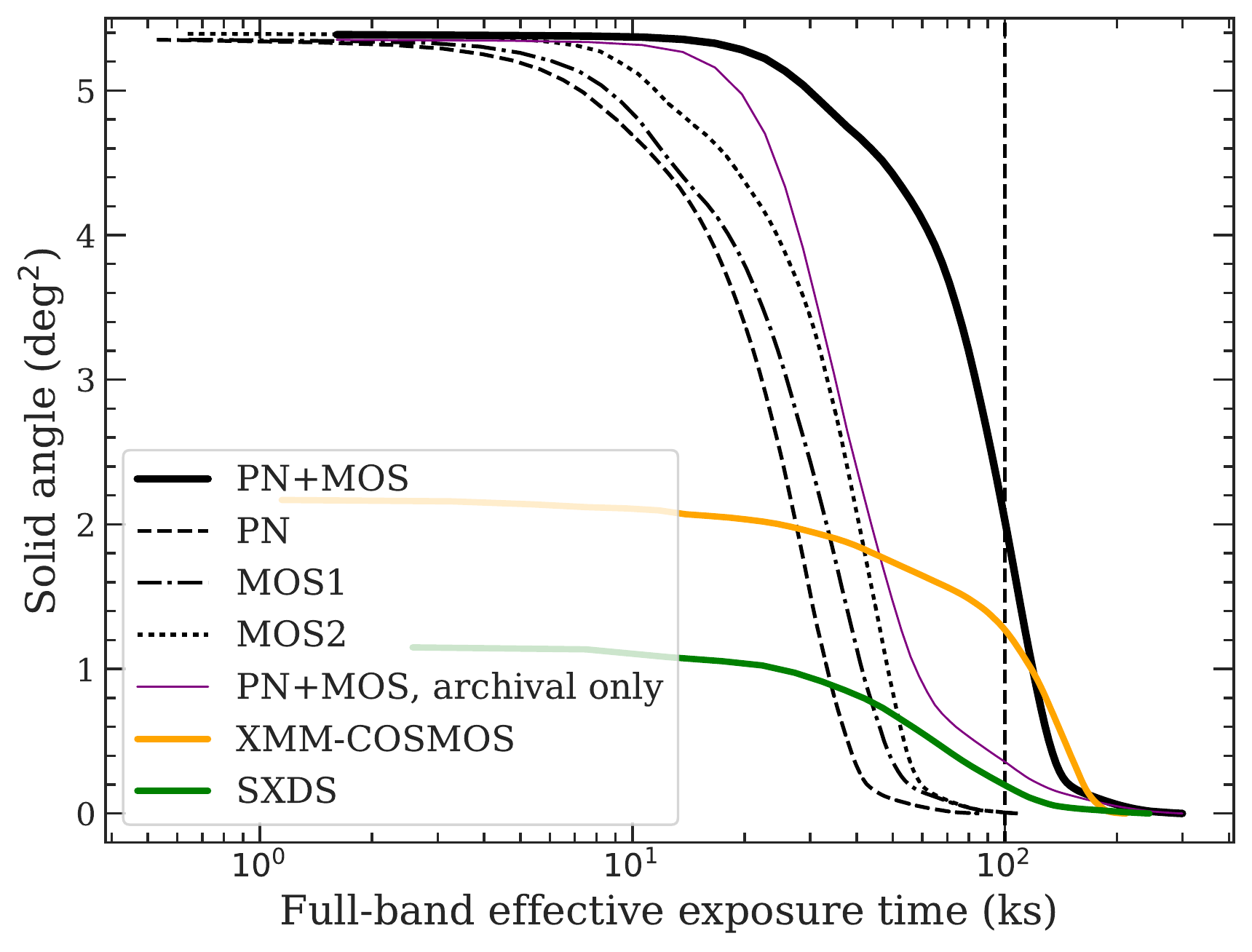}
    \caption{
    {\it Left} -- Full-band survey effective exposure map (PN + MOS). The 5.3 deg$^2$ survey region from which the \hbox{X-ray} source catalog is constructed is marked as the cyan rectangular box. Except for several regions with deep {\it XMM-Newton} follow-up observations of galaxy clusters, the {\it XMM-Newton} coverage in our survey region is generally uniform.
    {\it Right} -- The black solid line shows the cumulative survey solid angle as a function of full-band effective (i.e., vignetted) PN+MOS exposure for observations used in this work. Distributions for individual instruments are indicated as the dashed line (PN), dash-dotted line (MOS1), and dotted line (MOS2).
    For comparison, the cumulative survey solid angle 
    for the archival \xmm\ data in our survey region, in XMM-COSMOS, and in SXDS are shown as the thin purple line, thick orange line, and thick green line, respectively. 
    The dashed vertical line marks exposure time $=100$~ks.
}
    \label{fig:exposure}
\end{figure*}

We use the {\it XMM-Newton} Science Analysis System (SAS) 16.1.0\footnote{\url{https://www.cosmos.esa.int/web/xmm-newton/sas-release-notes-1610}.} and HEASOFT 6.21\footnote{\url{https://heasarc.gsfc.nasa.gov/FTP/software/ftools/release/archive/Release_Notes_6.21}.} for our data analysis.
The {\it XMM-Newton} Observation Data Files (ODFs) were processed with the SAS tasks {\sc epicproc} ({\sc epproc} and {\sc emproc} for PN and MOS, respectively) to create MOS1, MOS2, PN, and PN out-of-time (OOT) event files for each ObsID. 
For observations taken in mosaic mode or with unexpected interruptions due to strong background flares, we use the SAS task {\sc emosaic\_prep} to separate the event files into individual pseudo-exposures and assign pseudo-exposure IDs. 
For the mosaic-mode observations, we also determine the sky coordinates of each pseudo-exposure 
using the AHFRA and AHFDEC values in the attitude files created using the SAS task {\sc atthkgen}. 

For each event file, we create single-event light curves in time bins of 100~s for high (10--12~keV) and low (\hbox{0.3--10}~keV) energies using {\sc evselect} to search for time intervals without significant background flares (the ``good time intervals'', GTIs).
We first remove time intervals with 10--12~keV count rates exceeding $3\sigma$ above the mean, and then repeat the $3\sigma$ clipping procedure for the low-energy light curves. 
Since background flares usually manifest themselves as a high-count-rate tail in addition to the Gaussian-shape count-rate histogram, adopting the $3\sigma$ clipping rule can effectively remove the high-count-rate tail while retaining useful scientific data.
For a small number of event files with intense background flares, we filter the event files using the nominal count-rate thresholds suggested by the {\it XMM-Newton} Science Operations Centre.\footnote{\url{https://www.cosmos.esa.int/web/xmm-newton/sas-thread-epic-filterbackground}} 
We exclude 12 pointings with GTI $<$ $2$~ks from our analysis.
A total of 2.7~Ms (2.3~Ms) of MOS (PN) exposure remains after flare filtering, including 1.1~Ms (0.9~Ms) 
from AO--15 and 1.6~Ms (1.4~Ms) from the archival data.
The flare-filtered median PN exposure time of the full 5.3~deg$^2$ survey region is $\approx 45.8$~ks.
For the central $\approx 4.5$~deg$^2$ region covered by SERVS, the median PN exposure time is $48.5$~ks.
These values were not corrected for vignetting. 

After screening for background flares, we further exclude events in energy ranges that overlap with the instrumental background lines (Al K$\alpha$ lines at 1.45--1.54~keV for MOS and PN, which usually accounts for $\approx 10\%$ of the mean counts\footnote{\url{https://xmm-tools.cosmos.esa.int/external/xmm_user_support/documentation/uhb/epicintbkgd.html}.}; 
Cu lines at 7.2--7.6~keV and 7.8--8.2~keV for PN, which accounts for 30\% of the 2--10~keV counts\footnote{\cite{rana15hatlas}.}). 

From the flare-filtered, instrumental-line-removed event files, we construct images with a 
commonly adopted $4^{\prime\prime}$ pixel size using {\sc evselect} 
in the following bands: \hbox{0.5--2}~keV (soft), \hbox{2--10}~keV (hard), 
and \hbox{0.5--10}~keV (full).
For each image, we generate exposure maps with and without vignetting corrections using the SAS task {\sc eexpmap}. We set {\sc usefastpixelization=0} and {\sc attrebin=0.5} in order to obtain more accurate exposure maps. The exposure maps without vignetting-corrections are only used for generating maps of the instrumental background, which is not affected by vignetting
(see \S\ref{sec:mainx}).
Detector masks were also generated using the SAS task {\sc emask}. 
The distribution of vignetting-corrected exposure values across the 
XMM-LSS field and the PN+MOS1+MOS2 exposure map are presented in Fig.~\ref{fig:exposure}.

\begin{figure*}    
\includegraphics[width=0.45\textwidth]{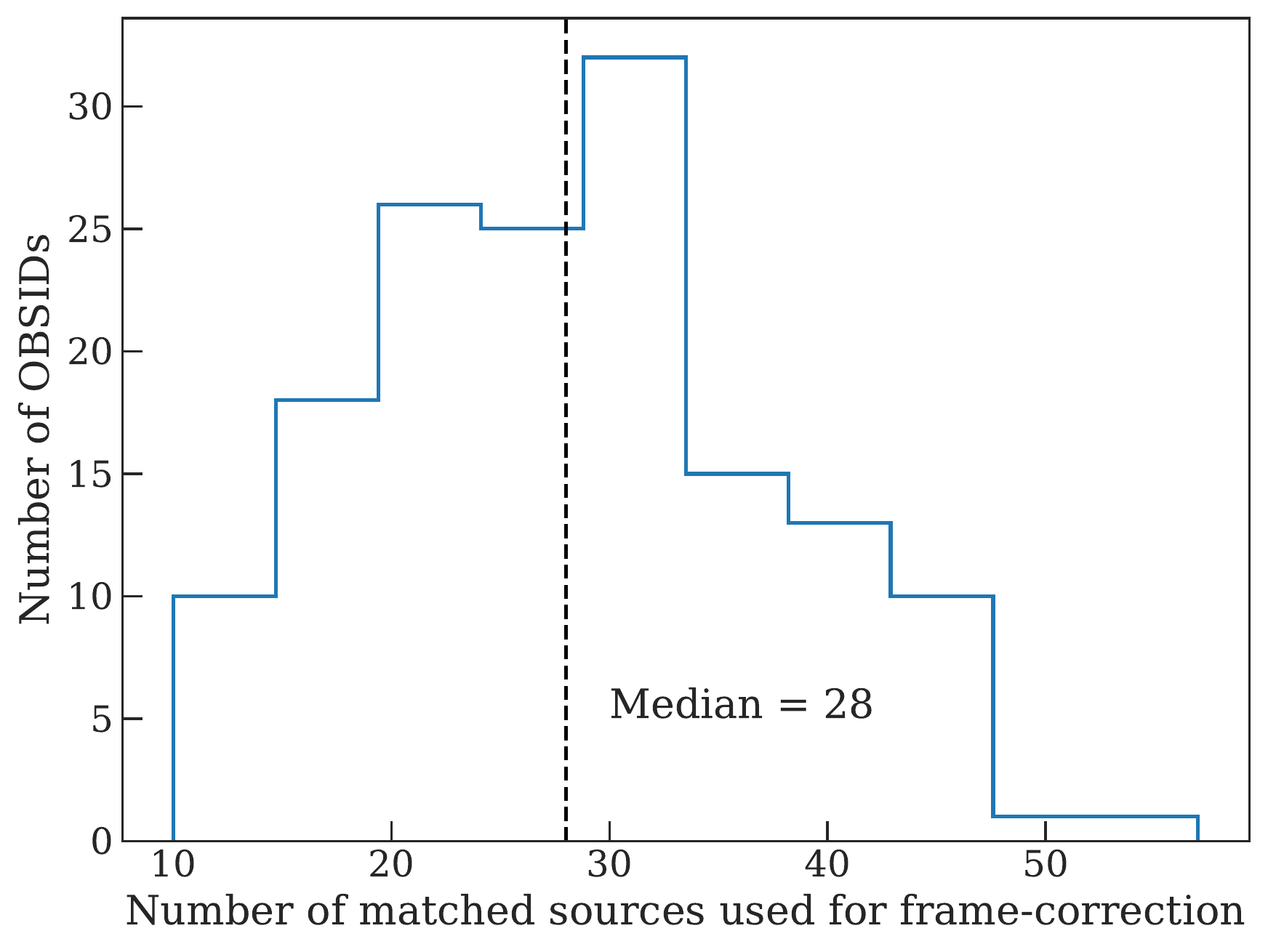}
\includegraphics[width=0.45\textwidth]{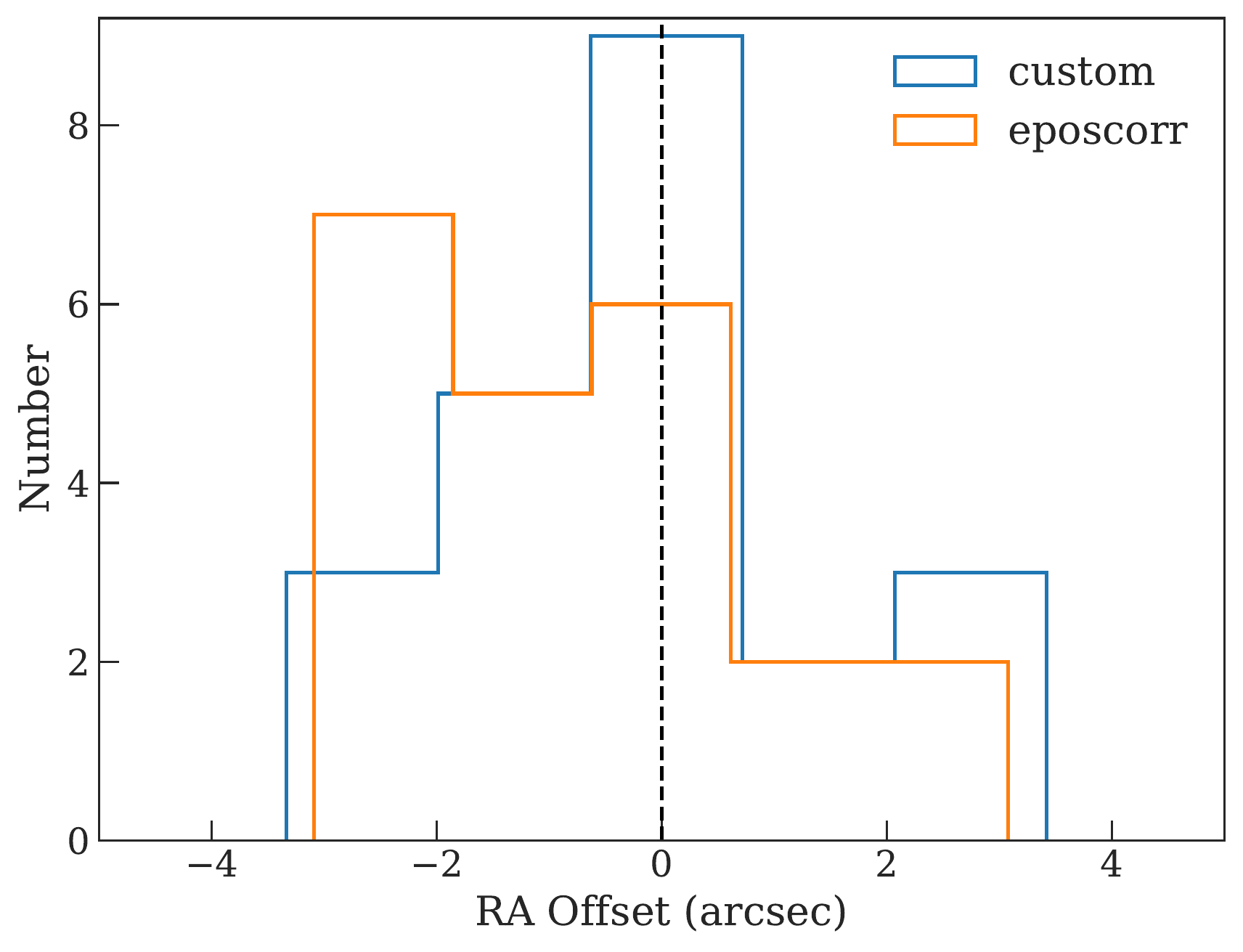}
\caption{
{\it Left}: Distribution of the number of X-ray sources used for finding the required angular offset. The median value is marked as the dashed line. 
{\it Right}:
Distributions of optical-to-X-ray separations in RA for the matched X-ray sources in ObsID 0037982201 
after the astrometric corrections. The results based on our iterative method are shown as the blue histogram, and the results based on the {\sc eposcorr} task are shown as the orange histogram. For the vast majority of the ObsIDs, the difference is small, but some have non-negligible differences and we choose the required astrometric correction based on comparing angular-offset distributions similar to the one shown here.
}
\label{fig:astrometry}
\end{figure*}

\section{The Main \hbox{X-ray} Source Catalog}\label{sec:mainx}

\subsection{First-pass source detection and astrometric correction}\label{subsec:astrometry}

The astrometric accuracy of {\it XMM-Newton} observations can be affected by the pointing uncertainties of {\it XMM-Newton}. This uncertainty is usually smaller than a few arcsec, 
but can be as large as $\approx 10^{\prime\prime}$ \citep[e.g.,][]{catxcosc07,cat2xmm,cat3xmm}. To achieve better astrometric accuracy and to minimize any systematic offsets between different {\it XMM-Newton} observations, we run an initial pass of source detection for each observation and then use the first-pass source list to register the {\it XMM-Newton} observations onto a common WCS frame. The first-pass source detection methods are outlined below: 
\begin{enumerate}[label={(\roman*)}]
    \setlength{\itemindent}{-1ex}
%\enumerate{
\item For the exposures taken by each of the three instruments for each observation, we generate a temporary source list using the SAS task {\sc ewavelet} with a low likelihood threshold ({\sc threshold=4}). {\sc ewavelet} is a wavelet-based algorithm 
that runs on the count-rate image generated using the image and vignetting-corrected exposure map extracted as described in \S\ref{subsec:dataprep}.

\begin{figure*}
\hspace{-0.5cm}
\includegraphics[width=\textwidth]{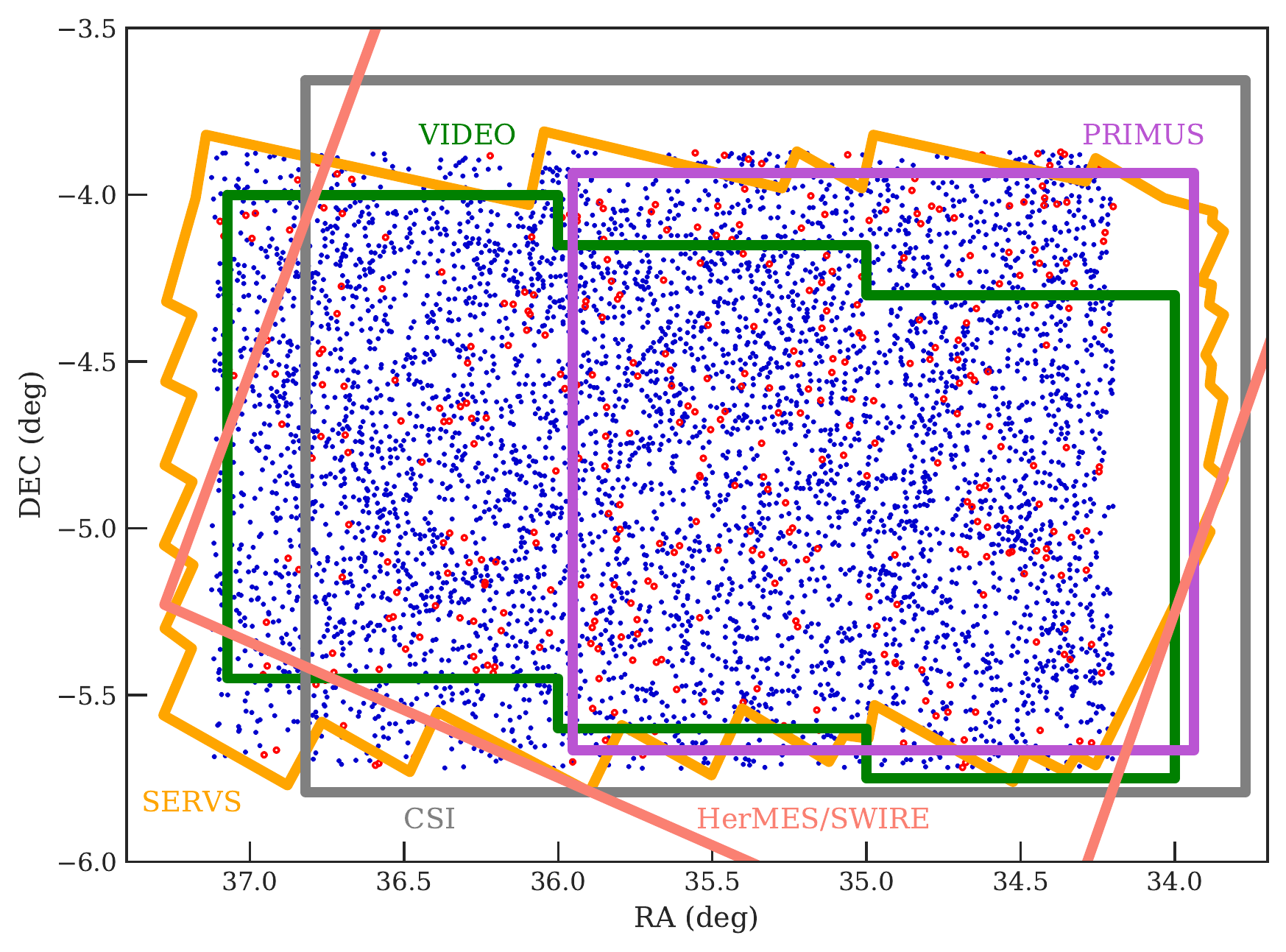}
\vspace{-0.2cm}
\caption{Spatial distribution of the 5242 sources detected in 
this work. We have identified reliable multiwavelength
counterparts (see Sec.~\ref{subsec:lrmatching} and 
Sec.~\ref{subsec:matchingcheck} for details) for 
93\% of the \xmm\ sources (blue dots), while the remaining 
7\% of sources are marked as open red circles. Some of the 
multiwavelength coverage of the XMM-LSS field is also shown
as labeled (see \S\ref{sec:mw} for details).
}
\label{fig:color}
\end{figure*}

\item We use the temporary source list as an input to generate background images 
using the SAS task {\sc esplinemap} with {\sc method=model}. 
This option fits the source-excised image with two templates: the vignetted exposure map, and the un-vignetted exposure map. The former represents the cosmic \hbox{X-ray} background with an astrophysical origin, while the latter represents the intrinsic instrumental noise. {\sc esplinemap} then 
finds the best-fit linear combination of the two templates and 
generates a background map. 
The details of this method are described in \cite{catxcosc07}. 
The background maps are used for the PSF-fitting based source detection task described in Step (iv).
\item We run {\sc ewavelet} again for each observation. This time the source list is generated by running {\sc ewavelet} on the exposure map and image coadded across the PN, MOS1, and MOS2 exposures (when available) with the default likelihood threshold ({\sc threshold=5}). 
\item For each {\sc ewavelet} source list, we use the SAS task {\sc emldetect} to re-assess the detection likelihood and determine the best-fit \hbox{X-ray} positions. {\sc emldetect} is a PSF-fitting tool which performs maximum-likelihood fits to the input source considering the {\it XMM-Newton} PSF, exposure values, and background levels of the input source on each image. {\sc emldetect} also convolves the PSF with a $\beta$-model brightness profile\footnote{\url{http://xmm-tools.cosmos.esa.int/external/sas/current/doc/emldetect/node3.html}.} for clusters and uses the result to determine if the input source is extended. 
Instead of running on the co-added image, {\sc emldetect} takes the image, exposure map, background map, and detector mask of each input observation into account. 
We use a stringent likelihood threshold ({\sc likmin}$=10.8$) to ensure that astrometric corrections are calculated based on real detections, and we only keep the point sources.
\item For the mosaic-mode observations (see Footnote 2), the multiple pointings under the same ObsID were already registered on the same WCS frame of the ObsID. Therefore, we do not correct the astrometry for each pseudo-exposure but only consider the astrometric offsets on an ObsID-by-ObsID basis. The source lists for the mosaic-mode observations were generated using the SAS task {\sc emosaic\_proc}, which is a mosaic-mode wrapper for procedures similar to (i)-(iv) described above. 
\end{enumerate}

For steps (iv) and (v), the source searching was conducted simultaneously on the images of the three EPIC cameras as the astrometric offsets between PN, MOS1, and MOS2 are negligible. For each ObsID, we cross-correlate the high-confidence {\sc emldetect} list of point sources (with the {\sc emldetect} flag EXT$=0$) with the optical source catalog culled from the 
Hyper Suprime-Cam Subaru Strategic Program Public Data Release 1 \citep[HSC-SSP;][]{cathscpdr1}, which is an ultra-deep optical photometric catalog with sub-arcsec angular resolution. The astrometry of HSC-SSP is calibrated to the Pan-STARRS1 $3\pi$ survey and has a $ \lesssim 0.05^{\prime\prime}$ astrometric uncertainty. More details of the HSC-SSP catalog can be found in \cite{cathscpdr1}, and it is also briefly discussed in \S\ref{sec:mw}.
For astrometric corrections, we limit the optical catalog to HSC sources with $i = 18-23$ to minimize possible spurious matches due to large faint source densities at $i > 23$ and matches to bright stars that might have proper motions or parallaxes.

The offset between each ObsID and the HSC catalog is calculated based on 
a maximum-likelihood algorithm similar to the SAS task {\sc eposcorr}. 
The major difference between our approach and {\sc eposcorr} is that we use 
an iterative optimization approach compared to the grid-searching algorithm adopted by {\sc eposcorr}. 
During each iteration, we cross-correlate the optical catalog with the \hbox{X-ray} 
catalog using a $10^{\prime\prime}$ search radius 
and exclude all matches with multiple counterparts (less than $5\%$ of our \hbox{X-ray} sources 
have more than one optical counterpart in the bright HSC-SSP catalog). 
The $10^{\prime\prime}$ search radius is motivated by both the positional accuracy and PSF size of {\it XMM-Newton}, and the largest separations between the \xmm\ and \chandra\ 
positions of the sources in the \chandra\ COSMOS Legacy Survey \citep{catcosmosm16}.
We then calculate the required astrometric corrections that maximize the cross-correlation likelihood.
After each iteration, we apply the best-fit astrometric offsets to the source list and 
next repeat the catalog cross-correlation steps and re-calculate the required additional corrections 
for the source list. 
The required astrometric corrections usually converge after \hbox{1--2} iterations.
For the purpose of frame correction, we adopt the \hbox{X-ray} positional uncertainties calculated based on the PSF-fitting likelihood ratios provided by {\sc emldetect} ($\sigma_{eml}$ hereafter). 
The positional uncertainty information is necessary because the required astrometric corrections 
should be weighted toward \hbox{X-ray} sources with better positions within each observation.
To avoid over-weighting sources with extremely small $\sigma_{eml}$, we also include a constant $0.5^{\prime\prime}$ systematic uncertainty when calculating the best-fit values for frame-correction.\footnote{We assume the systematic uncertainties to be $0.5^{\prime\prime}$ as suggested by \cite{cat2xmm}.} 
The median number of \hbox{X-ray} sources in an ObsID with only one HSC counterpart within $3^{\prime\prime}$ is 28.
See Fig.~\ref{fig:astrometry}-left for a histogram of the number of X-ray sources used for determining the required angular offsets.

The required frame-correction offsets calculated using our approach are less than $3^{\prime\prime}$ in both RA and DEC and are generally consistent with the results calculated using {\sc eposcorr}, with a median difference of $0.1^{\prime\prime}$. 
For demonstration purposes, we show the difference between our RA offsets and the {\sc eposcorr} RA offsets for ObsID 0037982201 in Fig~\ref{fig:astrometry}-right.
For two ObsIDs the difference between our offsets and the {\sc eposcorr} offsets are non-negligible ($> 0.5^{\prime\prime}$). 
We visually inspect the \hbox{X-ray} to optical angular offsets similar to the one shown in 
Fig~\ref{fig:astrometry}-right of these ObsIDs and conclude that our approach 
does improve the alignments between the optical and corrected \hbox{X-ray} images.
The event files and the attitude file for each ObsID are then projected onto the WCS frame of the HSC catalog by updating the relevant keywords using a modified version of {\it Chandra}'s
{\sc align\_evt} routine \citep[][]{rana13}. 
Since the sky coordinates for the event files of the mosaic-mode pseudo-pointings are derived based on the reference point centered at the nominal RA and DEC positions of the mosaic-mode ObsIDs, we also recalculate the sky coordinates for these event files with the SAS task {\sc attcalc} using the true pointing positions as the reference point, which is necessary for using regular SAS tasks for mosaic-mode pseudo-exposures.

\subsection{Second-pass source detection}\label{subsec:secondpass}
We re-create images, exposure maps, detector masks, and background maps using the frame-corrected event files and attitude files. 
We then run source-detection tasks for the second time considering all {\it XMM-Newton} observations listed in 
Table~\ref{tab:xmmdata}.
Similar to the approach used for the XMM-H-ATLAS survey \citep{rana15hatlas}, 
we divide the XMM-LSS field into a grid when running the second-pass source detection
because the number of images that can be processed by a single {\sc emldetect} thread is limited. 
We use a custom-built wrapper of relevant SAS tasks to carry out the
second-pass source detection, which is similar to the 
{\sc griddetect}\footnote{\url{https://github.com/piero-ranalli/griddetect}.} tool 
built for the XMM-H-ATLAS survey \citep{rana15hatlas}.

The cell sizes of the grid are determined by the number of {\sc ewavelet} sources. 
For each cell in the grid, we co-add the images and exposure maps for all observations with footprint inside the cell and run {\sc ewavelet} with a low detection threshold\footnote{{\sc threshold=4}.} on the co-added image and exposure map. 
For each cell, we only keep {\sc ewavelet} sources within the RA/DEC range of the cell plus 1$^\prime$ ``padding'' on each side of the cell.
We then use the {\sc ewavelet} list as an input for {\sc emldetect} to assess the detection likelihood. 
The {\sc emldetect} point-source list of the full XMM-LSS region is constructed from the union of the sources from all cells
after removing duplicates due to the ``padding''. 
We search for sources in three different bands: 
\hbox{0.5--2}~keV (soft), \hbox{2--10}~keV (hard), and \hbox{0.5--10}~keV (full). 
For each source, {\sc emldetect} computes a detection likelihood {\sc det\_ml},
which is defined as {\sc det\_ml}$= -\ln P$, where $P$ is the 
probability of a detected source being a random Poisson fluctuation of the background.
In practice, 
the spurious fractions of a source catalog derived 
based on simulations are known to 
differ from the values obtained with the simple {\sc det\_ml}$= -\ln P$ equation \citep[e.g.,][]{catxcosc07,capp09xcos,catsxdf,cat2xmm,lama16stripe82}. 
Since the source catalog is constructed based on a 
complex multi-stage source-detection approach, the relation between 
{\sc det\_ml} and the true spurious fraction may not be as straightforward as 
the simple {\sc det\_ml}$= -\ln P$ equation, especially 
in the low source count regime where even this simple relation fails.\footnote{See \url{http://xmm-tools.cosmos.esa.int/external/sas/current/doc/emldetect.pdf}.} 
Therefore, we do not adopt a single {\sc det\_ml} value for our source catalog. 
Instead, we use the {\sc det\_ml} value corresponding 
to the 1\% spurious fraction determined by simulations for each band (see the next subsection, 
\S\ref{subsec:simulation}, for details).
The {\sc det\_ml} thresholds with 1\% spurious fraction are 4.8, 7.8, and 6.2 for the soft, hard, and full bands, respectively. 
A total of 5242 sources satisfy this criterion in at least one of the three bands (see \S\ref{subsec:mainxcat}). 
We show the spatial distribution of the 5242 detected sources in Fig.~\ref{fig:color}.

\subsection{Monte Carlo simulations}\label{subsec:simulation}

\begin{figure*}    
\includegraphics[width=0.45\textwidth]{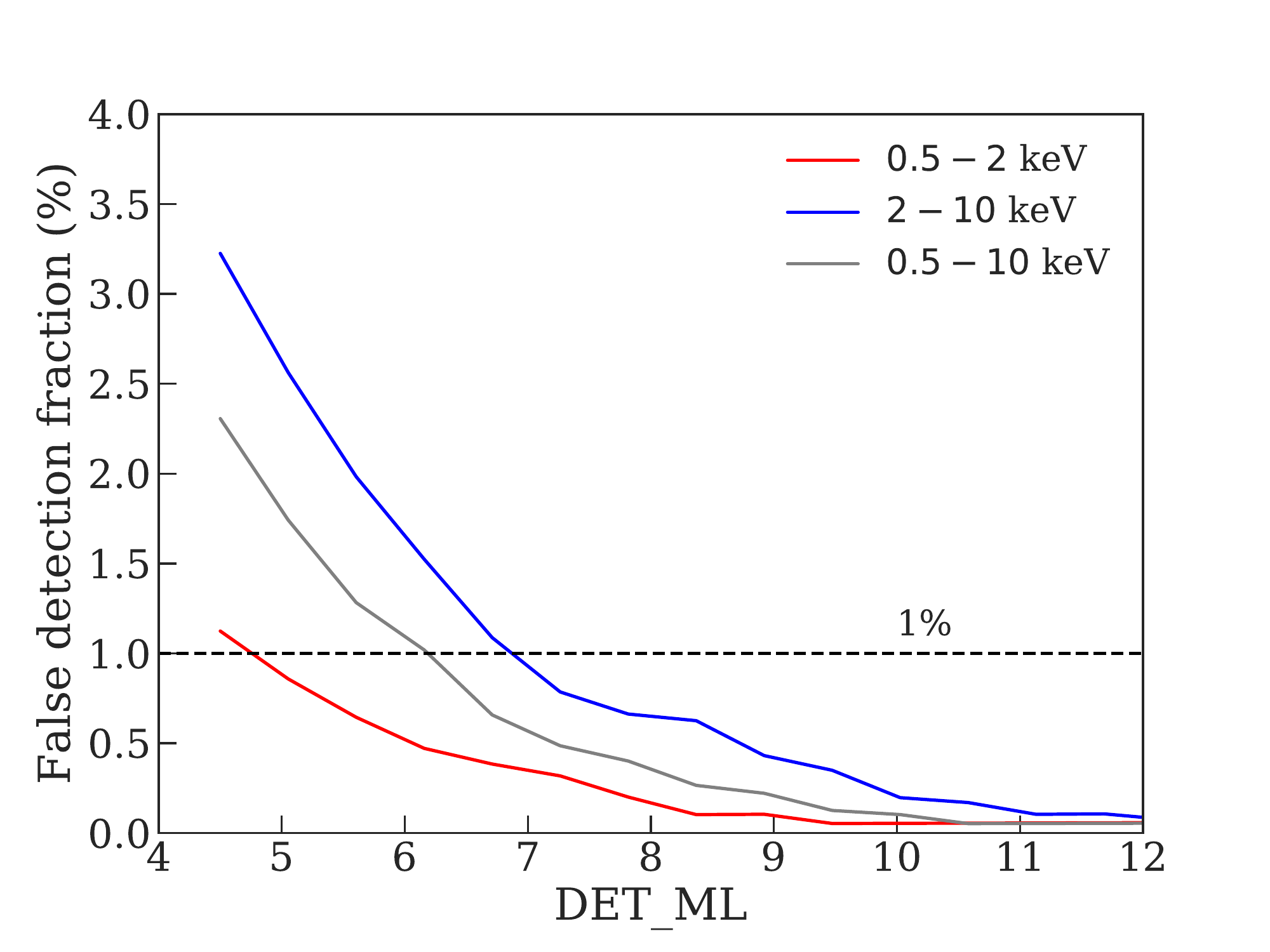}
\includegraphics[width=0.45\textwidth]{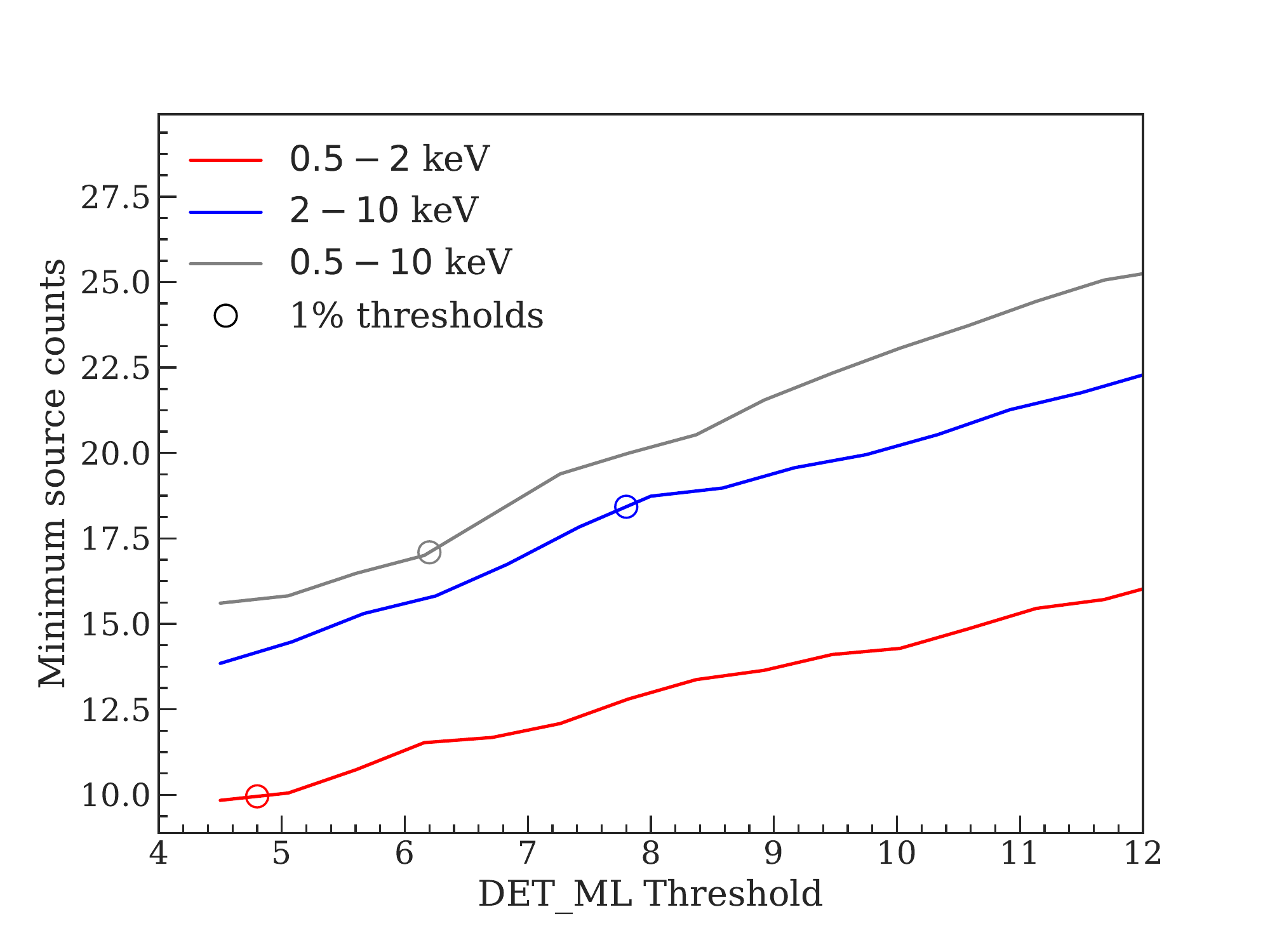}
\caption{
{\it Left}: The fraction of spurious sources detected at different {\sc det\_ml} based on simulations.
The detection threshold relevant to our catalog is marked as the horizontal dashed line.
{\it Right}: 
The minimum source counts for the detected sources (the median of all 20 simulations) with {\sc det\_ml} values above a given {\sc det\_ml} threshold. As expected, higher {\sc det\_ml} thresholds can only detect sources with higher numbers of counts. The {\sc det\_ml} values corresponding to the $1\%$ spurious fraction are marked as the open circles. 
}
\label{fig:detml}
\end{figure*}

To assess our survey sensitivity and catalog reliability, we perform Monte Carlo simulations of \hbox{X-ray} observations. 
For each simulation, we generate a list of mock \hbox{X-ray} sources by sampling from the $\log N-\log S$ 
relations reported in the XMM-COSMOS survey \citep[][for the \hbox{0.5--2}~keV and \hbox{2--10}~keV bands]{capp09xcos} 
and the {\it Chandra} Multiwavelength Project survey \citep[ChaMP;][for the \hbox{0.5--10}~keV band]{catchamp}. 
The maximum flux of the mock \hbox{X-ray} catalogs is set at $10^{-11}$~erg~cm$^{-2}$~s$^{-1}$. 
The minimum flux of the mock \hbox{X-ray} sources at each energy band is 
set as 0.5 dex lower than the minimum detected flux \citep[e.g.,][]{lama16stripe82}.
We randomly place the mock \hbox{X-ray} sources in the RA/DEC range 
covered by the {\it XMM-Newton} observations used in this work. 
We then use a modified version of the simulator written for the \xmm\ survey of the CDF-S \citep{rana13}, {\sc CDFS-SIM},\footnote{
\url{https://github.com/piero-ranalli/cdfs-sim}} to create mock event files. 
{\sc CDFS-SIM} converts \hbox{X-ray} fluxes to PN and MOS count rates with the same model used for deriving the ECFs, 
and it then randomly places \hbox{X-ray} events around the source location according to the count rates, the {\it XMM-Newton} PSFs at the given off-axis angle, and the real exposure maps. We extract images from the simulated event files using the same methods described in \S\ref{sec:mainx}. For each observation, the simulated image is combined with a simulated background, which is created by re-sampling the original background map according to Poisson distributions to create simulated images that mimic the real observations. For each energy band, a total of 20 simulations are created. 
We run the same two-stage source-detection procedures described in \S\ref{subsec:secondpass} on the simulated data products. For each simulation, we match the detected sources to the input sources within a $10^{\prime\prime}$ cut-off radius by minimizing the quantity $R^2$ \citep[Eq. 4 of][]{capp09xcos}:

\begin{equation}
\label{eq:rsq}
R^2 = \big(\frac{\Delta{\rm RA}}{\sigma_{\rm RA}} \big)^2 +  \big(\frac{\Delta{\rm DEC}}{\sigma_{\rm DEC}} \big)^2 + 
 \big(\frac{\Delta{\rm RATE}}{\sigma_{\rm RATE}} \big)^2.
\end{equation}
Here $\Delta{\rm RA}$ and $\Delta{\rm DEC}$ 
are the differences between the simulated RA/DEC positions 
and the RA/DEC positions obtained by running source detection on the simulated images. 
$\Delta{\rm RATE}$ is the difference between the simulated count rates
and the detected count rates. $\sigma_{\rm RA}$, $\sigma_{\rm DEC}$, and $\sigma_{\rm RATE}$ 
are the uncertainties of RA, DEC, and count rates of the detected sources.
Minimizing $R^2$ takes into account the flux and positional differences between the input catalog 
and the sources detected in the simulated images \citep[e.g.,][]{catxcosc07,rana15hatlas}. 
Detected sources without any input sources within the $10^{\prime\prime}$ radius are considered 
to be spurious detections. 

Fig.~\ref{fig:detml}-left presents the spurious fraction ($f_{\rm spurious}$) as a function of {\sc det\_ml} for the soft, hard, and full bands. For our catalog, we consider sources with 
$f_{\rm spurious}$ less than 1\% to be reliably detected. 
At this threshold, the corresponding {\sc det\_ml} values are 
4.8, 7.8, and 6.2 for the soft, hard, and full bands, respectively. 
The difference between the {\sc det\_ml} thresholds in the three bands are 
likely due to their different background levels.
For the full \hbox{X-ray} source catalog of 5242 sources, 
the $f_{\rm spurious}=1\%$ criterion translates to $\approx 52$ spurious detections.
For each source, we have also calculated a detection reliability parameter (defined as $1 - f_{\rm spurious}$)
for each band using the simulation results presented in Fig.~\ref{fig:detml}-left, which can be used for 
selecting sources with a desired reliability. 
We also display the minimum detected source counts (the median values of all 20 simulations) 
as a function of the {\sc det\_ml} threshold in Fig.\ref{fig:detml}-right. 
We test for source confusion following the methods described in \cite{hass98} and \cite{catxcosc07}.
For all the simulated sources that are detected (i.e., having {\sc det\_ml} values greater than the 1\% thresholds), we consider sources with 
observed fluxes ($S_{\rm out}$) that are larger than the simulated fluxes ($S_{\rm in}$)
by the following threshold to be ``confused'' sources:
$S_{\rm out}/(S_{\rm in} + 3\times S_{\rm out}^{\rm Err}) > 1.5$.
Here $S_{\rm out}^{\rm Err}$ is the statistical fluctuation of the observed fluxes.
The source confusion fractions are 0.14\%, 0.16\%, and 0.43\% in the soft, hard, and full bands, respectively. For the 5242 \hbox{X-ray} sources in this catalog, these fractions translate to $\approx 7-22$ sources with confusion.

\subsection{Astrometric accuracy}\label{subsec:poserr}
\begin{figure}
\includegraphics[width=0.45\textwidth]{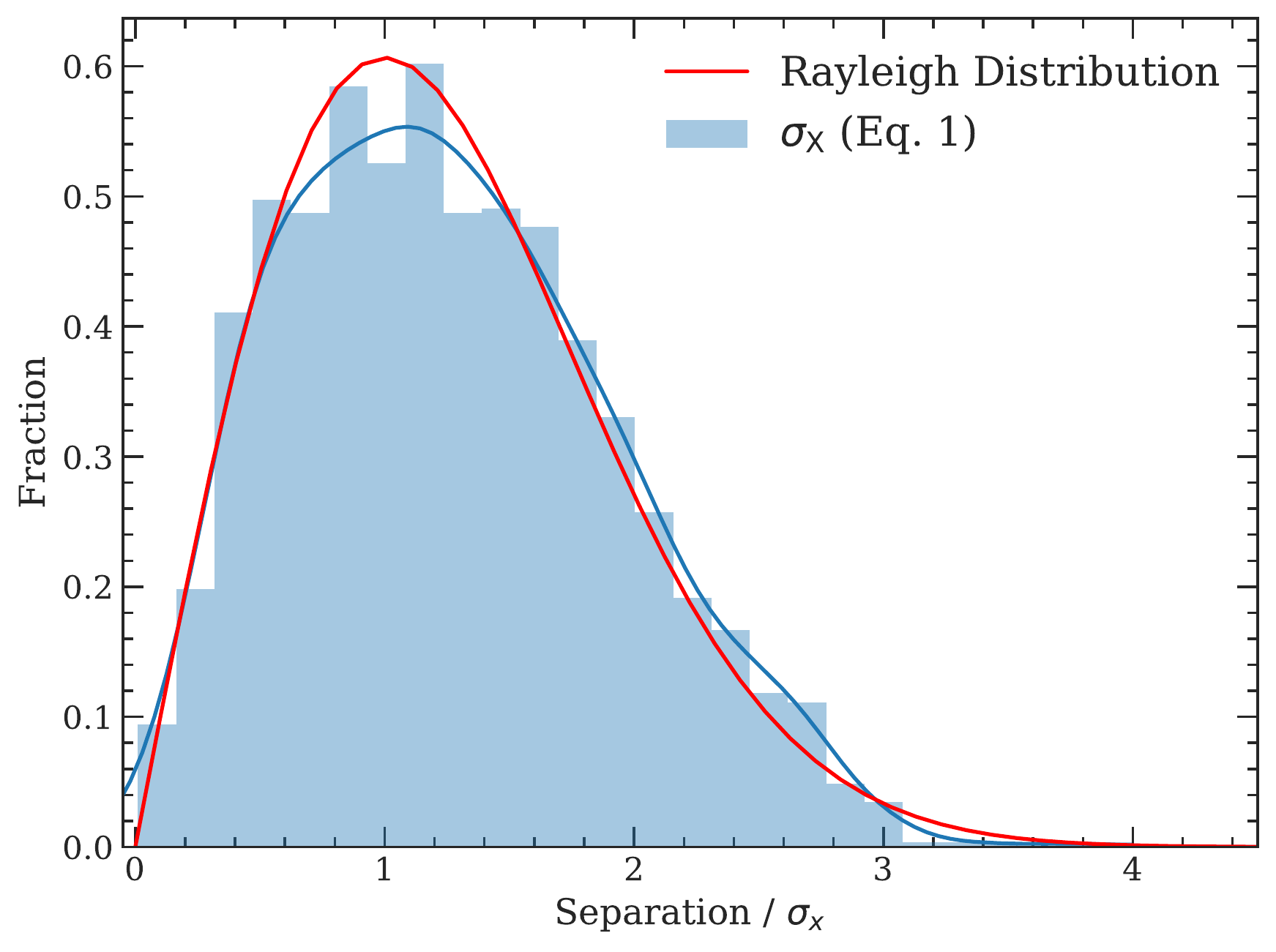}
\caption{
Histogram of the normalized full-band positional offsets, a dimensionless quantity
defined as the positional offsets normalized by the empirically derived positional uncertainty,
and comparison with the expected Rayleigh distribution, the solid red curve. 
The kernel-density estimation of the normalized positional offset distribution is shown as the solid blue curve.
The excellent agreement between the two distributions suggests that our empirically derived $\sigma_x$ values are reliable indicators of the true positional uncertainties.
}
\label{fig:rayleigh}
\end{figure}

We investigate the positional accuracy of the {\it XMM-Newton} sources by comparing the second-pass \hbox{X-ray} catalog with the HSC-SSP catalog. Similar to the frame-correction procedures described in \S\ref{subsec:astrometry}, we search for unique optical counterparts around the \hbox{X-ray} positions using a $3^{\prime\prime}$ search radius. 
For the 5199 \hbox{X-ray} sources detected in the full-band during the second-pass source-searching process, 
a total of 2434 \hbox{X-ray} sources are found to have only one $i = 18 - 23$ HSC counterpart within $3^{\prime\prime}$. 
We use the separations between the optical and \hbox{X-ray} positions of this subsample as a means to determine empirical \hbox{X-ray} positional uncertainties, 
which is a commonly adopted practice in \hbox{X-ray} 
surveys \citep[e.g.][]{cat2xmm,luo10cdfs,catcdfs4ms,xue16,catcdfs7ms}.

The \hbox{X-ray} positional accuracy is determined by how well the PSF-centroid location can
be measured, which usually depends on the number of counts of the detected source and 
the PSF size of the instrument (primarily dependent on the off-axis angle). 
For the vast majority of the \hbox{X-ray} sources presented 
in this work, the detected photons are from at least three different observations,
and hence the dynamical range of effective off-axis angle for 
each source detected on the coadded image is relatively small.
Thus, the \hbox{X-ray} positional uncertainty is mostly dependent on 
the number of counts available for detected sources. 
Using the angular separations between the 2434 \hbox{X-ray} sources and their unique optical counterparts, 
we derive an empirical relation between the number of \hbox{X-ray} counts, 
$C$,\footnote{An upper limit of 2000 is set on $C$ because the improvement of positional 
accuracy is not significant for larger source counts \citep[e.g.,][]{catcdfs7ms}.}
and the 68\% positional-uncertainty radius ($r_{68\%}$) for the full-band-detected \hbox{X-ray} sources, 
$\log_{10} r_{68\%} = -0.31^{+0.02}_{-0.01}\times\log_{10} C + 0.85$. 
The parameters are chosen 
such that $68\%$ of the sources have positional offsets smaller than the empirical relation.

For this work, 
we define the \hbox{X-ray} positional uncertainty, $\sigma_x$, 
to be the same as the uncertainties in RA and DEC where
$\sigma_{\rm RA} = \sigma_{\rm DEC} = \sigma_x$.
Under this definition, $\sigma_x$ is $r_{68\%}$ divided by a factor of $1.515$ 
\citep[e.g., Eq. 21 and \S4.2 of][]{pine17match}. The factor 1.515 is determined 
by integrating the Rayleigh distribution until the cumulative probability reaches 
0.68. For reference, 90\%, 95\%, and 99.73\% uncertainties correspond to 
$2.146\sigma_x$, $2.448\sigma_x$, and $3.439\sigma_x$, respectively. Because the 
separations in both RA and DEC behave as a univariate normal distribution with 
$\sigma_{\rm RA}$ and $\sigma_{\rm DEC}$, respectively,\footnote{Here we consider 
the positional uncertainties of the HSC-SSP catalog to be negligible compared to 
the {\it XMM-Newton} positional uncertainties.} the angular separation should 
therefore follow the joint probability distribution function of the uncertainties 
in the RA and DEC directions. 
Since we assume $\sigma_{\rm RA} = \sigma_{\rm DEC}$, the angular separation 
between an optical source and an \hbox{X-ray} source 
should follow the univariate Rayleigh distribution with the scaling parameter $\sigma_x$, where
$\sigma_x = \sigma_{\rm RA} = \sigma_{\rm DEC}$ \citep[see \S4 of][for details]{pine17match}.

For each energy band, we repeat the same process to find the best-fit relation for $\sigma_x$ using the following equation:
\begin{equation}\label{eq:counts}
\log_{10} \sigma_{x} = \alpha\times\log_{10} C + \beta.
\end{equation}
Given the PSF size and positional accuracy of {\it XMM-Newton}, 
it is possible for \hbox{X-ray} sources to have angular separation from optical sources 
larger than $3^{\prime\prime}$, and the positional uncertainties derived 
based on counterparts found within the $3^{\prime\prime}$ search radius 
can be underestimated. 
Therefore, we adopt an iterative process.
For each iteration, 
we use  the derived $\sigma_x$ to identify reliable matches using the 
likelihood-ratio matching method described in \S\ref{subsec:lrmatching}.
We then re-derive Eq.~\ref{eq:counts} using the reliable matches, 
and the updated astrometric uncertainties are used for running 
likelihood-ratio matching again. 
This is a stable process, as the parameters converge after 2--3 iterations. 
The average positional uncertainties ($\sigma_x$) 
for our soft-band, hard-band, and full-band \hbox{X-ray} catalogs are 1\farcs35, 1\farcs37, 
and 1\farcs31, respectively. 
The standard deviations of the positional uncertainties 
are 0\farcs37, 0\farcs25, and 0\farcs30 for the soft, hard, and full bands, respectively.
Fig.~\ref{fig:rayleigh} presents a comparison of the normalized separation (Separation/$\sigma$) between the full-band \hbox{X-ray} sources and their bright optical counterparts with $\sigma$ derived using Eq.~\ref{eq:counts}, $\sigma_x$. 
%, and the $\sigma$ calculated by {\sc emldetect}, $\sigma_{eml}$.
The agreement between the Rayleigh distribution and the Separation/$\sigma_x$ distribution of our sample 
demonstrates that our empirically derived $\sigma_x$ values are reliable indicators of the true positional uncertainties. 
As for $\sigma_{eml}$, previous studies have reported 
that some on-axis sources with large numbers of counts
can have unrealistically low $\sigma_{eml}$ values, 
therefore an irreducible systematic uncertainty should be added to 
$\sigma_{eml}$ for the normalized separation to follow a Rayleigh distribution \citep[e.g.,][]{cat2xmm}, but the nature of this systematic uncertainty remains unclear.
For this work, we use $\sigma_x$ as the positional uncertainties of our \hbox{X-ray} catalog, but $\sigma_{eml}$ is also included in the final catalog for completeness.

\subsection{The main \hbox{X-ray} source catalog}\label{subsec:mainxcat}

\begin{figure}    
\includegraphics[width=0.45\textwidth]{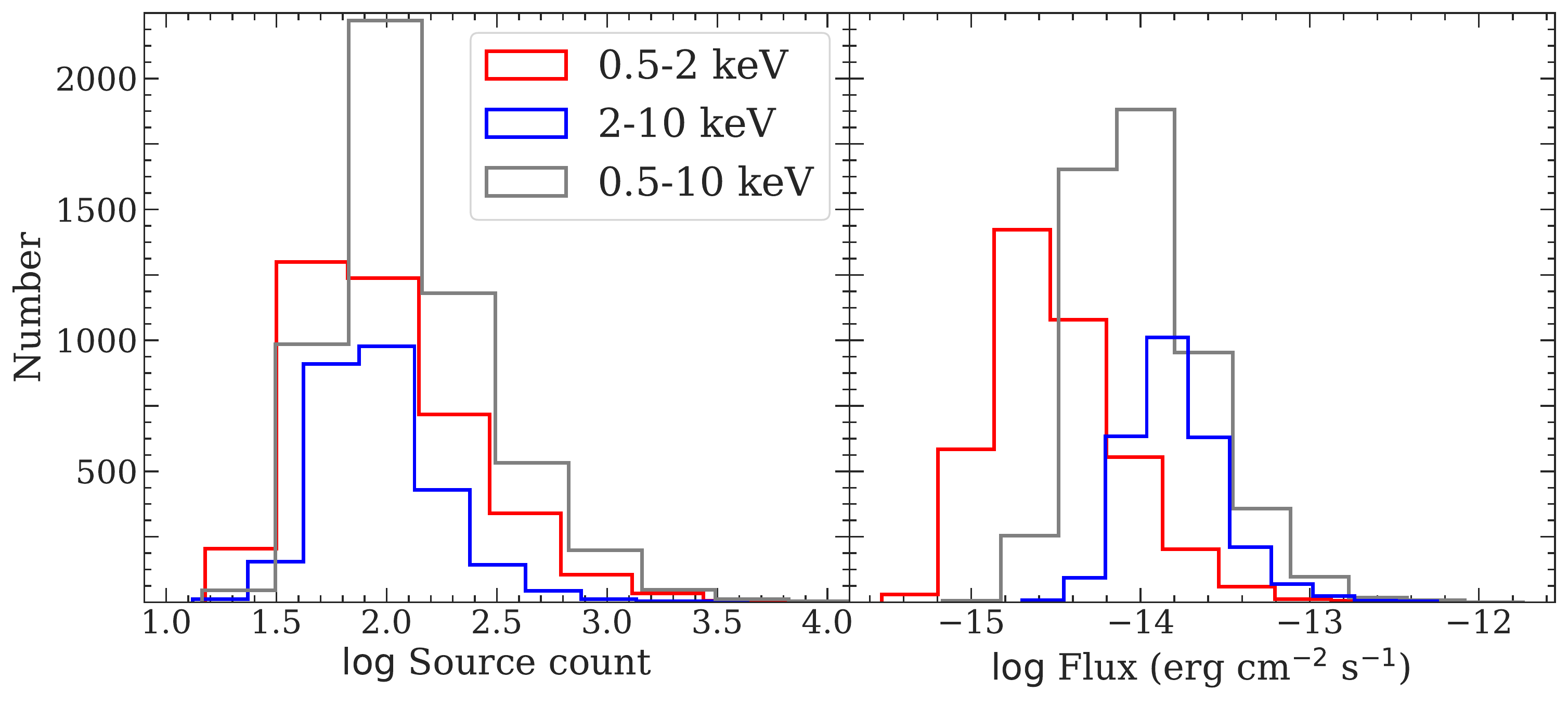}
\caption{
{\it Left} -- Source-count distributions for the sources 
detected in the soft (red), hard (blue), and full (gray) bands.
{\it Right} -- Flux distributions of the sources detected in the three bands. Colors are the same as in the left panel.
}
\label{fig:fluxes}
\end{figure}

We detect  3988, 2618, and 5199 point sources with $f_{\rm spurious} \leq 1\%$ 
in the \hbox{0.5--2}~keV, \hbox{2--10}~keV, and \hbox{0.5--10}~keV bands, respectively.
The details of the main \hbox{X-ray} source catalog are reported in Table~\ref{tab:mainxtab} of Appendix~\ref{sec:catcols}.  
The extended sources (identified by the  {\sc EXT > 0} flag of {\sc emldetect}) are not included, as the properties of the extended \hbox{X-ray} emission are beyond the scope of this work.\footnote{There are 68, 11, and 77 sources identified as {\sc EXT $>$ 0} by {\sc emldetect} in the \hbox{0.5--2}~keV, \hbox{2--10}~keV, and \hbox{0.5--10}~keV bands, respectively.
The properties of the extended sources will be reported in a separate work.
}
We combine catalogs from the three energy bands using a similar approach to that adopted by the {\it XMM-Newton} Serendipitous Source Catalogue.
We consider two sources from different catalogs to be the same if their angular separation is smaller than any of the following quantities:
(1) 10$^{\prime\prime}$, (2) distance to the nearest-neighbor in each catalog, or (3) quadratic sum 
of the $99.73\%$ positional uncertainties from both bands.
The final source catalog is the union of the sources detected in the three energy bands. 
We check for potential duplicate sources by visually inspecting all sources with distance to the nearest-neighbor (DIST\_NN) less than 10$^{\prime\prime}$, 
and only one set of sources is found to be duplicated, resulting in a total of 5242 unique sources. 
There are 2967 sources with more than 100 PN+MOS counts in the full-band, 
and 126 sources with more than 1000 \hbox{X-ray} counts. 
A unique \hbox{X-ray} source ID is assigned to each of the 5242 sources at this stage. 
Visual inspection of the image in each band suggests that no apparent sources were missed by 
our detection algorithm.

We also derive the count rate (vignetting-corrected) to flux energy conversion factors 
(ECFs) assuming a power-law spectrum with photon index $\Gamma=1.7$, 
which is typical for distant \hbox{X-ray} AGNs found in {\it XMM-Newton} 
surveys with comparable sensitivities 
(e.g.,  XMM-COSMOS, \citealt{main07xcos} and XMM-H-ATLAS, \citealt{rana15hatlas})
and Galactic absorption, $N_{\rm H} = 3.57\times10^{20}$ cm$^{-2}$.
The energy ranges are those where the removed instrumental lines are excluded when deriving the ECFs.
Since the archival observations and the AO-15 observations were carried out in different epochs between \hbox{2000--2017}, 
we compute the ECFs by taking the slight temporal variations in the EPIC instrumental calibrations into account.
In detail, we make use of the ``canned'' response files of 14 different epochs for MOS and 3 different epochs for PN available at the \xmm\ SOC website.\footnote{
\url{https://www.cosmos.esa.int/web/xmm-newton/epic-response-files}.}
The effective ECF for each detected source is the exposure-time-weighed average of all relevant observations. 
For all \hbox{X-ray} sources, the mean conversion factors for (PN, MOS1, MOS2) are $(6.23, 1.78, 1.76)$, $(1.15, 0.43, 0.43)$, and $(2.84, 0.88, 0.87)$ counts~s$^{-1}/10^{-11}$erg~cm$^{-2}$~s$^{-1}$, in the \hbox{0.5--2}~keV, \hbox{2--10}~keV, and \hbox{0.5--10}~keV bands, respectively. We note that temporal variations in the ECFs are $<1\%$ for all three bands \citep[e.g.,][]{mate09,cat3xmm}.
For each source detected by {\sc emldetect}, the flux from each EPIC camera is calculated separately using the corresponding ECF. The final flux of the source is the error-weighted mean of the fluxes from the three EPIC cameras, when available.
The median fluxes for the soft, hard, and full bands are 
$2.9\times10^{-15}$, $1.5\times10^{-14}$, and $9.4\times10^{-15}$ erg~cm$^{-2}$~s$^{-1}$, respectively. 
The source-count and flux distributions of the sources detected in the three energy bands are displayed in Fig.~\ref{fig:fluxes}.

\begin{figure}    
\includegraphics[width=0.45\textwidth]{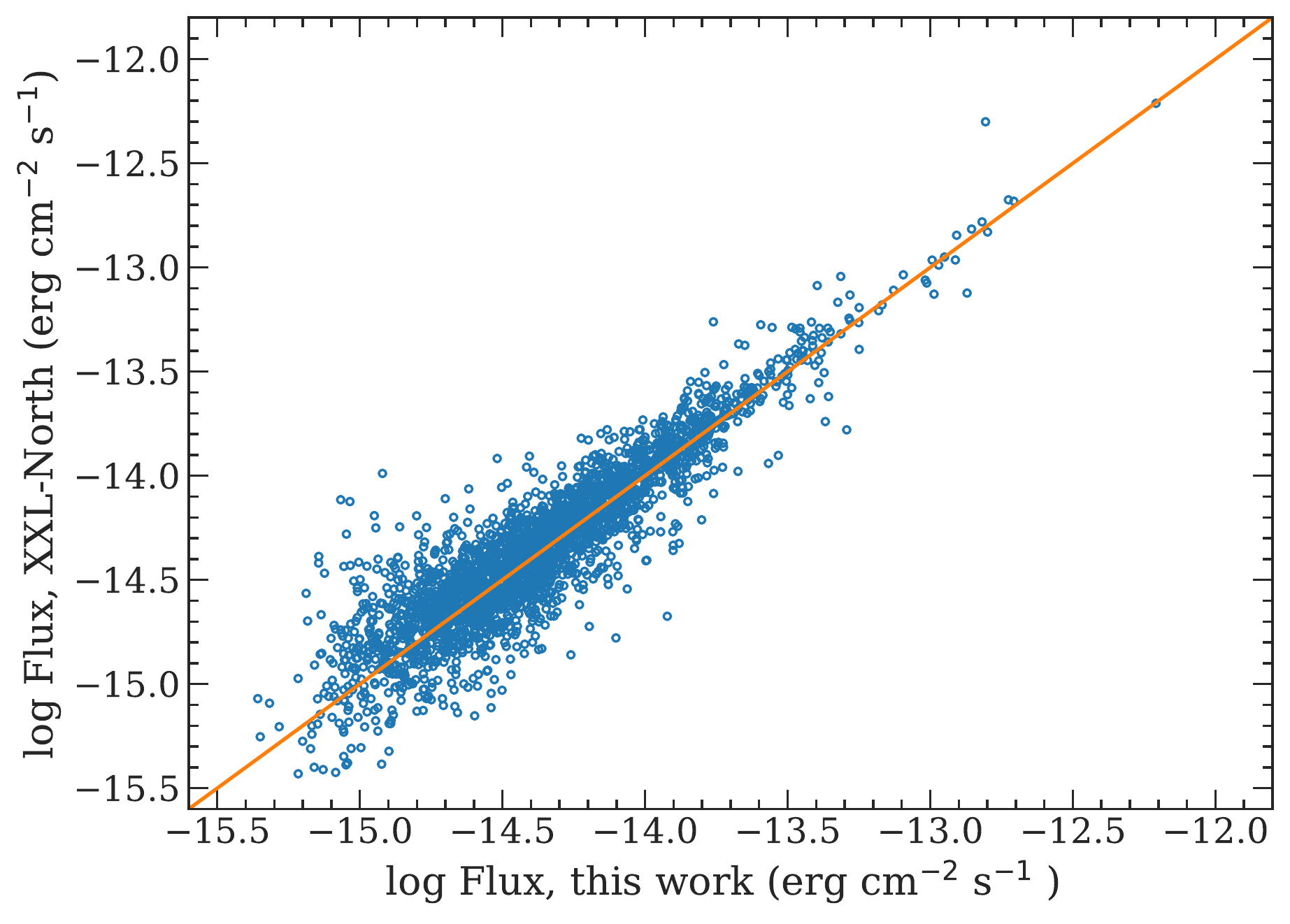}
\caption{
Comparison of the soft-band \hbox{X-ray} fluxes of our \hbox{X-ray} sources and those of the {\it XMM-Newton} counterparts identified in the XMM-XXL-North  source catalogs \citep{liu16xxl} within a 10$^{\prime\prime}$ radius. As expected, almost all of the XMM-XXL-North  \hbox{X-ray} sources in our catalog region 
can be matched to a counterpart in our \hbox{X-ray} source catalog with comparable flux.
}
\label{fig:compflux}
\end{figure}

For sources that are detected in fewer than three bands, we calculate the source-count upper limits using the mosaicked background map of the band in which the source is not detected. The mosaicked background map of each band is generated by summing the background maps from all individual observations (see \S\ref{subsec:astrometry}).
According to the Poisson probability set by the {\sc emldetect} detection likelihood threshold ($P_{\rm Random}$, the probability of the detected source being a random Poisson fluctuation due to the background), we can calculate
the minimum required total counts ($m$ in the following equation) 
required to exceed the expected number of background counts, $B$,
using the regularized upper incomplete $\Gamma$ function 
(which is equivalent to Eq. 2 of \citealt{civa16} if $m$ is a positive integer): 

\begin{equation}
\label{eq:invg}
P_{\rm Random} = \frac{1}{\Gamma(m)} \int_{B}^{\infty} t^{m -1} e^{-t} dt
\end{equation}
The upper limits are those corresponding to the {\sc det\_ml} values
with a 1\% spurious fraction: $P_{\rm Random} = 8.2\times10^{-3}$ 
for the soft band, $P_{\rm Random} = 4.1\times10^{-4}$ for the hard band, 
and $P_{\rm Random} = 2.0\times10^{-3}$ for the full band.
For each non-detected source in each band, we determine the 
background counts by summing the background map within the circle with 70\% encircled energy fraction (EEF). We then calculate $m$ by solving Eq.~\ref{eq:invg} using the {\sc Scipy} function 
{\sc scipy.special.gammainccinv}.\footnote{This quantity is the inverse function of Eq. 2.}
Since $m$ is the required {\it total} counts to exceed random background fluctuations 
at the given probability, the flux upper limit is calculated based on the following equation, which is similar to Equation 2 of 
\cite{capp09xcos} and Equation 2 of \cite{civa16}:

\begin{equation}
\label{eq:uplim}
S = \frac{m - B}{t_{\rm exp}\times {\rm EEF} \times {\rm ECF}}.
\end{equation}
Here EEF corrects for PSF loss and is 0.7, and $t_{\rm exp}$ is the median exposure time within the 70\% EEF circle. 
The flux upper limits are calculated as the exposure-time-weighted mean of the three EPIC detectors.

For each source detected in either the soft or the hard band (or both), 
we calculate its hardness ratio (HR), defined as 
$(H - S) / (H + S)$, where $H$ and $S$ are the source counts weighted by the 
effective exposure times in the hard and the soft bands, respectively. 
The source counts are the default output of {\sc emldetect}, 
which is the sum of the counts from all three EPIC detectors.\footnote{
Not all sources have data from all three EPIC detectors because one of the chips of MOS1 is permanently damaged, and some sources happen to fall on the chip gaps in one of the detectors. The exposure times for these sources are set to $-$99 in Table~\ref{tab:mainxtab} for the relevant detector.
} 
The three EPIC detectors have different energy responses, 
and the hardness ratios reported here did not take these into account.
We report this value in our catalog for direct comparison with previous {\it XMM-Newton} studies. 
The uncertainties on HR are calculated based on the count uncertainties from the 
output of {\sc emldetect} using the error-propagation method described in \S1.7.3 of Lyons (1991).
For sources not detected in either the soft or the hard band, 
we calculate the limits of their HRs assuming each non-detection has net counts $=m-B$,
where $m$ is the count upper limits calculated using Eq.~\ref{eq:invg} and $B$ is the background counts.
The HR uncertainties for these sources are set to $-$99.

We also report the hardness ratios independently for PN, MOS1, and MOS2,
calculated using the Bayesian Estimation of Hardness Ratios (BEHR) code
\citep{toolbehr} assuming the recommended indices for the $\Gamma$-function priors ({\sc softidx$=1$} and {\sc hardidx$=1$}). BEHR is designed to determine HRs for low-count sources in the regime of Poisson distributions.
It also computes uncertainties using Markov chain Monte Carlo methods for sources including those with non-detections in either the soft or hard band. 
Since our sources are usually detected over multiple exposures, we scale the HRs by setting the {\sc softeff} and {\sc hardeff} parameters in BEHR to account for the effective exposure times using Eq. (6)  of \citet{geor11xmm}. 
Qualitatively, the BEHR hardness ratios for sources that are detected in both the soft and hard bands are consistent with those calculated using the simple approach described in the previous paragraph. For the sources with non-detections in either the soft or hard band, we quote the default 68\% upper or lower bounds calculated with BEHR. As expected, these limits are almost always weaker than the HR limits obtained by assuming the non-detections have 99\% source count upper limits given by Eq.~\ref{eq:invg}.

\begin{figure}
\DeclareGraphicsExtensions{.png}
\includegraphics[width=\columnwidth]{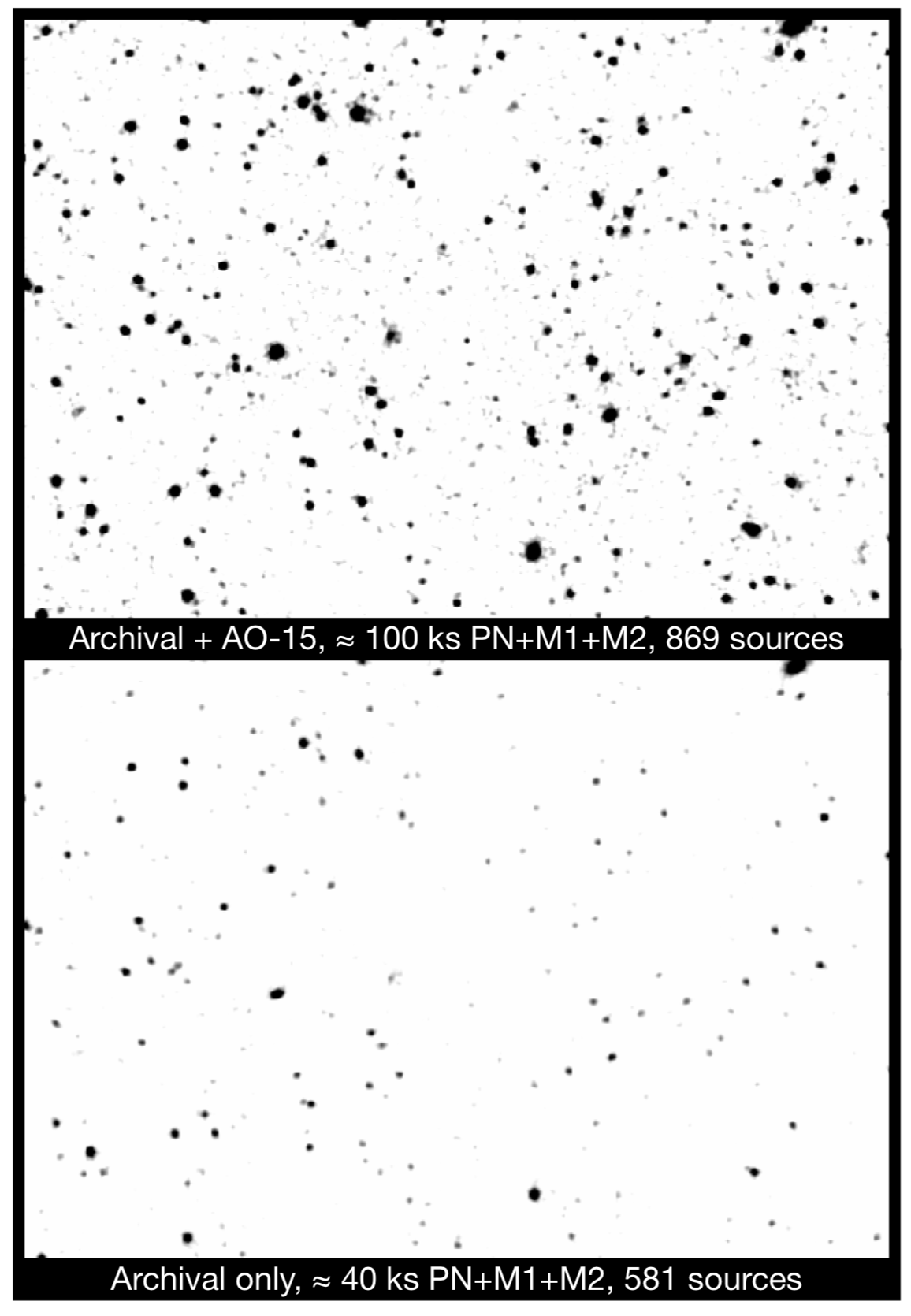}
\vspace{-0.5cm}
\caption{{\it Top} -- Background-subtracted, smoothed, and coadded PN+MOS image in the \hbox{0.5--10}~keV band for a $0.8\times0.6$ deg$^2$ region centered at RA$=35.580^{\circ}$, DEC=$-4.965^{\circ}$. This image is created using both archival data and the new AO-15 data, and a total of 869 sources are detected in this region. 
{\it Bottom} -- Same as the top image, but only the archival data are included. The two images are matched in color scale. In the \protect\cite{liu16xxl} catalog, only 581 sources can be found in this region. The typical vignetting-corrected exposure 
times are shown at the bottom of both panels. The exposure time of the full survey region is shown in 
Fig.~\ref{fig:exposure}.}
\label{fig:first}
\end{figure}
As a comparison,  
a total of 2861 \hbox{X-ray} sources from XMM-XXL-North  \citep{liu16xxl} % and 714 from SXDS are 
are found to have a counterpart within the 10$^{\prime\prime}$ radius in our \hbox{X-ray} catalog.\footnote{The 10$^{\prime\prime}$ search radius is approximately 3 times 
the quadratic sum of the largest positional uncertainties in both catalogs.} 
For these matched sources, we show a comparison between the soft-band \hbox{X-ray} fluxes reported in the XMM-XXL-North  catalog and those in our catalog in Fig.~\ref{fig:compflux}. 
As expected, the majority of the archival sources detected in our catalog have archival 
soft-band fluxes consistent with those in our catalog. 
The small scatter in the measured fluxes is expected as the XMM-XXL-North catalog
adopts a different source-detection method, background-subtraction approach, 
and energy conversion factors. 
Since the SXDS observations were also used for constructing the
XXL-North \citep{liu16xxl} catalog, the 2861 sources matched to 
the XMM-XXL-North  catalogs are considered to be 
matched to all available archival sources, and we conclude that 
the other 2381 \hbox{X-ray} sources in our catalogs are new sources.
We include the IDs from the \cite{liu16xxl} catalog for
these matched sources in our catalog (Table~\ref{tab:mainxtab}). 

In our source-detection region, 172 sources from the original \citet{liu16xxl} catalog do not have a counterpart in our point-source catalog.
Of these 172 sources, 150 can be associated with extended sources or 
sources deemed unreliable based on our {\sc det\_ml} criteria
(see \S\ref{subsec:simulation}).
The remaining sources comprise $< 1\%$ of the XMM-XXL-North  catalog in our source-detection region.
Visual inspection suggests that the vast majority of these sources might be spurious detections, 
but we cannot rule out the possibility 
that some sources are missed in our catalog due to \hbox{X-ray} variability \citep[e.g.,][]{yang16var,falo17,paol17,zhen17}. Also, the XMM-XXL-North  catalog adopted a different 
source-detection approach (see \S2 of \citealt{liu16xxl} for details).
The properties of sources that exhibit strong \hbox{X-ray} variability will be presented in a separate work.
Fig.~\ref{fig:first} shows the background-subtracted, \hbox{0.5--10}~keV PN+MOS image (see \S\ref{sec:mainx} for the details of the data analysis) from a $\approx 0.5$ deg$^{2}$ region in XMM-LSS generated using the combined AO-15 and archival data. An image produced using only the archival data is also displayed for comparison, demonstrating the improved source counts with the additional AO-15 observations.

\begin{table}
\tiny
    \caption{\label{tab:senscurve}
    Sensitivity curves. 
    Column 1: Soft-band flux.
    Column 2: Soft-band survey solid angle.
    Columns 3--4: Similar to Columns 1--2 but for the hard band.
    Columns 5--6: Similar to Columns 1--2 but for the full band.
    This table is available in its entirety online.
    }
    \begin{tabular}{cccccc}
    $\log S_{\rm 0.5-2 keV}$ & $\Omega_{\rm 0.5-2 keV}$ &
    $\log S_{\rm 2-10 keV}$ & $\Omega_{\rm 2-10 keV}$ & 
    $\log S_{\rm 0.5-10 keV}$ & $\Omega_{\rm 0.5-10 keV}$ \\
    (cgs) & (deg$^2$) & 
    (cgs) & (deg$^2$) & 
    (cgs) & (deg$^2$)\\
    \hline
    (1) & (2) & (3) & (4) & (5) & (6) \\
    \hline
$-$14.78 & 4.828 & $-$13.93 & 4.652 & $-$14.38 & 3.421 \\
$-$14.77 & 4.862 & $-$13.92 & 4.694 & $-$14.37 & 3.583 \\
$-$14.76 & 4.898 & $-$13.91 & 4.737 & $-$14.36 & 3.727 \\
$-$14.75 & 4.931 & $-$13.90 & 4.778 & $-$14.35 & 3.855 \\
$-$14.74 & 4.960 & $-$13.89 & 4.815 & $-$14.34 & 3.976 \\
$-$14.73 & 4.991 & $-$13.88 & 4.852 & $-$14.33 & 4.081 \\
$-$14.72 & 5.016 & $-$13.87 & 4.885 & $-$14.32 & 4.182 \\
$-$14.71 & 5.044 & $-$13.86 & 4.918 & $-$14.31 & 4.262 \\
... & ... & ... & ... & ...& ...\\
\hline
    \end{tabular}    
\end{table}

\subsection{Survey sensitivity, sky coverage, and $\log N - \log S$}\label{subsec:Sens}

\begin{figure*}    
\includegraphics[width=0.54\textwidth]{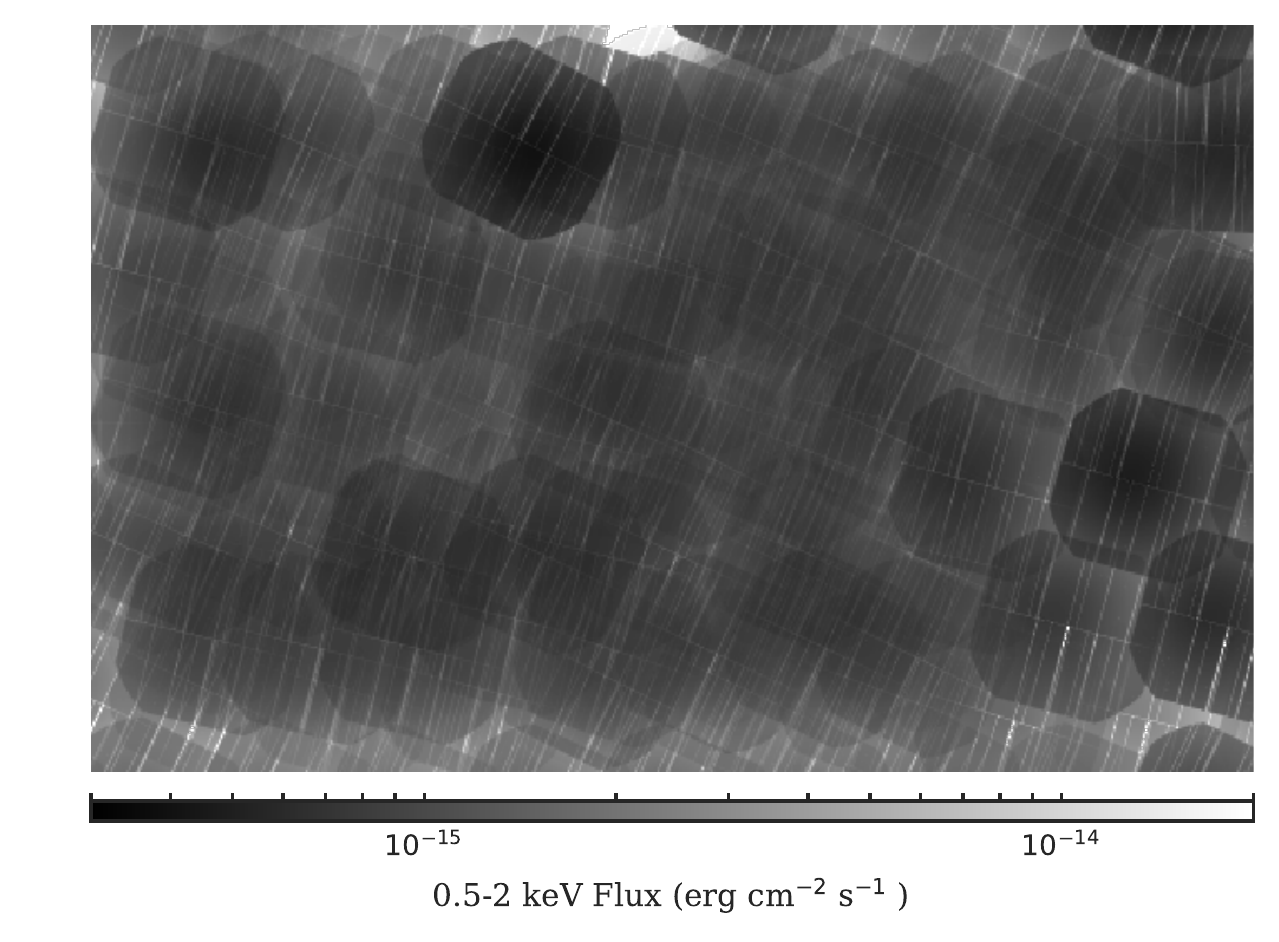}
\includegraphics[width=0.42\textwidth]{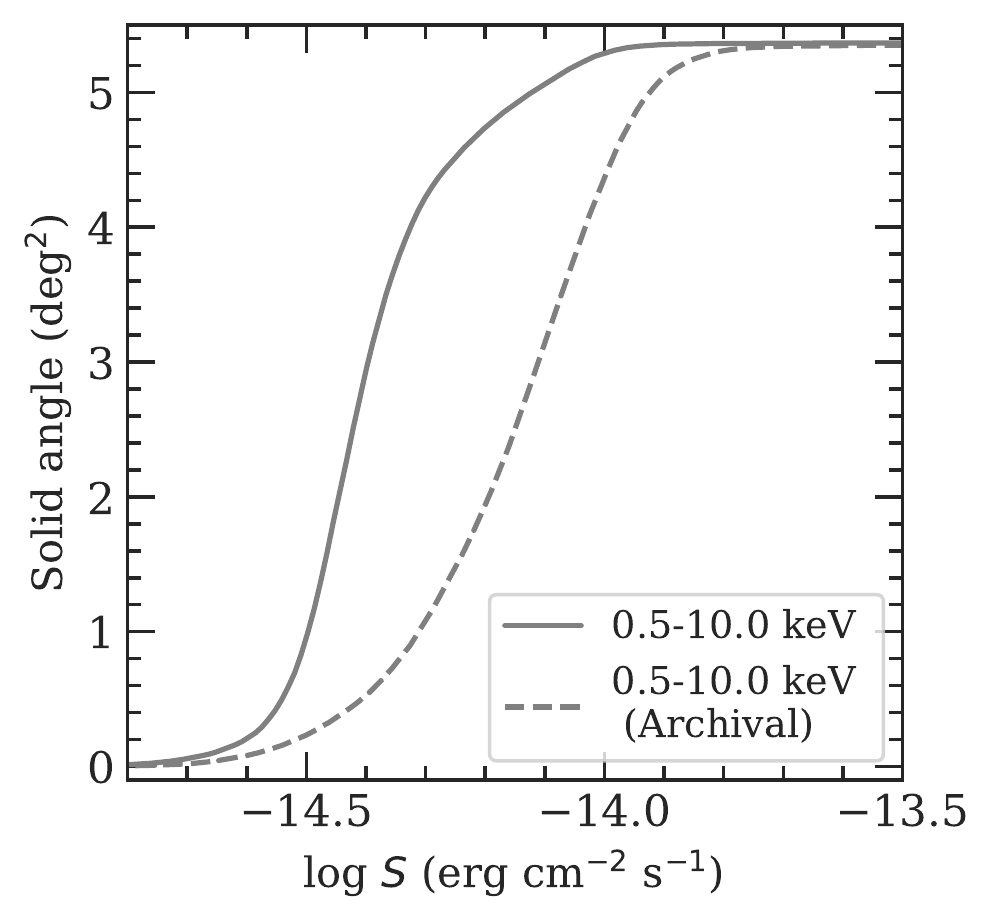}
%\DeclareGraphicsExtensions{.png}
\caption{
{\it Left} -- Soft-band sensitivity map of the source-detection region (the same as the cyan box shown in 
Fig.~\ref{fig:exposure}).
{\it Right} -- 
Comparison of the full-band sky coverages between this work (solid line) and the archival \xmm\ observations (dashed line),
demonstrating the improved and more uniform sensitivity across the wide field enabled by the new data.
}
\label{fig:sens}
\end{figure*}

\begin{figure}    
\includegraphics[width=0.4\textwidth]{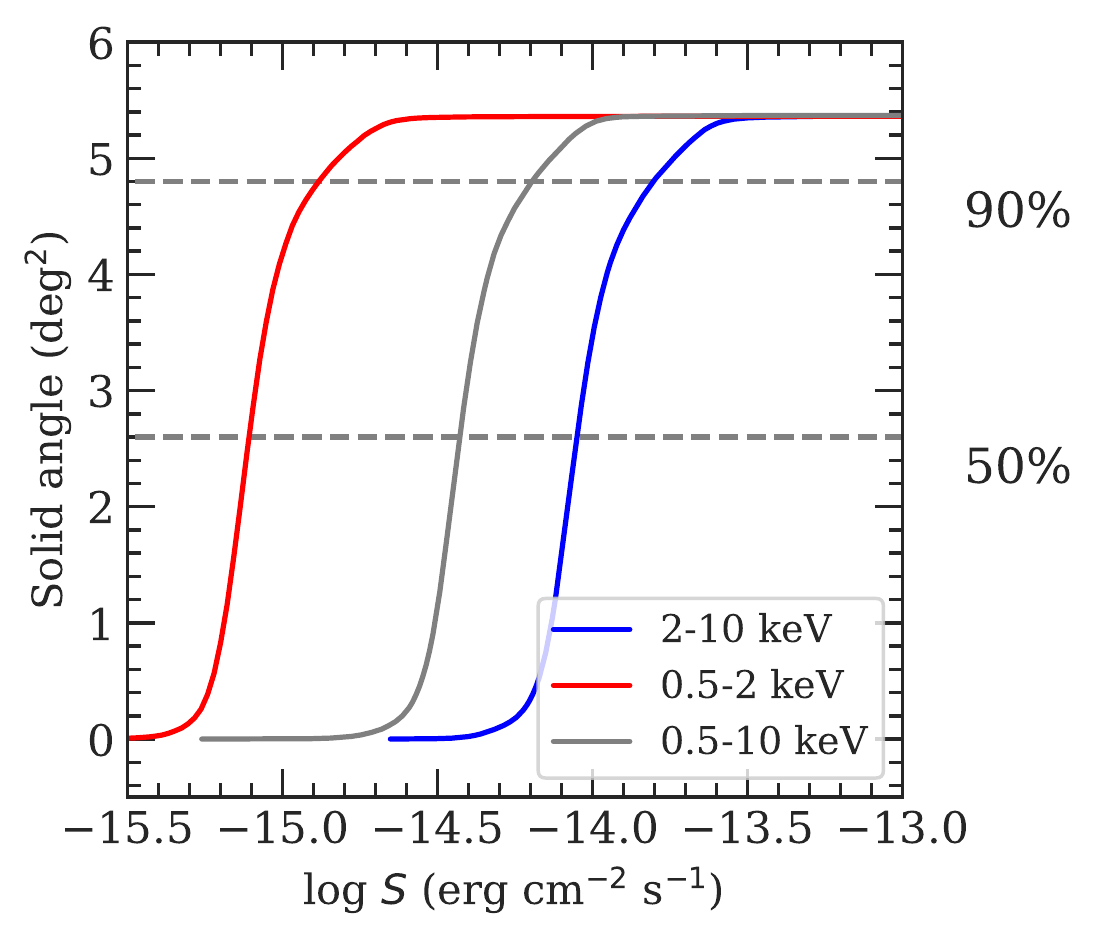}
\caption{
Sky coverage in the soft, hard, and full bands of 
our \hbox{X-ray} survey in XMM-LSS. 
The sensitivity curves
were calculated with {\sc det\_ml}$= 4.8$, 7.8, and 6.2 for the soft, hard, and full bands. 
These {\sc det\_ml} values correspond to $1\%$ spurious fraction based on extensive simulations (see \S\ref{subsec:simulation}.) 
}
\label{fig:sky}
\end{figure}

We create sensitivity maps of our survey region in different bands 
using the background and exposure maps generated as described in 
\S\ref{subsec:dataprep}. 
The mosaicked background and exposure maps are binned to $5\times5$ pixels ($20^{\prime\prime}\times20^{\prime\prime}$). 
For each pixel of the binned, mosaicked background map, 
the minimum required source counts to exceed the random background fluctuations
are calculated using Eq.~\ref{eq:invg}. The sensitivity is then calculated using Eq.~\ref{eq:uplim} with the corresponding EEF and ECF values. 
According the sensitivity maps, our survey has flux limits of 
$1.7\times10^{-15}$,
$1.3\times10^{-14}$, and
$6.5\times10^{-15}$~erg~cm$^{-2}$~s$^{-1}$ over 90\% of its area in the soft, hard, and full bands, respectively, reaching the desired depth-area combination.
We also compared the sensitivity maps with the detected sources, and find that the
spatial distribution of the fluxes of our sources largely obey the sensitivity maps.
The soft-band sensitivity map is presented in 
Fig.~\ref{fig:sens}-left. We also generated a soft-band sensitivity map using only the archival data.
To visualize the improvement upon the archival data, 
we compare the full-band sky coverage obtained from all available \xmm\ data in our survey region
with the sky coverage obtained using only the archival data. 
Fig.~\ref{fig:sens}-right demonstrates the improved survey depth and uniformity
with the new \xmm\ observations. 
The sensitivity curves corresponding to the {\sc det\_ml} thresholds 
in the soft, hard, and full bands are shown
in Fig.~\ref{fig:sky} and presented in Table~\ref{tab:senscurve}.

\begin{figure*}
\includegraphics[width=1.0\textwidth]{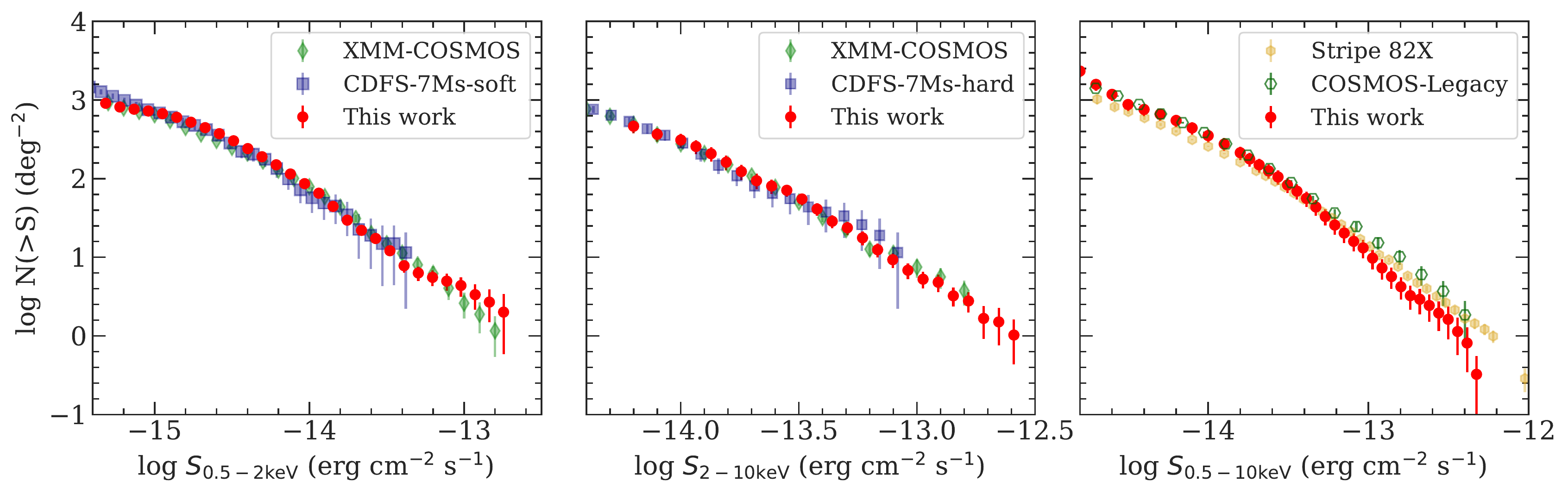}
\vspace{-0.5cm}
\caption{
The $\log N - \log S$ relations for our catalog 
in the soft band (left), hard band (middle), and full band (right). 
For comparison, a few $\log N - \log S$ relations from surveys spanning a wide range of area and sensitivity
are also shown (XMM-COSMOS, \citealt{capp09xcos}; Stripe-82X, \citealt{lama16stripe82}; 
COSMOS-Legacy, \citealt{civa16}; and CDF-S 7Ms, \citealt{catcdfs7ms}; 
the energy range and power-law photon index differences have been corrected).
The $\log N - \log S$ relations of our survey are generally consistent with those 
of previous studies.
}
\label{fig:lognlogs}
\end{figure*}

We calculate the $\log N - \log S$ relations of our survey using the sky coverage curves described above and the following equation: 

\begin{equation}
N(>S) = \sum\limits_{i=1}^{N_s} \frac{1}{\Omega_i}.% \quad ({\rm deg}^{-2}).
\end{equation}
Here $N(>S)$ represents the total number of detected sources with fluxes larger than $S$,
and $\Omega_i$ is the sky coverage associated with the flux of the $i$th source.
The $\log N - \log S$ relations of our survey are shown in Fig.~\ref{fig:lognlogs}, 
along with the $\log N - \log S$ relations for a selection of surveys spanning a wide range of area and sensitivity (CDF-S 7Ms, \citealt{catcdfs7ms}; XMM-COSMOS, \citealt{capp09xcos}; 
COSMOS-Legacy, \citealt{civa16}; and Stripe 82X, \citealt{lama16stripe82}). 
The flux differences caused by different choices of power-law indices and/or 
slight differences in energy ranges have been corrected assuming a $\Gamma=1.7$ power-law spectrum adopted in this work.
Considering factors such as different spectral models and/or methods of generating survey 
sensitivity curves, our $\log N - \log S$ relations are consistent with the relations 
reported in the literature within the measurement uncertainties. 

\section{Multiwavelength counterpart identifications}\label{sec:mw}
The XMM-LSS region is one of the most extensively observed extragalactic fields. 
The publicly available multiwavelength observations in the XMM-LSS region utilized in this work are 
SERVS \citep{catservs},
SWIRE \citep{catswire}, 
VIDEO \citep{catvideo},
the CFHTLS-wide survey \citep{catcfhtls}, and 
the HSC-SSP survey \citep{cathscpdr1}.

We focus on identifying the correct counterparts for our \hbox{X-ray} sources in four deep optical-to-near-IR (OIR) catalogs:
SERVS, VIDEO, CFHTLS, and HSC-SSP.
SERVS is a post-cryogenic {\it Spitzer} IRAC survey in the near-IR 3.6 and 4.5 $\mu$m bands with $\approx 2 \mu$Jy survey sensitivity limits and $\approx 5$~deg$^2$ solid-angle coverage in the XMM-LSS region. 
We make use of the highly reliable two-band SERVS catalog built using {\sc SExtractor}, obtained from the {\it Spitzer} Data Fusion catalog \citep{vacc15}, which has $\approx 4\times10^5$ sources. 
The Spitzer Data Fusion catalog has already integrated data from SWIRE, which include photometry in all four IRAC bands and the photometry in MIPS 24, 70, and 160 $\mu$m. 
A total of $82\%$ of the \hbox{X-ray} sources have at least one SERVS counterpart candidate within their 99.73\% positional-uncertainty radius 
($r_{99\%}$ hereafter, which is equivalent to 3.44$\sigma_x$), which is calculated based on the quadratic sum of the 99.73\% \hbox{X-ray} positional uncertainties and the corresponding OIR positional uncertainties. 

VIDEO is a deep survey in the near-infrared {\it Z, Y, J, H}, and $K_s$ bands with $\approx 80\%$ completeness at $K_s < 23.8$. In the XMM-LSS region, VIDEO covers a  $4.5$~deg$^{2}$ area 
(\hbox{$\approx 85\%$} of our \hbox{X-ray} survey region) with a total of $\approx 5.7\times10^5$ sources; $79\%$ of the \hbox{X-ray} sources 
have at least one VIDEO counterpart candidate within $r_{99\%}$.

The CFHTLS-W1 survey covers the entirety of our \hbox{X-ray} data, with an $80\%$ completeness limit of $i^\prime=24.8$. We select the CFHTLS sources in the RA/DEC ranges marginally larger 
(1$^\prime$) than our source-detection region.
We limit the CFHTLS sources to those with $SNR > 5$ in the $i^\prime$-band. 
The total number of sources in the $i^\prime$-band selected catalog is 
$\approx 8.1\times 10^5$. A total of 90\% of the \hbox{X-ray} sources in our catalog 
have at least one CFHTLS counterpart candidate within $r_{99\%}$.

The XMM-LSS field is entirely encompassed by the 108~deg$^2$ HSC-SSP wide survey. 
The limiting magnitude in the {\it i}-band for the wide HSC-SSP survey is 26.4.
Inside the XMM-LSS field, HSC-SSP also has ``ultra-deep'' ($\approx 1.77$~deg$^2$) and ``deep'' ($\approx 5$ deg$^2$) surveys, which overlap with the SXDS and XMDS regions, respectively. 
We focus only on the wide survey because in the currently available data release it is only 0.1 mag shallower than the deep survey in the {\it i}-band, and the uniform coverage is important for determining the background source density when matching to the \hbox{X-ray} catalog (see \S\ref{subsec:lrmatching}). 
We select the {\it i}-band detected HSC-SSP sources in the RA/DEC ranges slightly larger 
than our source-detection region.\footnote{We select sources with the {\sc detect\_is\_primary} and {\sc idetected\_notjunk} flags set as {\sc True}, 
and {\sc centroid\_sdss\_flags} set as {\sc False}. According to the HSC-SSP example script for selecting ``clean objects'', 
we also exclude the HSC sources with {\sc flags\_pixel\_edge}, {\sc flags\_pixel\_saturated\_center}, 
{\sc flags\_pixel\_cr\_center}, {\sc flags\_pixel\_bad } flags in the {\it i}-band to avoid unreliable {\it i}-band sources.} 
The total number of HSC-SSP sources in our source-detection region is $\approx 3.1\times10^6$, and $ \approx 93\%$ of the \hbox{X-ray} sources in our main catalog have at least one HSC-SSP counterpart candidate within $r_{99\%}$. 

Although CFHTLS is not as deep as HSC-SSP in the {\it g, r, i}, and {\it z} bands, 
it has complementary {\it u$^*$}-band photometry. Including photometry 
from both optical surveys also ensures that we will 
minimize the risk of missing an optical counterpart due to 
bad photometry caused by artifacts such as satellite tracks in either survey. 

Since there are small systematic offsets in the astrometry of each catalog, 
we match SERVS, VIDEO, and CFHTLS to the HSC-wide catalog, and correct for the small offsets between
each catalog to the HSC-wide catalog to maximize the counterpart matching accuracy.
In the RA direction, the adopted corrections are 0\farcs020, 0\farcs027, 
and 0\farcs026 for SERVS, VIDEO, and CFHTLS, respectively.
For DEC, the adopted corrections are $-$0\farcs009, $-$0\farcs006, $-$0\farcs008
for SERVS, VIDEO, and CFHTLS, respectively.

\subsection{The likelihood-ratio matching method}\label{subsec:lrmatching}
\begin{figure}
\includegraphics[width=\columnwidth]{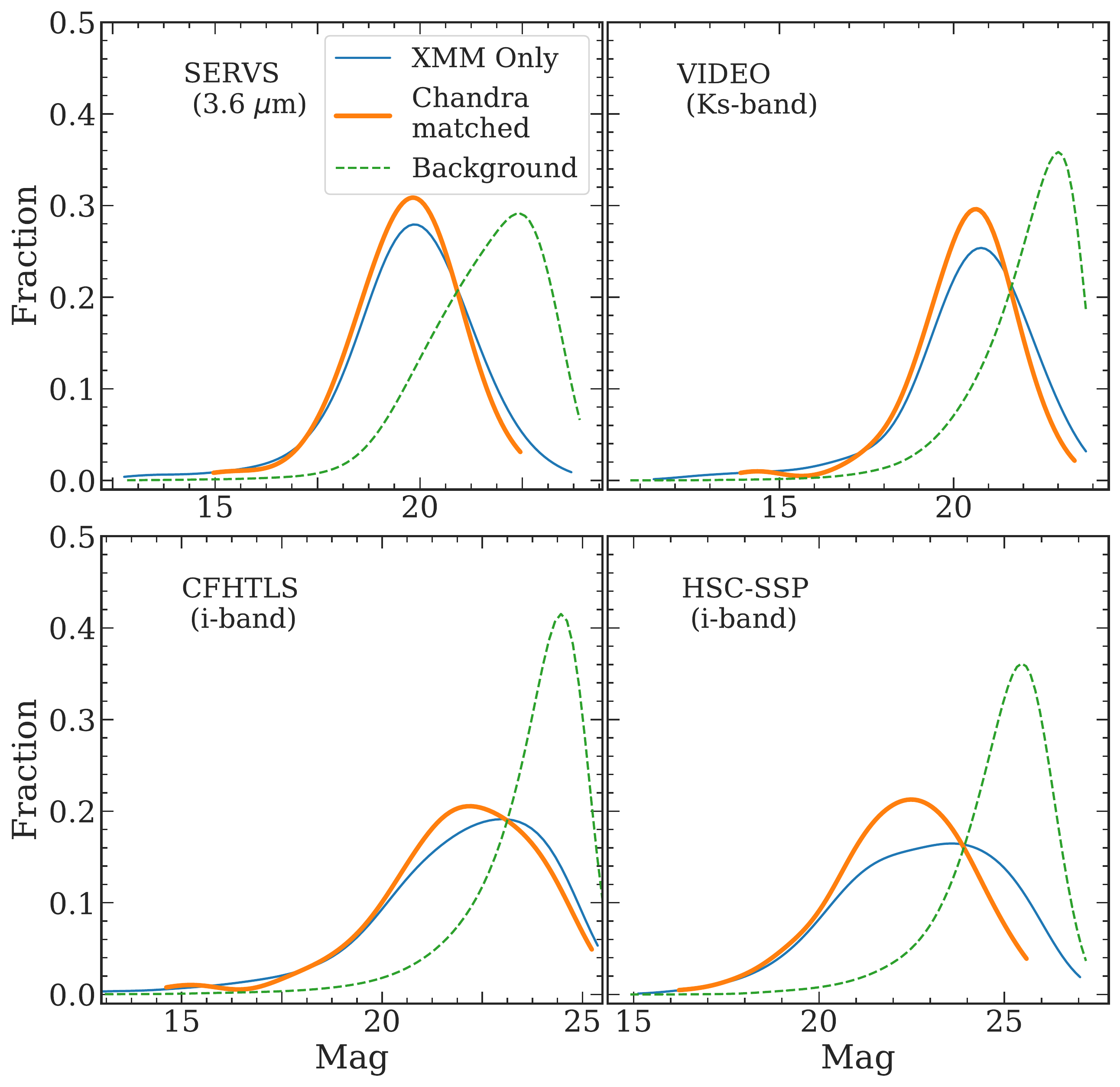}
\caption{
Kernel-density estimations of the magnitude distributions (solid lines) for the expected counterparts in SERVS (top-left), VIDEO (top-right), CFTHLS (bottom-left), and HSC-SSP (bottom-right).
We show the distributions obtained using the full {\it XMM-Newton} catalog ($q(m)_{\it XMM-Newton}$),
and the distributions obtained using the {\it Chandra} sources in the XMM-LSS field ($q(m)_{\it Chandra}$).
The magnitude distributions of the background, unrelated sources are also displayed in each panel as the dashed curves.
This figure demonstrates that $q(m)_{\it Chandra}$ significantly improves upon the 
background-dominated $q(m)_{\it XMM-Newton}$ for the deep OIR catalogs in the bottom panels
(in particular, the most-probable magnitude values).
}
\label{fig:realm}
\end{figure}

To match reliably the \hbox{X-ray} sources to the OIR catalogs with much higher source densities, we employ the 
likelihood-ratio method (LR hereafter) similar to previous \hbox{X-ray} surveys, \cite[e.g.,][]{catcosmosb07,luo10cdfs,catcdfs4ms,xue16,catcdfs7ms}. The likelihood ratio is defined as the ratio between the probability that the source is the correct counterpart, and the probability that the source is an unrelated background object \citep{suth92}:

\begin{equation}\label{eq:LR}
LR = \frac{q(m)f(r)}{n(m)}.
\end{equation}
Here $q(m)$ is the magnitude distribution of the expected counterparts in each OIR catalog, 
$f(r)$ is the probability distribution function of the angular separation between \hbox{X-ray} and OIR sources, 
and $n(m)$ is the magnitude distribution of the background sources in each OIR catalog. 

We calculate the background source magnitude distributions using OIR sources between 10$^{\prime\prime}$ and 50$^{\prime\prime}$ from any sources in our \hbox{X-ray} catalog.

As discussed in \S\ref{subsec:poserr}, the probability distribution function of the angular separation should follow the Rayleigh distribution:

\begin{equation}\label{eq:poserr}
f(r) = \frac{r}{\sigma_x^2} \exp^\frac{-r^2}{2\sigma_x^2}.
\end{equation}
Note that Eq.~\ref{eq:poserr} is different from the two-dimensional Gaussian 
distribution function that maximizes at $r=0$, and thus the $LR$ values calculated 
in this work are not directly comparable to previous works that adopted a Gaussian $f(r)$.

In practice, for an \hbox{X-ray} source with a total of $Nc$ counterpart candidates within the search radius, 
the matching reliability for the $i$-th counterpart candidate $MR_i$, can be determined using
the following equation:
\begin{equation}\label{eq:reliabiliy}
MR_i = \frac{LR_i}{\sum_{k=0}^{N_c} LR_k + (1 - Q)}
\end{equation}
Here $Q$ is the completeness factor, which is defined as $Q = \int_{-\infty}^{m_{\rm lim}}  q(m)$,
where $m_{\rm lim}$ is the limiting magnitude of the OIR catalog being used for matching.
For each counterpart candidate, $MR$ is equivalent to the relative matching probability 
among all possible counterpart candidates. See Eq. 5 of \cite{suth92} and \S2.2 of \cite{luo10cdfs} for details.

\begin{figure*}    
\includegraphics[width=0.50\textwidth]{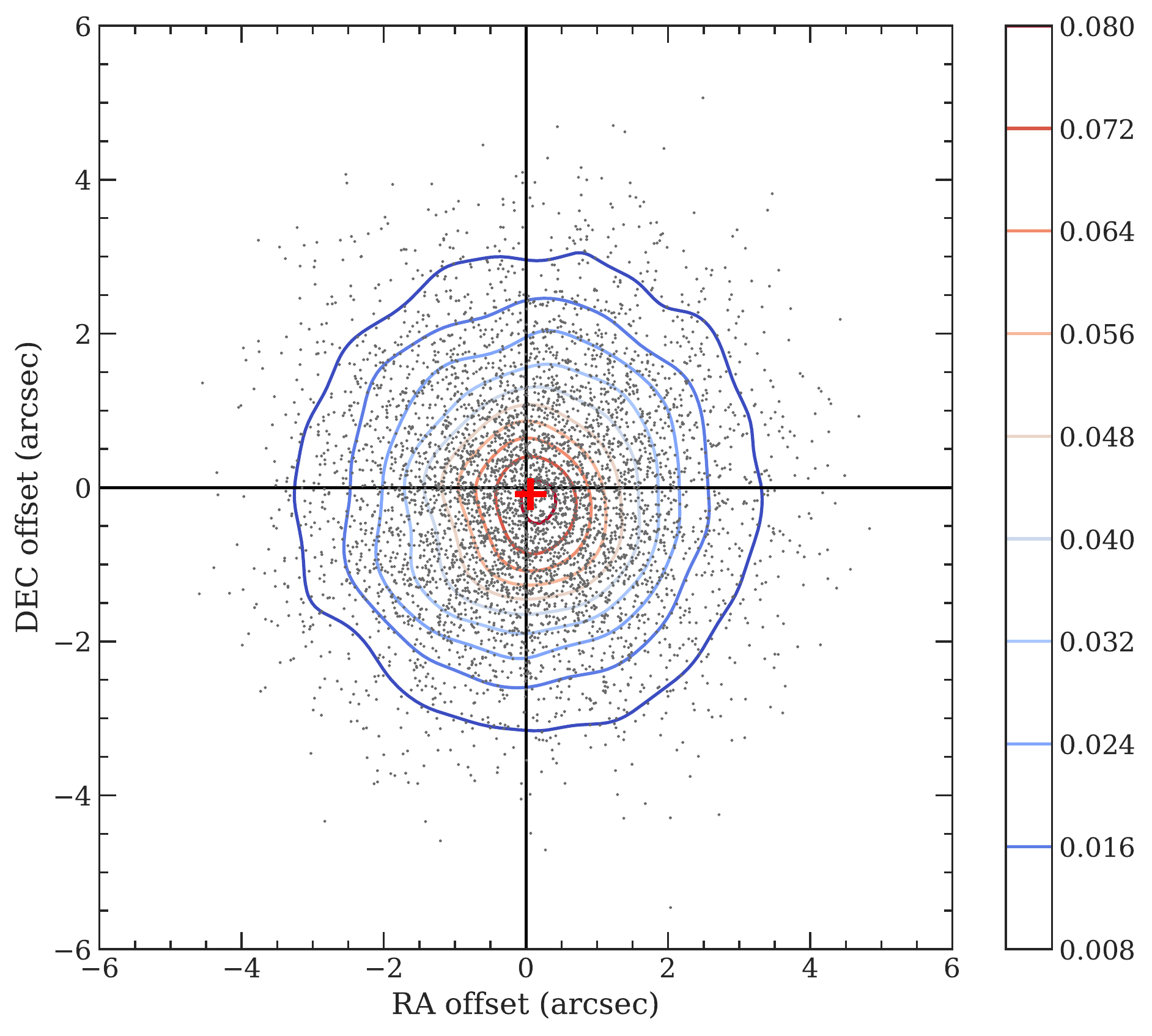}
\includegraphics[width=0.44\textwidth]{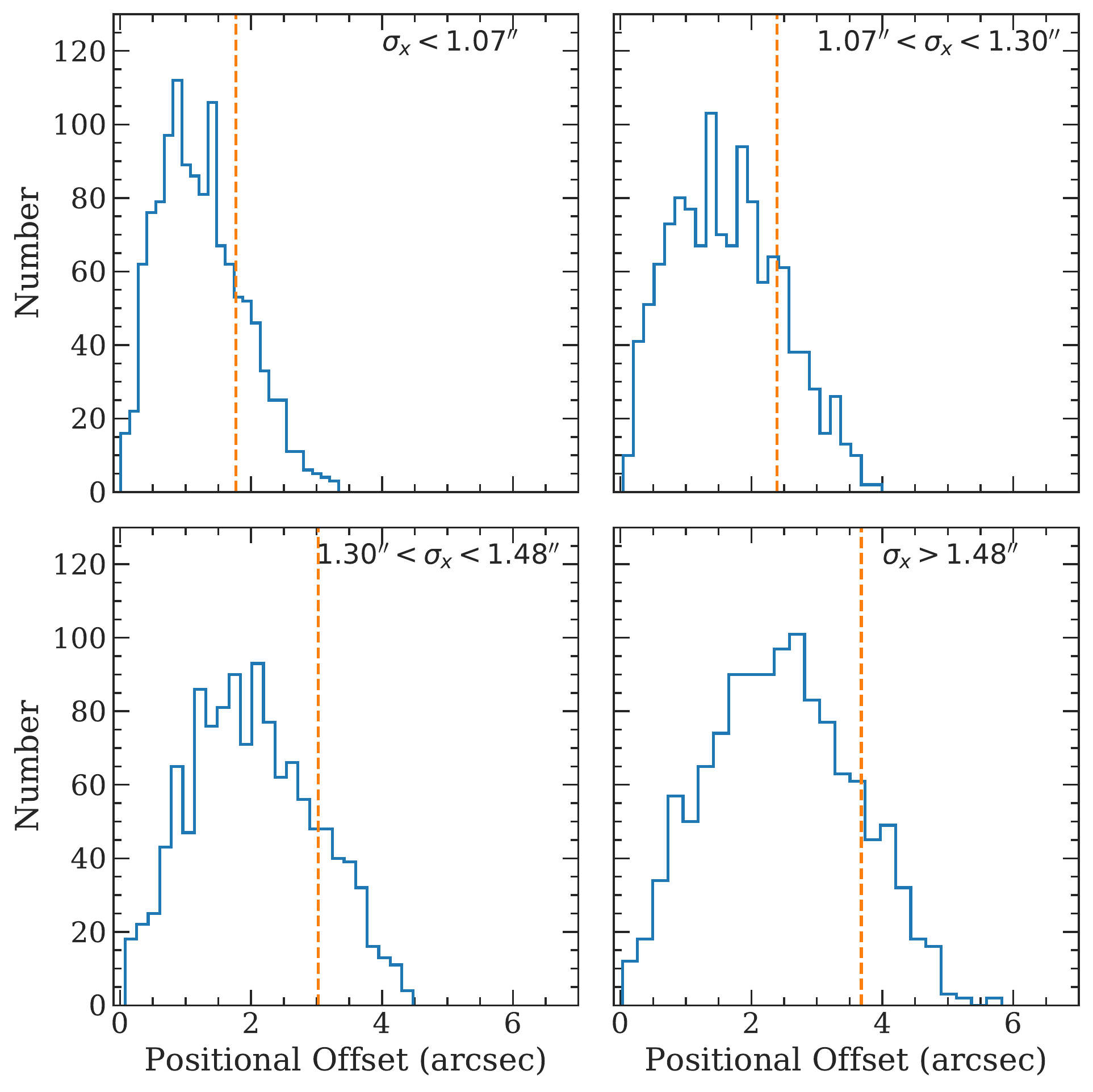}
\caption{{\it Left}: Distribution of the OIR-to-X-ray positional offsets in the RA vs. DEC plane for the 4858 \xmm\ sources with reliably matched OIR counterparts.
The contours represent the isodensity levels of the points. 
The mean positional offsets are $<$ 0.1$^{\prime\prime}$ in both the RA and DEC directions (the red cross).
{\it Right}: Histograms of positional offsets for the 4858 reliably matched sources, divided into four bins based on their positional uncertainties. In each panel, we also mark the median 68\% positional offset value ($r_{\rm 68\%}$) as the vertical dashed line.}
\label{fig:offset_grid}
\end{figure*}    

Due to the relatively large positional uncertainties of {\it XMM-Newton} and the high source 
densities of the OIR catalogs, deriving an accurate magnitude distribution of the expected counterparts, $q(m)$,
using \xmm\ data is challenging. 
Therefore, we obtain $q(m)$ for our \hbox{X-ray} sources by first matching 
our {\it XMM-Newton} catalog to the \chandra\ Source Catalog 2.0 \citep[CSC 2.0;][]{catcsc} 
to take advantage of the higher angular resolution and positional accuracy of {\it Chandra}. 
We derive the positional uncertainties of the \chandra\ sources in our survey region 
using the same empirical approach described in \cite{catcdfs4ms} 
by selecting CSC sources in the RA/DEC range of our catalog, 
and matching them onto HSC-SSP using a $1.5^{\prime\prime}$ radius. 
We select CSC sources that are uniquely matched to our \hbox{X-ray} catalogs within 
the $95\%$ uncertainties (\chandra\ and {\it XMM-Newton} positional uncertainties are added in quadrature). 
A total of 223 sources in our {\it XMM-Newton} catalog are matched 
to a unique \chandra\ source in the CSC. 
We match these \chandra\ sources to the four OIR catalogs using Eq.~\ref{eq:LR}, 
with $q(m)$ derived using the iterative approach described in \cite{luo10cdfs}, 
which determines the $LR$ threshold by optimizing the matching reliability and completeness. 
The $q(m)$ derived from the CSC sources, $q(m)_{\it Chandra}$, 
is then used as the expected magnitude distribution for OIR counterparts of our \xmm\ sources.  
The \xray\ flux distributions in the soft, hard, and full bands of the {\it Chandra}-matched 
subsample are similar to those of our entire {\it XMM-Newton} catalog, 
and therefore $q(m)_{\it Chandra}$ should be consistent with the intrinsic magnitude 
distributions of the real OIR counterparts of our full \hbox{X-ray} catalog. 
The counterpart-matching processes are run on four different OIR catalogs: 
SERVS, VIDEO,
CFHTLS, and HSC-SSP. The details of the filters and apertures of the photometry in each OIR catalog
can be found in Appendix \ref{sec:catcols}, where we give the descriptions of the columns reported in the source catalog (Columns 128--187 of Table~\ref{tab:mainxtab}).
For illustration, Fig.~\ref{fig:realm} shows the magnitude distributions of 
the background sources and the distributions of the expected counterparts 
derived using CSC sources.

For comparison, we also obtain $q(m)$ for the full {\it XMM-Newton} catalog
without using the \chandra\ positions, $q(m)_{\it XMM-Newton}$.
We again use the \cite{luo10cdfs} iterative method, but with a $3^{\prime\prime}$ initial 
search radius. $q(m)_{\it XMM-Newton}$ is also plotted in Fig.~\ref{fig:realm}. 
It is evident that for ultra-deep OIR catalogs such as HSC-SSP and CFHTLS, 
$q(m)_{\it XMM-Newton}$ is skewed toward the faint background sources 
compared to the {\it Chandra}-matched subsample. 
For the other catalogs, we find no qualitative difference between 
$q(m)_{\it Chandra}$ and $q(m)_{\it XMM-Newton}$, 
but we still use $q(m)_{\it Chandra}$ for consistency.

We next compute the $LR$ values for all OIR sources within a 10$^{\prime\prime}$ radius (i.e., the counterpart ``candidates'') of the 
\hbox{X-ray} sources using Eq.~\ref{eq:LR}.
For each OIR catalog, 
we choose the $LR$ thresholds ($LR_{\rm th}$) such that the reliability and 
completeness parameters are maximized 
(see Eq. 5 of \citealt{luo10cdfs} for details).
%\citep[see Equation 5 of][for details]{luo10cdfs}. 
Counterparts with $LR > LR_{\rm th}$ are considered to be reliably matched. 
A summary of the results is reported in Table~\ref{tab:matching}.
For each OIR catalog, we list the number of all \hbox{X-ray} sources with at least one OIR counterpart candidate
within $r_{99\%}$ of the \hbox{X-ray} sources, $N_{\rm All}$,
and the number of \hbox{X-ray} 
sources with at least one reliably matched source with $LR > LR_{\rm th}$, $N_{\rm Reliable}$.

\begin{table*}
\scriptsize
    \caption{\label{tab:matching}
    Summary of $LR$ counterpart-matching results for each OIR catalog,
    with an additional summary row for the combined results from all OIR catalogs considered.
    The columns in the summary row
    are the same as those for individual OIR catalogs except for Column 7).
    Column 1: Catalog name. 
    Column 2: Survey magnitude limit for each catalog in AB. 
    Column 3: Survey area. 
    Column 4: Positional uncertainty for each OIR catalog.
    Column 5: $LR$ threshold. 
    Column 6: Total number of \hbox{X-ray} sources with at least one counterpart within the 10\arcsec\ search radius in each catalog. 
    Column 7: Average number of OIR sources within $r_{99\%}$ of the \hbox{X-ray} sources (if the \hbox{X-ray} source is within the coverage of the OIR catalog). 
    Here the summary row shows the total number of \hbox{X-ray} sources with at least one OIR counterpart within $r_{99\%}$. 
    Column 8: Total number of \hbox{X-ray} sources with at least one counterpart with $LR > LR_{\rm th}$.
    The summary row displays the number of all \hbox{X-ray} sources with at least one $LR > LR_{\rm th}$ counterpart from any of the four OIR catalogs, 
    plus the 23 sources with only one unique counterpart within $r_{99\%}$ from all OIR catalogs considered (see \S\ref{subsec:lrmatching} for details).
    \hbox{X-ray} sources having only one unique OIR counterpart in all OIR catalogs considered within $r_{99\%}$, but the $LR$ values do not exceed the reliability thresholds in all OIR catalogs. 
    Columns 9--11: See \S\ref{subsec:matchingcheck} for details. 
    Column 9: The fraction of \hbox{X-ray} sources in the ``associated population'' based on the results of Monte Carlo simulations.
    Column 10: False-matching rates determined using Monte Carlo simulations.
    Column 11: Fraction of the \hbox{X-ray} sources having identical reliable counterparts found based on their
    \chandra\ and \xmm\ positions. Based on sources in regions where there is overlapping {\it XMM-Newton}
    and {\it Chandra} coverage.
    For the summary row, Columns 9--11 are calculated as the weighed sum (based on the number of primary counterparts from each catalog) of the results from all four OIR catalogs.
    }
    \begin{tabular}{ccccccccccc}
        \hline
        Catalog & 
        Limiting Magnitude & 
        Area &
        $\sigma$ &
        $LR_{\rm th}$&
        $N_{\rm All}$ & 
        $\overline{N_{99\%}}$ & 
        $N_{\rm Reliable}$ &
        $f_{\rm AP}$ &
        False Rate &
        Identical Fraction 
       \\
        & 
        & 
        deg$^2$ & %(10$^5$/deg$^2$) & 
        &
        & 
        & 
        &
        & 
        & (Simulation)
        & ({\it Chandra}) 
       \\
        (1) & (2) & (3) & (4) & (5) & (6) & (7) & (8) & (9) & (10) & (11) 
       \\
        \hline
        SERVS                     & $3.6\mu$m $< 23.1$ & 5.0  & 0.5\arcsec\ &
        %9.6 &  
        0.32 &
        4689 & 
        1.0 &
        3948  &  96.8\% &  4.2\% & 97.3\%\\
        VIDEO                     & $Ks < 23.8 $       & 4.5  & 0.3\arcsec\ &
        %11.3 & 
        0.25  &
        4380 & 
        1.3 &
        3827  &  86.3\% &   8.0\% & 94.4\%\\
        CFHTLS-wide              & $i < 24.8 $         & 5.4  & 0.2\arcsec\ &
        %12.3 &  
        0.22 &
        5185 & 
        1.5 &
        4207  &  75.6\% & 15.6\% & 90.8\%\\
        HSC-SSP                  & $i < 26.5$         &  5.4 & 0.1\arcsec\ &
        %43.9    & 
        0.25 &  
        5124    & 
        2.3 &
        4317 &  78.6\%   & 18.4\% & 87.3\%\\
        \hline
        Summary &N/A & N/A & 
        N/A     & N/A &  5237    & 
        5147 &
        4858 &  93.1\%  & 5.8\% & 97.1\% \\
        \hline
    \end{tabular}    
\end{table*}

Motivated by the spurious-matching rates of different OIR catalogs (see \S\ref{subsec:matchingcheck} for the cross-matching reliability analysis), 
we first select a ``primary'' counterpart for each \hbox{X-ray} source from, in priority order, SERVS, VIDEO, CFHTLS, and HSC-SSP.
After selecting the primary OIR counterpart, we associate different OIR catalogs with each other using a simple nearest-neighbor algorithm.
Thanks to the much smaller positional uncertainties of the OIR catalogs, we adopt a constant search radius of 1$^{\prime\prime}$ for the OIR catalog associations, which is the approach used by the {\it Spitzer} Data Fusion database \citep{vacc15}.

Using this approach, 4832 ($\approx 93\%$) \hbox{X-ray} sources have at least one robust counterpart with $LR > LR_{\rm th}$. We consider an additional 26 \hbox{X-ray} sources without any counterpart candidates having $LR > LR_{\rm th}$ to have ``acceptable'' matches because there is only one unique counterpart in all four OIR catalogs within $r_{\rm 99\%}$. When considering both the $LR > LR_{\rm th}$ counterparts and the acceptable counterparts, 4858 \hbox{X-ray} sources in our catalog are considered to have reliable OIR counterparts ($93\%$). Of these sources, 3968 are matched to SERVS as the primary counterpart, 
367 are from VIDEO, 
386 are from CFHTLS, 
and 137 are from HSC. 

Besides the 4858 \hbox{X-ray} sources with reliable/acceptable counterparts, most of the remaining 384 sources have 
$f_{\rm spurious}\leq 0.05\%$ in at least one 
band,
and thus they are unlikely to be spurious \xray\ detections. 
289 of these 384 sources still have at least one OIR counterpart candidate 
within the $r_{99\%}$ circle. Therefore, 5147 \hbox{X-ray} sources have at least 
one OIR counterpart candidate within $r_{\rm 99\%}$. Of the other 95 sources, 
90 still have at least one OIR counterpart candidate within the $10^{\prime\prime}$ counterpart-searching radius. 
We still select counterparts for these sources and the properties of these counterparts are included in the main \hbox{X-ray} catalog. 
However, only the previously mentioned 4858 sources are considered to be reliably matched and are flagged in the catalog.
We find 5 sources that are completely ``isolated'', 
i.e., no counterpart candidates were found within a 10$^{\prime\prime}$ search radius. 
Visual inspection of these sources shows that all of them coincide with a bright star, 
thus making the pipeline OIR photometry unavailable. 

\begin{figure*}
\DeclareGraphicsExtensions{.png}
\includegraphics[width=\textwidth]{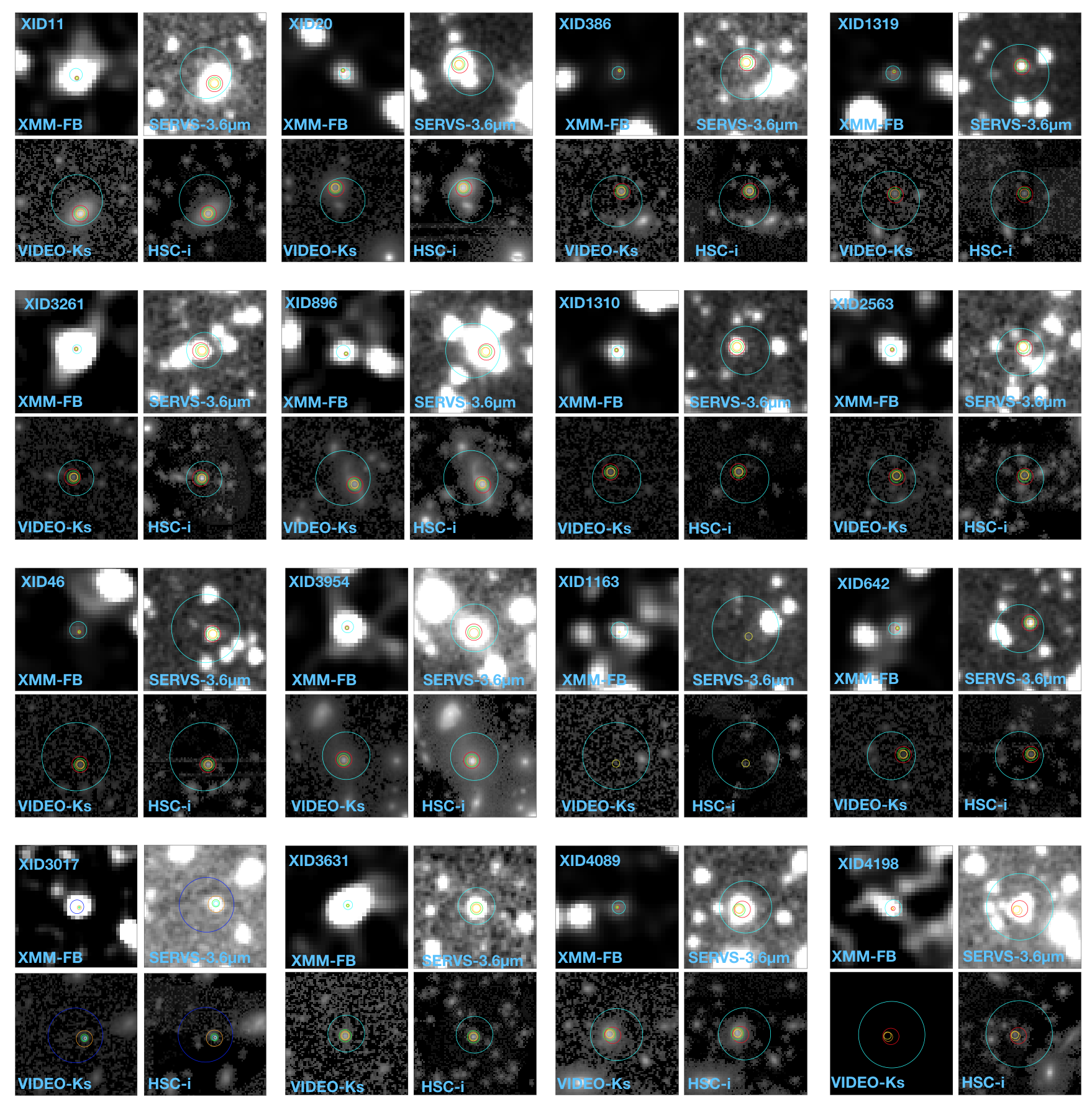}
%\vspace{-1.5cm}
\caption{Postage-stamp images for 16 randomly selected \hbox{X-ray} sources in our catalog. 
For each source, we show 
(1) Full-band \hbox{X-ray} image smoothed with a Gaussian filter (upper-left panel). The unique source ID from Table A is shown in this panel. 
(2) 3.4$\mu$m mid-IR image from SERVS (upper-right panel) 
(3) {\it Ks}-band near-IR image from VIDEO (lower-left panel) 
(4) {\it i}-band optical image from HSC-SSP (lower-right panel), re-sampled 
to a 0\farcs4 pixel size.
Due to the large pixel size, 
the \hbox{X-ray} image for each source is set at $2^{\prime} \times 2^{\prime}$.
For the OIR images, the sizes are set at $0.5^{\prime} \times 0.5^{\prime}$.
In each image, the \hbox{X-ray} position is marked as the cyan circle with the $r_{99}$ radius. 
The position of the most-probable mid-IR SERVS counterpart is marked as the red circle with a 1\farcs5 radius. 
The positions of VIDEO, CFHTLS, and HSC-SSP counterparts are marked as 
green, orange, and yellow circles with a 0\farcs9, 0\farcs6, and 0\farcs3 radius, respectively.
The size of the OIR counterpart circles are 3 times the positional uncertainty values reported
in Table~\ref{tab:matching}. The entire set of postage-stamp images is available in the electronic version.
}
\label{fig:postage}
\end{figure*}

Fig.~\ref{fig:offset_grid} presents the positional offsets between the \hbox{X-ray} sources and the reliably matched sources. 
The small median positional offsets in the RA and DEC directions demonstrate the quality of our astrometry, and the histograms of the positional offsets for sources binned in different $\sigma_x$ show that our empirically derived positional uncertainties are reliable.
For each source, we also generate postage-stamp images at \hbox{X-ray}, mid-IR, near-IR, and optical wavelengths. 
For illustration, we show a random collection of 16 \hbox{X-ray} sources with reliable counterparts in Fig.~\ref{fig:postage}.

For the 4335 \hbox{X-ray} sources with primary counterparts from SERVS or VIDEO (regardless of matching reliabilities), 
269 of them have no optical counterparts in CFHTLS and HSC-SSP. 
Visual inspection suggests that most of these sources are genuinely optically-faint.
For 33 of the 269 sources, the optical counterpart is a bright star (or in the vicinity of one),
and the photometry is unavailable from the CFHTLS or HSC-SSP catalogs due to saturation. 
There are also 1217 \hbox{X-ray} sources without a VIDEO counterpart,
of which 787 are not in the footprint of VIDEO. 
For the remaining 430 \hbox{X-ray} sources without VIDEO photometry, visual inspection 
suggests that most of them are indeed NIR-faint, except for these 42 sources 
that either coincide with a bright star or are located on artifacts such as satellite tracks. 
To obtain useful OIR information for sources without reliable optical or NIR photometry, 
we search for counterparts in several additional OIR surveys with footprint in 
our \hbox{X-ray} catalog region, including 
the Sloan Digital Sky Survey \citep{catsdssmain} Data Release 12 \citep[SDSS,][]{catsdssdr12},  
the Two Micron All Sky Survey \citep[2MASS,][]{cat_2mass}, and the 
UK Infrared Telescope Deep Sky Survey (the Deep Extragalactic Survey layer, UKIDSS-DXS; \citealt{catukidss}). 
For our \hbox{X-ray} sources catalog, we only search for counterparts in these catalogs
that are within 1$^{\prime\prime}$ of the OIR positions of the primary counterparts. 
With the supplementary catalogs, we recover the optical photometry 
for the 33 sources that do not have pipeline photometry from CFHTLS and HSC-SSP. 
We also identify an additional 333 sources with NIR photometry from 2MASS or 
UKIDSS-DXS. The basic properties of counterparts in these supplementary catalogs are also reported
in the final source catalog (Table~\ref{tab:mainxtab}).

There are also $1034$ sources with multiple counterparts having $LR > LR_{\rm th}$ 
and $LR > 0.5 LR_{\rm primary}$ in various OIR catalogs. 
For these sources, we select a ``secondary'' counterpart based on the 
following priority order:
(i) 235 best matches from VIDEO;
(ii) 48 second-best matches from SERVS;
(iii) 79 second-best matches from VIDEO; 
(iv) 290 best matches from CFHTLS; 
(v) 223 best matches from HSC; 
(vi) 79 second-best matches from CFHTLS; and
(vii) 80 second-best matches from HSC.
Finally, there are 25 \hbox{X-ray} sources with three reliable counterparts; these tertiary counterparts are from VIDEO (4),
CFHTLS (5) and HSC (16).

For the 1034 \hbox{X-ray} sources with secondary 
and/or tertiary counterparts, 869 of them have a SERVS source as the primary counterpart. 
Due to the larger PSF size of {\it Spitzer} IRAC ($\approx 2^{\prime\prime}$ at [3.6$\mu m$]) 
compared to the other OIR catalogs used in this work, 
it is possible that some of these secondary/tertiary counterparts from 
VIDEO, CFHTLS, or HSC-SSP are blended with the primary counterparts in the {\it Spitzer} image. 
Among these 1034 \hbox{X-ray} sources, a total of 318 of them 
are matched to a primary SERVS counterpart which appears to be 
two sources separated by $< 2^{\prime\prime}$ 
in higher angular resolution bands. 
These counterparts are flagged in our final catalog. 
Excluding these 318 \hbox{X-ray} sources with potentially blended SERVS counterparts, 
the vast majority ($\approx 85\%$) of \hbox{X-ray} sources with secondary 
and/or tertiary counterparts have a primary counterpart with $MR > 0.9$, 
suggesting that these additional counterparts are unlikely to be true counterparts 
of the \hbox{X-ray} sources. 
For completeness, these secondary and tertiary counterparts are also 
reported in our final catalog in Table~\ref{tab:mainxtab}.

\subsection{Counterpart identification reliability}\label{subsec:matchingcheck}
We assess the reliability of the $LR$ matching results using the 
Monte Carlo simulation approach described in \cite{broo07} and \cite{catcdfs4ms}. 
Compared to the simple estimation based on matching 
OIR catalogs to a random \hbox{X-ray} catalog,
the \cite{broo07} method usually provides 
a more realistic assessment of the
matching reliability.
As described in \cite{broo07} and \cite{broo11}, 
we consider our \hbox{X-ray} sources to 
consist of two different intrinsic populations, the ``associated population'' and the
``isolated population''.
The associated population is comprised of
\hbox{X-ray} sources that do have a real counterpart in the corresponding OIR catalog,
and the \hbox{X-ray} sources that should not have any OIR counterparts belong to the 
isolated population.

For the associated population, 
counterpart-matching procedures can produce three different outcomes:
(1) an \hbox{X-ray} source is matched to its correct counterpart (correct match, or CM), 
(2) an \hbox{X-ray} source is matched to an incorrect counterpart (incorrect match, or IM), and
(3) no counterparts were recovered (false negative, or FN).
The spurious fraction of the associated population is defined as $N_{\rm IM} / (N_{\rm IM} + N_{\rm CM})$.
For the isolated population, there are two possible matching results:
(1) no counterparts are found (true negative, or TN), and
(2) an OIR source is identified as a counterpart (false positive, or FP).
The spurious fraction of the isolated population is defined as 
the number of FPs divided by the size of the \hbox{X-ray} catalog.
By definition, the spurious matches for these two populations are intrinsically different.
The chance for the \hbox{X-ray} sources in the isolated population
to have a counterpart is mostly determined by the source surface density of 
the OIR catalog being matched. 
On the other hand, since \hbox{X-ray} sources in the associated population must have a real OIR counterpart 
within a reasonable search radius, the spurious fraction is essentially determined 
by how well the $LR$ matching method can discern a real counterpart from background sources. 

In order to estimate the fractions of \hbox{X-ray} sources in both populations for our catalog, 
we simulate each population separately. 
The details of the simulation procedure  
can be found in the appendix of \cite{broo07} and \S5 of \cite{broo11}.
A brief summary of the simulations is given below:
(1) For the ``associated population'', 
we remove all OIR sources considered to be a match in \S\ref{subsec:lrmatching},
then move the position of each OIR source by 1$^{\prime}$ in a random direction.
We then generate fake OIR ``counterparts'' for each \hbox{X-ray} source in our catalog 
based on the \hbox{X-ray} and OIR positional uncertainties, and the expected magnitude distributions derived in \S\ref{subsec:lrmatching}.
(2) For the ``isolated population'', we create mock \hbox{X-ray} sources 
that are at least 20\arcsec\ away from any real \hbox{X-ray} sources.

A total of 100 simulations are carried out for each population, 
and we run the $LR$ matching procedures on each simulation 
as described in \S\ref{subsec:lrmatching}.
The simulations of the isolated populations
usually produce a much higher spurious fraction (i.e., the number of false-positives divided by the size of the \hbox{X-ray} catalog).
For the SERVS, VIDEO, CFHTLS, and HSC-SSP catalogs, the median spurious fractions of the isolated populations are
19\%, 24\%, 30\%, and 40\%, respectively. 
For the associated populations, the spurious fractions 
(defined as $N_{\rm IM} / (N_{\rm IM} + N_{\rm CM}$)) for SERVS, VIDEO, CFHTLS, and HSC-SSP 
are 3\%, 5\%, 7\%, and 9\%, respectively. 

For the $LR$ matching results with the real data, \hbox{X-ray} 
sources that were not reliably matched to any counterparts 
(with a total number of $N_{\rm negative}$) should contain a mixture 
of the FNs of the associated population and the TNs of the isolated population.
Therefore, we can use the median FN and TN from simulations to estimate 
the fraction of \hbox{X-ray} sources in the associated population ($f_{\rm AP}$):

\begin{equation}
N_{\rm negative} = N_{\rm FN} \times f_{\rm AP} + N_{\rm TN} \times (1 - f_{\rm AP}).
\end{equation}
With $f_{\rm AP}$, we can estimate the expected number of \hbox{X-ray} sources that have 
a spurious match as the weighted sum of the numbers of IM and FP. 
The false-matching rate, $f_{\rm False}$, should therefore be: 

\begin{equation}
f_{\rm False} = (N_{\rm IM} \times f_{\rm AP} + N_{\rm FP} \times (1-f_{\rm AP})) / (N_{\rm positive}).
\end{equation}
Here we consider $N_{\rm positive}$ as the combination of both the ``reliable'' 
and ``acceptable'' matches reported in Table~\ref{tab:matching}. 

We carry out simulations for each OIR catalog.
The values of $f_{\rm False}$ and $f_{\rm AP}$ for each OIR catalog 
are also reported in Table~\ref{tab:matching}. 
Due to the high $f_{\rm AP}$ values, the false-matching rates 
of our matching results are mostly determined by the 
spurious fractions of the associated populations,
which are much lower than those %obtained by simply running the shift-and-rematch simulations 
of the isolated populations.
Adopting the \hbox{{\it Chandra}}-matched counterpart magnitude density, $q(m)_{\it Chandra}$, does reduce the false-matching rates 
compared to those derived using $q(m)_{\it XMM-Newton}$. For the SERVS and VIDEO catalogs, the improvements are marginal ($< 0.5\%$), while the improvements for CFHTLS and HSC-SSP are more significant ($\approx 2\%$ and $6\%$,  respectively).

We further scrutinize the $LR$ matching reliabilities by making use of the 223 CSC sources 
and their multiwavelength matching results described in \S\ref{subsec:lrmatching}.
We assess the reliability of the matching results of these \chandra\ 
sources using the Monte Carlo method above, and measure false-match 
fractions of 
0.9\%, 1.4\%, 2.8\%, and 3.3\%, 
for SERVS, VIDEO, CFHTLS, and HSC-SSP, respectively. 
For each catalog, we also directly compare the reliable matches obtained with 
{\it XMM-Newton} and \chandra\ positions;
97\%, 94\%, 91\%, and 87\% of the reliable \chandra\ matching results and the reliable {\it XMM-Newton} results are the same for the SERVS, VIDEO, CFHTLS, and HSC catalogs, respectively.
The high ``identical fractions'' between the matching results 
obtained using \chandra\ positions and \xmm\ positions 
are slightly lower than the false-matching rates 
calculated based on the Monte Carlo simulation 
because we only compare \hbox{X-ray} sources with reliable counterparts at 
the \chandra\ and \xmm\ positions in each catalog. 
Similar to what was done for the full \xmm\ catalog, we also select ``primary'' counterparts for the \chandra\ sources using the same priority orders. 85\%, 10\%, 1\%, and 4\% of the \chandra\ sources have their ``primary'' counterparts from SERVS, VIDEO, CFHTLS, and HSC-SSP, respectively. When comparing the primary counterparts of these \chandra\ sources and the primary counterparts of the corresponding \xmm\ sources, $\approx 97\%$ are identical, demonstrating that the matching results of the \xmm\ catalog are highly reliable.

\subsection{Supplementary multiwavelength matching results
with the {\sc NWAY} Bayesian catalog matching method}\label{subsec:nwaymatching}
\begin{figure*}
\includegraphics[width=\textwidth]{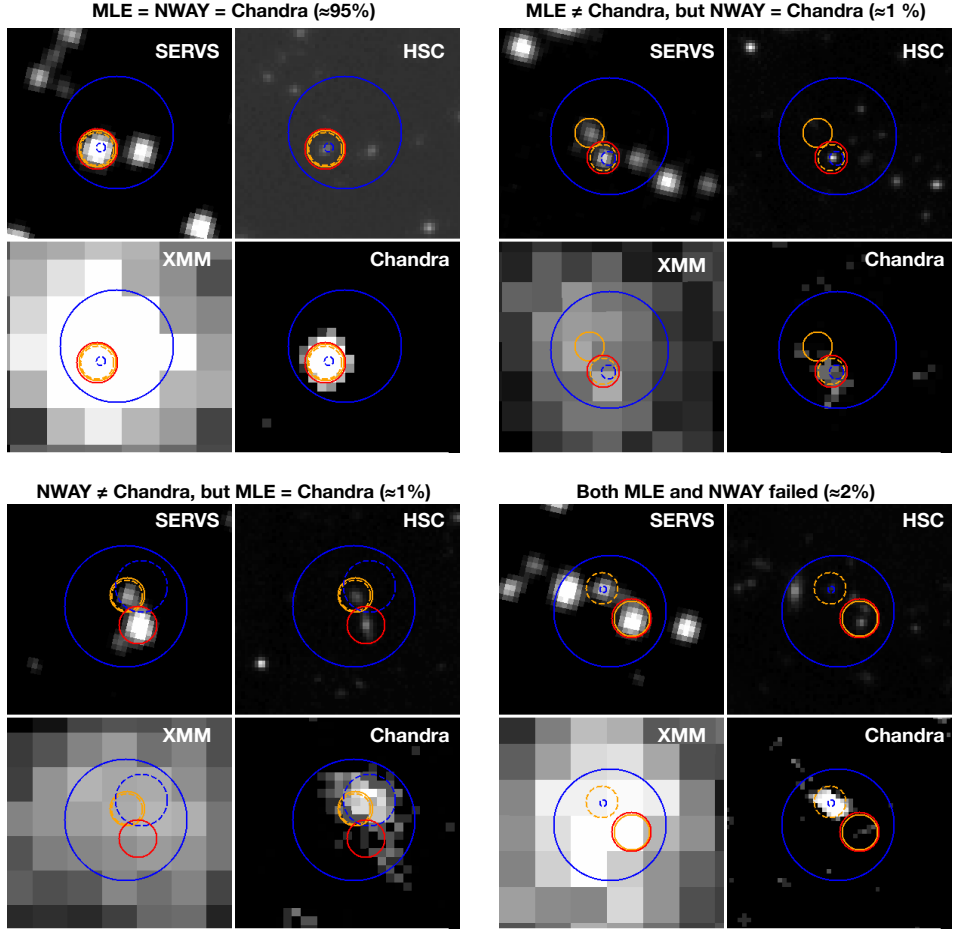}
\vspace{-1.0cm}
\caption{Illustrations of the comparison between the matching results using {\it XMM-Newton} positions or \chandra\ positions for four \hbox{X-ray} sources in our sample. The panel for each source shows 
images from SERVS [3.6$\mu$m](top-left), HSC {\it i}-band (top-right), {\it XMM-Newton} \hbox{0.5--10}~keV (bottom-left), and \chandra\ 0.5--7~keV (bottom-right).
\hbox{X-ray} positions are marked as blue circles with a 99.73\% error radius, with the {\it XMM-Newton} positions indicated using solid lines and the \chandra\ positions identified using dashed lines. 
SERVS counterparts identified with the $LR$ method are marked as orange circles with a $2^{\prime\prime}$ radius, solid lines are the counterparts of the {\it XMM-Newton} positions, and dashed lines are the counterparts of the \chandra\ positions. SERVS counterparts of the {\it XMM-Newton} positions identified using {\sc NWAY} are shown as the red circles. 
For the vast majority of {\it XMM-Newton} sources with \chandra\ counterparts from CSC, our counterpart-matching results are identical to the results obtained using \chandra\ coordinates and positional uncertainties. }
\label{fig:chandravsxmm}
\end{figure*}

We supplement the $LR$ matching results with the Bayesian catalog matching tool  
{\sc NWAY} \citep{toolnway}.\footnote{\url{https://github.com/JohannesBuchner/nway}.} 
The fundamental difference between the Bayesian approach and the likelihood-ratio approach is that 
the former makes use of the distance and magnitude priors from multiple catalogs simultaneously to select the most-probable counterpart in all catalogs considered. 
{\sc NWAY} also allows cases in which counterparts can be absent, and the matching results were computed considering all possible combinations.
The details of the  {\sc NWAY} matching methodology are described in Appendix \ref{sec:nwaycols} of \cite{toolnway}. 

{\sc NWAY} computes three quantities for deciding the most-probable match, 
$p\_single$, $p\_any$ and $p\_i$, where each possible counterpart has a different $p\_single$ value based on its distance from the {\it XMM-Newton} position. This value could be weighted by the priors supplied (e.g., $q(m)$ and $n(m)$ in Eq.~\ref{eq:LR} are similar to a magnitude prior). 
In our case, $p\_single$ is the posterior probability for a counterpart to be correctly associated with the \hbox{X-ray} source based on the angular separation from the \hbox{X-ray} position weighted by the magnitude-distribution prior, and the surface densities of the \hbox{X-ray} and OIR catalogs. 
For each \hbox{X-ray} source, $p\_single$ of all possible counterparts is considered to compute a single $p\_any$ value, 
which represents the posterior probability of the \hbox{X-ray} source having any correct counterparts 
(i.e., $p\_any = 0$ if there are no OIR counterparts within the search radius of the \hbox{X-ray} source).
The last quantity, $p\_i$, is the relative probability of a possible counterpart being the correct match.
For an \hbox{X-ray} source with multiple possible counterparts, the counterpart with the highest $p\_i$ ($p\_i_{\rm Best}$)
is considered to be the most-probable match and is assigned the {\sc match\_flag = 1} flag by {\sc NWAY}. 
Counterparts with $p\_i$ higher than 50\% of $p\_i_{\rm Best}$ are also flagged 
by {\sc NWAY} as {\sc match\_flag = 2}. 

Similar to our $LR$ approach, we make use of the \chandra\ sources in the XMM-LSS field 
to compute the priors of the expected counterparts. 
We use the ``auto'' functionality of {\sc NWAY} with a 1.5$^{\prime\prime}$ search radius for defining the ``real'' counterparts. In addition to the magnitude priors, we include an additional prior based on the {\it Spitzer} IRAC color from SERVS, [$3.6\mu$m]/[$4.5\mu$m]. 
Since the majority of our \hbox{X-ray} sources are expected to be AGNs, the distinct [$3.6\mu$m]/[$4.5\mu$m] 
mid-IR color of luminous AGNs (see Fig.~\ref{fig:iraccolor}) provides additional discerning power. 
For a small number of sources, this additional prior is useful for discerning two adjacent SERVS sources with comparable magnitudes (see the top-right panel of Fig.~\ref{fig:chandravsxmm} for illustration).  

After computing the magnitude and IRAC color priors using the {\it Chandra} sources, we run {\sc NWAY} on the full \hbox{X-ray} catalog with a search radius of 10$^{\prime\prime}$. 
Based on the results from $LR$ matching (\S\ref{subsec:lrmatching}), we do not assume a completeness prior because only 0.01\% of the \hbox{X-ray} sources are completely isolated.
All four OIR catalogs are considered simultaneously. We report the multiwavelength matches with {\sc match\_flag$=$1,2} in Table~\ref{tab:nwaytab} 
supplementary to the $LR$ matching results.

Since {\sc NWAY} matches all four OIR catalogs simultaneously, we cannot determine the spurious-matching rates for the
``associated'' and ``isolated'' populations as we did for estimating the spurious-matching rates for $LR$ results using Monte Carlo simulations (see \S\ref{subsec:matchingcheck}).
\cite{toolnway} suggest that the {\sc NWAY} matching reliability can 
be determined by a $p\_any$ threshold, which is chosen based on re-running {\sc NWAY} on randomly shifted ``fake'' \hbox{X-ray} catalogs. 
However, this approach is equivalent to estimating the spurious matching 
rates for the ``isolated'' population using the \cite{broo07} method, which is usually much higher than the results obtained with the two-population approach (see \citeauthor{broo07} 2007, \citeauthor{catcdfs4ms} 2011, and \S\ref{subsec:matchingcheck} for details). 
Therefore, we do not adopt any $p\_any$ thresholds for the {\sc NWAY} matching results. 
The {\sc NWAY} matching results can still be assessed by investigating 
the CSC-matched subsample of 223 \hbox{X-ray} sources; 
the difference between the matching results obtained using \chandra\ and \xmm\ positions with {\sc NWAY} are similar to the $LR$ results described in \S\ref{subsec:matchingcheck}.

We also use the 223 {\it Chandra}-detected subsample 
as a baseline for comparing matching results obtained using the {\sc NWAY} or $LR$ methods. 
We focus only on comparing the SERVS counterparts, 
as the vast majority of $LR$ matching results are decided based on the primary counterparts from SERVS. 
We confirm that all {\it Chandra} sources have the same SERVS matching results using $LR$ and {\sc NWAY}.
Therefore, we can use the \chandra\ results obtained with $LR$ to assess the matching reliability of both $LR$ and {\sc NWAY} matching results with \xmm\ positions.
Examples of such comparisons are shown in Fig.~\ref{fig:chandravsxmm}.
$96\%$ of the sources have the same matching results from LR, {\sc NWAY}, and {\it Chandra}.
A small fraction (two sources) of $LR$ matching results do not agree with those of \chandra\ 
but could be recovered by {\sc NWAY}. 
On the other hand, two of the {\sc NWAY} matching results do not agree with the \chandra\ results but could be identified by LR. Five of the \chandra\ sources have different SERVS counterparts than both the $LR$ and {\sc NWAY} results.
{\it Chandra} and OIR images of these sources suggest that they are either two \hbox{X-ray} sources blended due to the \xmm\ PSF,
or there are multiple OIR counterparts with very similar magnitudes and distances to the \hbox{X-ray} position, and thus it is not surprising neither $LR$ nor NWAY could successfully recover the correct counterparts. 
As demonstrated in Fig.~\ref{fig:chandravsxmm} (bottom-left), 
these five sources 
have multiple counterparts with comparable magnitudes and similar spatial separations from the \xmm\ position.
This result suggests $LR$ and {\sc NWAY} perform similarly for finding SERVS counterparts. 
For the two \hbox{X-ray} sources with different LR and {\sc NWAY} counterparts, 
their \hbox{X-ray} fluxes are relatively low (with a median full-band flux of $7.5\times10^{-15}$ {erg~cm$^{-2}$~s$^{-1}$}, which is $\approx 44\%$ of the median flux of the full \hbox{X-ray} catalog). 
This is expected as fainter \hbox{X-ray} sources have larger positional uncertainties, which leads to higher numbers of counterpart candidates.

When further scrutinizing the $96\%$ of sources with identical SERVS counterparts from 
LR, {\sc NWAY}, and \hbox{{\it Chandra}}, 
we find that {\sc NWAY} 
occasionally (for $\sim 10\%$ of the \hbox{X-ray} sources) 
considers the best-fit combination to be the one with 
counterparts in some of the other OIR catalogs
being ``absent''.
For instance, one of the \xray\ sources has a reliable SERVS counterpart
identified by both {\sc NWAY} and $LR$. For the SERVS counterpart, 
there is only one VIDEO source within the 0.5$^{\prime\prime}$ positional error circle of SERVS.
For the $LR$ approach described in \S\ref{subsec:lrmatching}, 
the VIDEO source is assigned to the correct SERVS counterpart.
However, {\sc NWAY} does not consider this VIDEO source
to be among the most-probable combination of counterparts from all four OIR catalogs
that were being matched simultaneously. 
This result is likely due to how {\sc NWAY} computes $p\_i$. 
When multiple OIR catalogs are taken into account simultaneously, 
$p\_i$ represents the relative probability of counterparts 
from {\it all} OIR catalogs being the correct match.
In this example, the VIDEO counterpart has an unlikely 
magnitude according to the VIDEO magnitude prior; therefore,
including the VIDEO source as a correct match would result in a lower $p\_i$
compared to the case where the VIDEO source is excluded from the matched counterparts.
Similar mismatches are found when comparing the {\sc NWAY} and $LR$ matching results for the full {\it XMM-Newton} catalog. 
Note that the {\sc p\_any} values for these sources are generally lower
(with a median of 0.16) compared to the sources without such problems (their median {\sc p\_any}$=0.98$),
but sources with {\sc p\_any}$>0.98$ can still have this behavior.
{\sc NWAY} does not have this behavior when no magnitude or color priors are used;
however, without the inclusion of magnitude and color priors,
{\sc NWAY} can only rely on the distance-based priors, thereby losing critical discerning powers for matching \xmm\ sources to the dense OIR catalogs. 
Further corroborating the Bayesian method's effectiveness of counterpart-matching 
with multiple OIR catalogs is beyond the scope of this work. Therefore, 
we list the {\sc NWAY} matching results ``as-is'' in Table~\ref{tab:nwaytab}, 
and we consider only the $LR$ matching results listed in Table~\ref{tab:mainxtab} 
when exploring the multiwavelength properties of the \hbox{X-ray} sources reported in this work. 
The matching results obtained using {\sc NWAY} are shown in Table~\ref{tab:nwaytab},
and the descriptions of this table's columns are listed in Appendix \ref{sec:nwaycols}. 
Only the counterparts with {\sc match\_flag$\geq 1$} are included.

\section{Redshifts}\label{sec:redshifts}
\subsection{Spectroscopic redshifts}\label{subsec:zspec}
\begin{figure}
\includegraphics[width=0.48\textwidth]{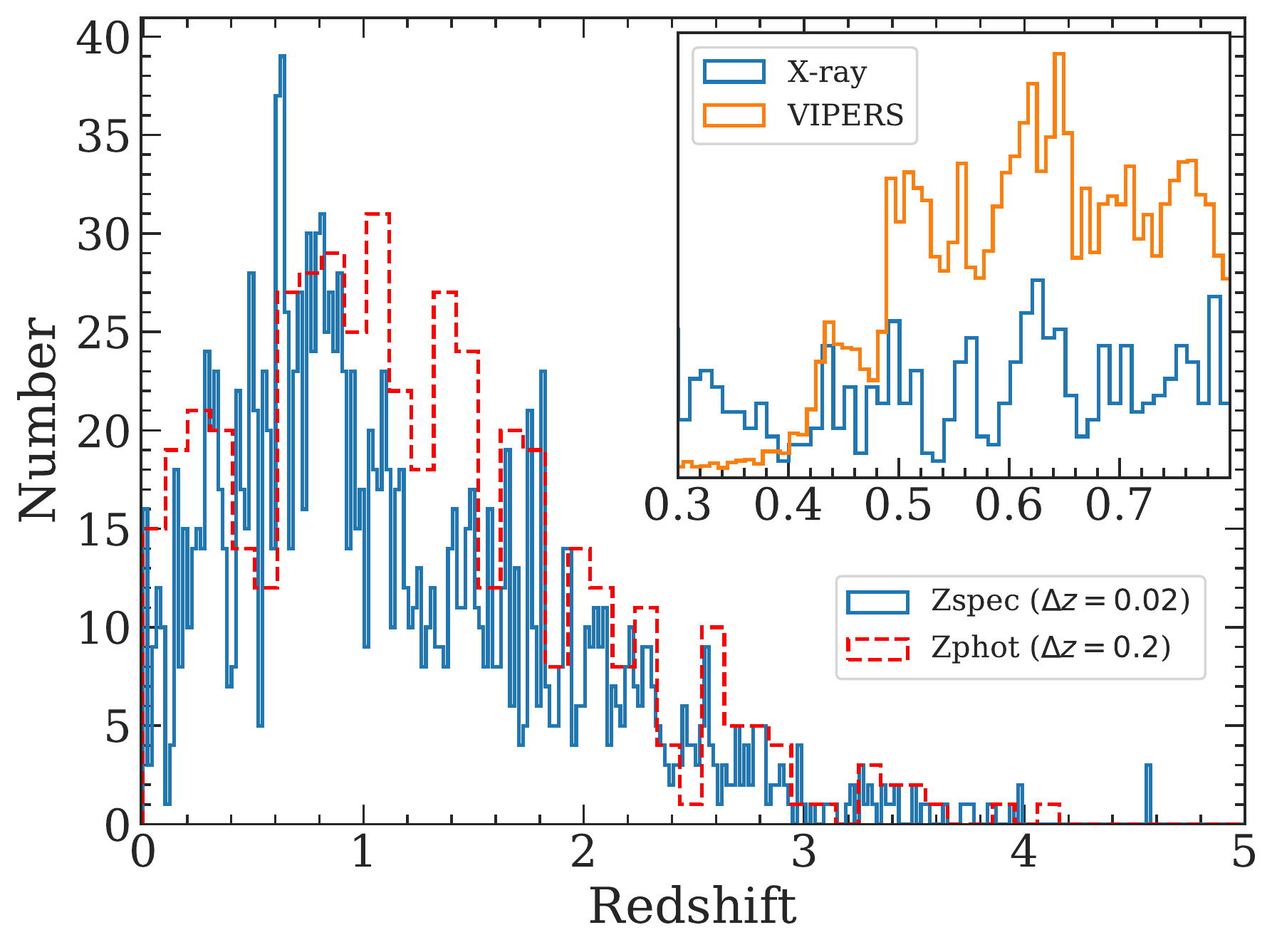}
\caption{Distribution of the redshifts in bins of $\Delta z=0.02$ for the 1782 \hbox{X-ray} sources with spectroscopic-redshift measurements from the literature. 
The photometric-redshift distribution for the 2105 sources with high-quality photometric redshifts is also plotted as the red dashed histogram in bins of $\Delta z=0.2$.
The redshift spikes are likely associated with large-scale structure filaments \citep[e.g.,][]{catcdfs7ms,xue17review}. A comparison between the normalized redshift distribution of the \hbox{X-ray} sources and that of the $i$-band selected galaxies from the VIPERS survey is also shown in the insert
with $\Delta z=0.01$ bins, which suggests that some of the redshift spikes (e.g., $z\approx 0.6$) of \hbox{X-ray} sources overlap with those of the general galaxy population. 
}
\label{fig:spike}
\end{figure}

\begin{figure}
\includegraphics[width=0.48\textwidth]{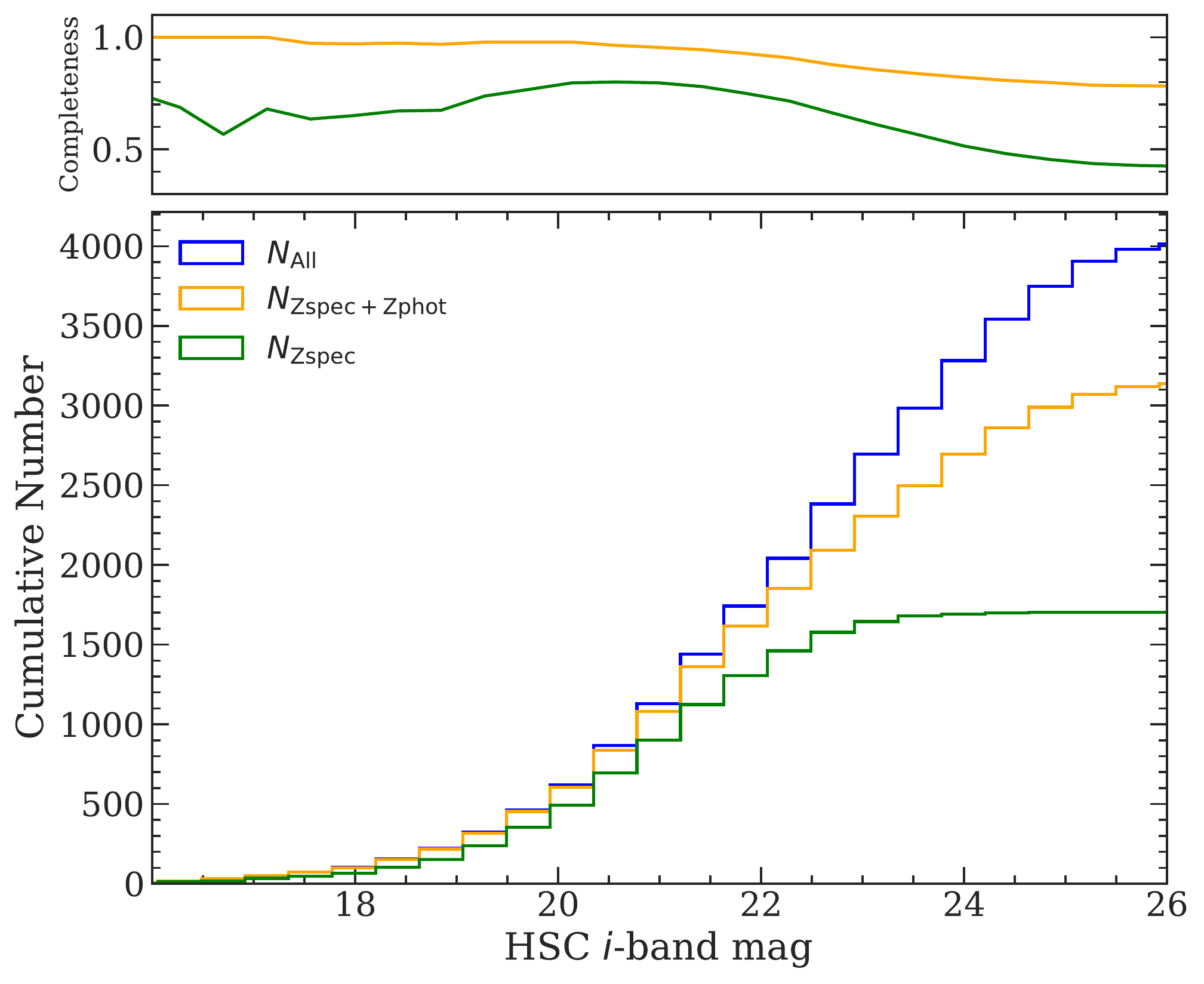}
\caption{
The blue histogram shows the cumulative distribution of the HSC {\it i}-band magnitudes 
for 4012 {\it i}-band-detected \hbox{X-ray} sources in the $4.5$~deg$^{2}$ subfield 
covered by the N18 forced-photometry catalog (see \S\ref{subsec:zphot}).
The cumulative distribution of the sources with spectroscopic redshifts is indicated by the 
green histogram, and the cumulative distribution of the sources with high-quality 
photometric or spectroscopic redshifts is indicated by the orange histogram.
The top panel shows the fraction of sources with spectroscopic redshifts (i.e., the green histogram divided by the blue histogram) as a function of {\it i}-band magnitude. The fraction of sources with good photometric or spectroscopic redshifts as a function of {\it i}-band magnitude is shown as the orange curve. 
}
\label{fig:redshift}
\end{figure}

\begin{table*}
\scriptsize
    \caption{\label{tab:redshift}
Redshift catalogs used in this work. 
Column 1: Redshift survey name. 
Column 2: Survey instrument.
Column 3: Survey sensitivity.
Column 4: Targeting fields.
Column 5: Survey area.
Column 6: Total number of redshifts matched the main \hbox{X-ray} catalog.
Column 7: Total number of redshifts assigned to the \hbox{X-ray} sources in the main catalog.
Column 8: Reference
}
    \begin{tabular}{cccccccl} % four columns, alignment for each
        \hline
        Catalog & Instrument & Survey sensitivity & Targeting fields & Area & 
        $N_{\rm matched}$ & $N_{\rm assigned}$ & Reference \\
        (1) & (2) & (3) & (4) & (5) & (6) & (7) & (8) \\
        \hline
        SDSS   & BOSS                             & $r \lesssim 22.5$ & 
        XMM-XXL-North  & 25 deg$^2$ & 1075 & 1075 & \cite{catsdssboss}; \\
        &&&&&&&\cite{menz16xxl}\\
        PRIMUS & IMACS                               & $i \lesssim 23.5$ & 
        XMM-LSS & 2.9 deg$^2$ & 749 & 347 & \cite{catspeczprimus}\\
               & (Low-Dispersion Prism)   & & & & & &\\
        VIPERS & VIMOS                            & $i \lesssim 22.5$  & 
        XMM-LSS & 7.8 deg$^2$ & 332 & 161 & \cite{catspeczvipers} \\
        UDS Compilation & Multiple instruments    & N/A      & 
        UDS & 0.8 deg $^2$  & 302 & 72 &    N/A (see Footnote 25) \\
        CSI    & IMACS                            & [3.6$\mu$m]$_{\rm AB} \lesssim 21$ &
        XMM-LSS & 6.9 deg$^2$ & 516 & 68 & \cite{catspeczcsi};\\
               & (Uniform-Dispersion Prism) & & & & & &\cite{catspeczcsi2}\\
        VVDS   & VIMOS                            & $17.5 \lesssim i \lesssim 24.5$ & 
        XMDS+SXDS & 3 deg$^2$ & 81 & 38 & \cite{catspeczvvds} \\
        UDSz   & VIMOS/FORS2                      & $K < 23$ & 
        UDS & 0.5 deg$^2$ & 22 & 15 & \cite{catspeczudsz1};\\
        &&&&&&&\cite{catspeczudsz2}\\
        3D-HST & WFCS G141 Grism                  &  $JH_{\rm IR} \lesssim 24$ &
        UDS & 191.2 arcmin$^2$ & 15 & 6  & \cite{catspecz3dhst}; \\
        &&&&&&& \cite{catspecz3dhst2}\\
        \hline
    \end{tabular}    
\end{table*}

The XMM-LSS region is covered by a number of spectroscopic redshift (spec-z) surveys that target galaxies with various optical magnitude constraints: the PRIsm MUlti-Object Survey \citep[PRIMUS;][]{catspeczprimus}, 
the VIMOS Public Extragalactic Redshift Survey \citep[VIPERS;][]{catspeczvipers}, and
the VIMOS VLT Deep Survey \citep[VVDS;][]{catspeczvvds}. 
As part of the SDSS Baryon Oscillation Spectroscopic Survey (SDSS-BOSS) program, 3042 \hbox{X-ray} sources found in the XMM-XXL-North  field (25 deg$^2$) with $r < 22.5$ were all observed by the SDSS-BOSS \citep{catsdssboss,menz16xxl}. 
Also, there are three other redshift surveys in the XMM-LSS region that target near-IR selected galaxies, including the spectroscopic observations of the UKIDSS Ultra-Deep Survey \citep[UDSz;][]{catspeczudsz1,catspeczudsz2}, 
the 3D-HST Survey \citep{catspecz3dhst,catspecz3dhst2} in the UDS region, and the Carnegie-Spitzer-IMACS Redshift Survey (CSI; \citealt{catspeczcsi}). 
We list the properties of each redshift catalog in Table~\ref{tab:redshift}.

We adopt the same nearest-neighbor matching criterion with a 1$^{\prime\prime}$ matching radius to associate these redshifts to each OIR catalog. The redshift for each \hbox{X-ray} source is determined by the coordinates of its primary OIR counterpart. In cases where redshifts from different catalogs do not agree with each other, we choose redshifts using the following ordering (ranked by the spectral resolution at {\it r}-band and reliability): SDSS, VVDS, VIPERS, UDSz, PRIMUS (reliable), CSI (reliable), 3D-HST, PRIMUS (acceptable), and CSI (acceptable). 
In addition to these redshift surveys, 
we include the compilation of $\approx 4000$ 
publicly available but unpublished redshifts in the UDS field.\footnote{
These redshifts were obtained with Subaru FOCAS, AAT~2dF, VLT VIMOS, and AAOmega, 
and the full redshift catalog is available at \url{http://www.nottingham.ac.uk/~ppzoa/UDS_redshifts_18Oct2010.fits}, see \url{http://www.nottingham.ac.uk/astronomy/UDS/data/data.html} for an overview of this compilation.} An additional 72 \hbox{X-ray} sources have 
spec-zs culled from this catalog. 
We also search for publicly available spec-zs for all of our counterparts 
not included in the aforementioned redshift catalogs in the NASA Extragalactic Database (NED), but no additional secure redshifts were found.

Of the 5242 sources in our main \hbox{X-ray} source catalog, 1782 have spec-zs ranging from $0 < z < 4.57$. Fig.~\ref{fig:spike} presents the redshift histogram in bins of $\Delta z=0.02$. There are several redshift ``spikes'' indicative of large-scale structures containing 
\xray\ AGNs (e.g., Fig.~9 of \citealt{catcdfs7ms} and Fig.~20 of \citealt{xue17review}).
Notably, the \hbox{X-ray} source redshift spike at $0.6 < z < 0.7$ appears to coincide with one of the major large-scale structures seen in the VIPERS redshift survey 
(see Fig.~14 of \citealt{catspeczvipers} and the insert panel of Fig.~\ref{fig:spike}). The cumulative histogram of the {\it i}-band magnitudes of the sources with spec-zs is shown in Fig.~\ref{fig:redshift} as the green histogram.

\subsection{Photometric redshifts}\label{subsec:zphot}
\begin{figure*}    
\includegraphics[width=\columnwidth]{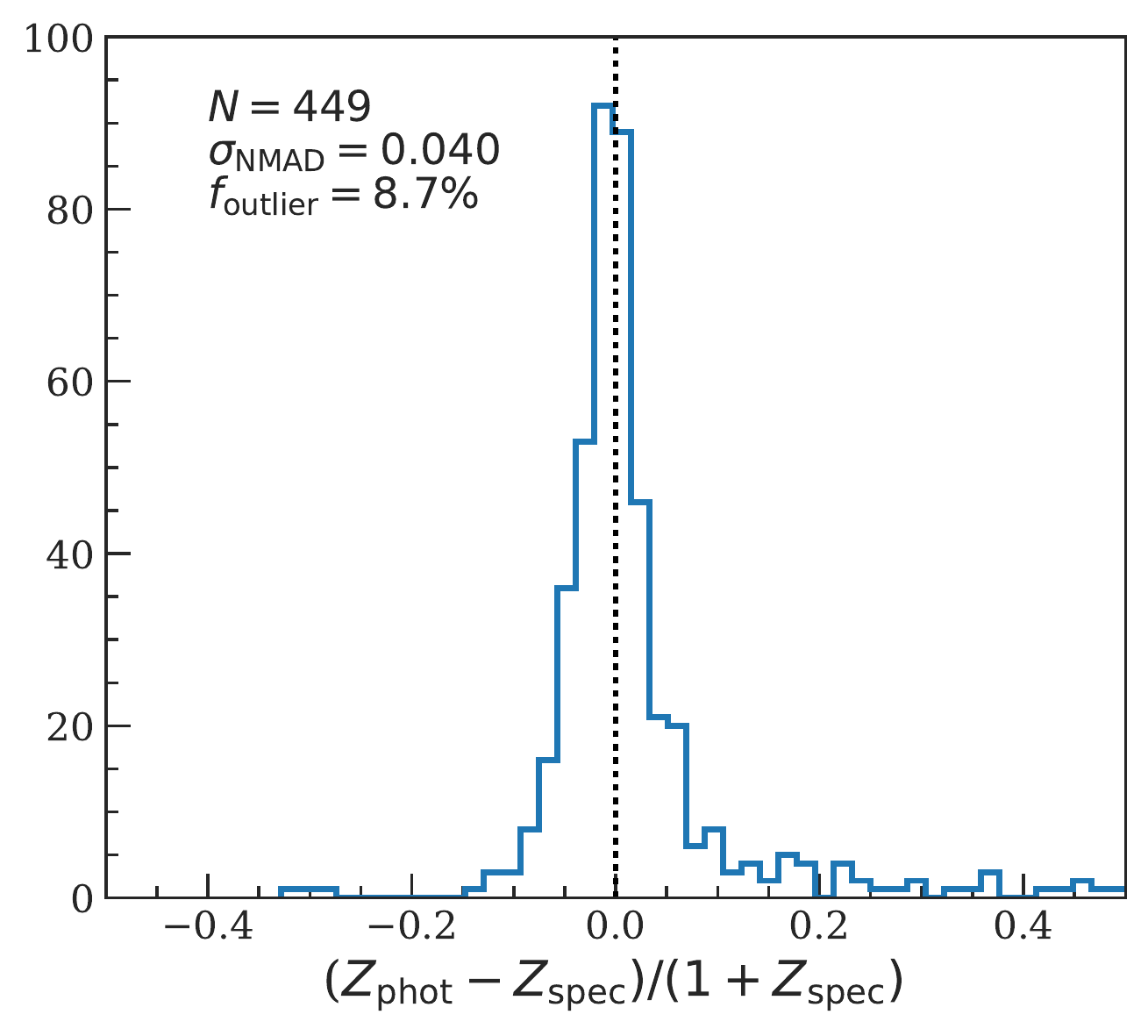}
\includegraphics[width=\columnwidth]{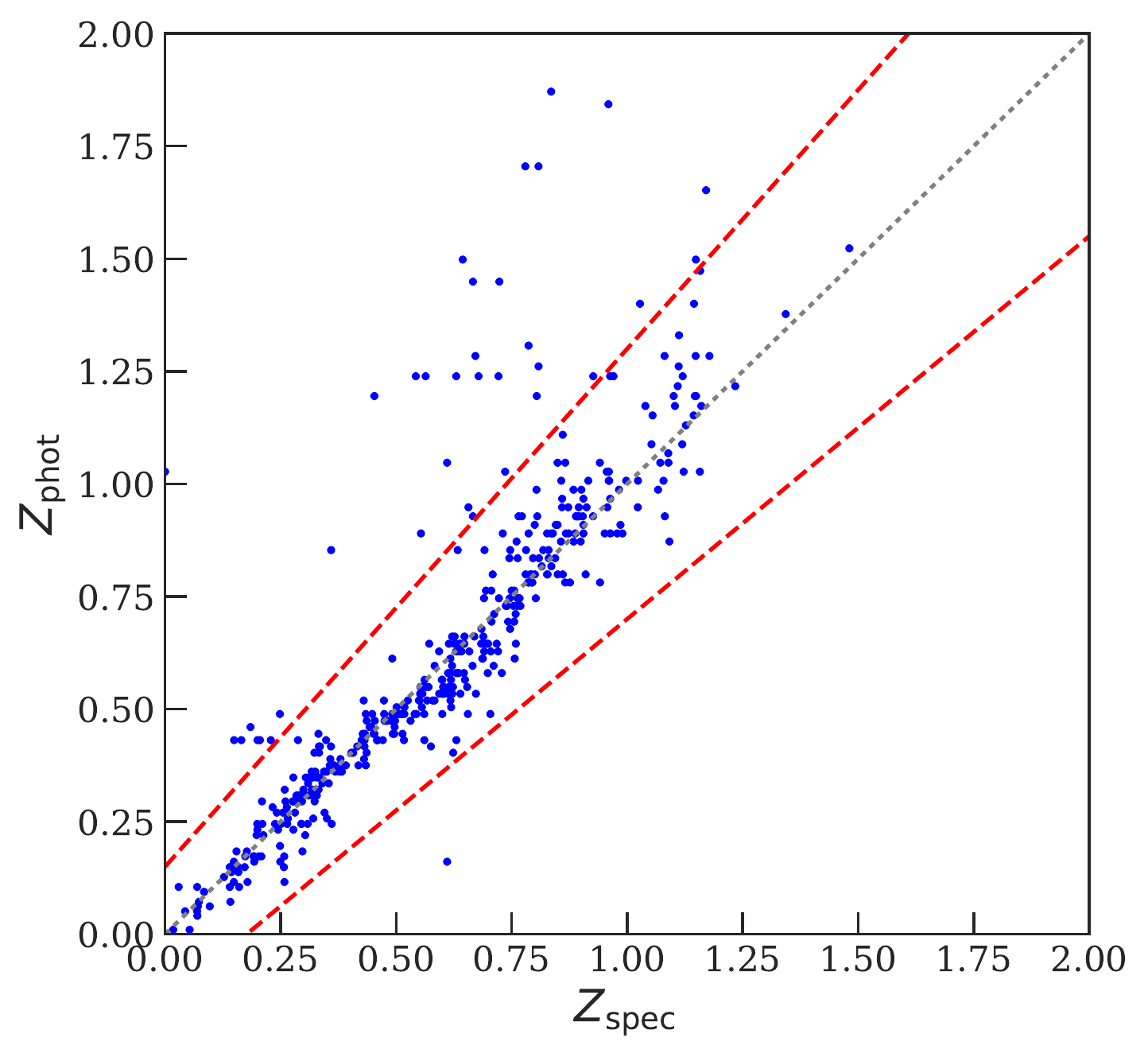}
\caption{
Spectroscopic and photometric redshifts for the \ngpz\ non-broad-line AGN sources with high-quality photo-zs and spec-zs. The left panel shows the histogram of the fractional difference between the photo-zs and the spec-zs. 
The right panel shows the direct comparison between the photo-zs and the spec-zs. 
The black dotted lines in both panels mark the $z_{\rm spec} = z_{\rm phot}$ relation.
In the right panel, the red dashed lines mark the $\mathopen|\Delta z\mathclose|/(1+z_{\rm spec}) = 0.15$ 
thresholds for outliers. 
}
\label{fig:photoz}
\end{figure*}    

High-quality photometric redshifts (photo-zs) for AGNs are not yet available for our full 
survey region, but they are available in a number of smaller subfields. 
In particular, in a $\approx 1$ deg$^2$ area within the XMM-LSS region,
\citet{nyla17} have presented a ``forced-photometry'' catalog using the 
Tractor image-modeling code \citep{tooltractor}.
The forced-photometry technique employs source-position and surface-brightness 
profile priors from the high-resolution fiducial band of the VIDEO survey to model 
and fit the fluxes of lower-resolution bands.
\cite{nyla17} demonstrated that their multi-band forced photometry of 
mixed resolution optical and IR surveys using the Tractor led to a statistically 
significant improvement in photometric-redshift accuracy compared to 
position-matched multi-band catalogs (see \S5.2 of \citealt{nyla17} for details).
For this work, we make use of a similar forced-photometry catalog for 
the full 4.5~deg$^2$ area with VIDEO and SERVS coverage (Nyland et al. 2018, in preparation; N18 hereafter). The N18 
catalog is similar to the \citet{nyla17} catalog,
except the image cutout width for each source has increased by a factor of two 
(from 10$^{\prime\prime}$ to 20$^{\prime\prime}$) and the sky noise and sky level 
are now calculated in each 
image cutout using iterative sigma clipping. 
Also, N18 used IRAC data from the SERVS DeepDrill survey (P.I. Mark Lacy), which 
expands upon the coverage of the SERVS project by providing deep IRAC imaging to 
microJy-depth of the four predefined Deep Drilling Fields for the LSST. 
In the XMM-LSS field, the DeepDrill data more than double the footprint of the 
SERVS post-cryogenic data, thus leading to higher-quality data along the edges of 
the SERVS coverage where there is overlap with the VIDEO data. 
Thus, our IRAC 3.6 and 4.5~$\mu$m photometry is based on the DeepDrill data.
We make use of the 13-band photometry 
from {\it u$^{\prime}$} to IRAC 4.5$\mu$m to derive photo-zs for 
the \hbox{X-ray} sources in this region using the methods described in \cite{yang14photoz}.
The photometric bands include CFHTLS $u$-band; HSC-SSP {\it g, r, i, z}, and {\it y} bands (wide layer); VIDEO {\it Z, Y, J, H}, and {\it Ks} bands (DR5); and {\it Spitzer} 3.6~$\mu$m and 
4.5~$\mu m$ bands from the SERVS DeepDrill survey. 

We match the N18 catalog to the coordinates of the 
primary counterparts of the \hbox{X-ray} sources that are considered to be reliable 
matches using a 1$^{\prime\prime}$ matching radius.
We exclude the 930 \hbox{X-ray} sources that are classified as broad-line AGNs (see \S\ref{sec:class} for details) 
according to their optical spectra due to their much higher photometric-redshift uncertainties.
A total of 3418 \hbox{X-ray} sources satisfy these criteria.
Of these sources, $\approx 38\%$ of them are detected (i.e., with Tractor measured signal-to-noise-ratio $>5$) 
in all 13 bands. 
The 25th, 50th, and 75th percentiles of the number of 
bands with detection for the 3418 \hbox{X-ray} sources are 10, 11, and 13.
Since the flux uncertainties in N18 do not account for 
uncertainties in the PSF homogenization processes, we adopt an additional 3\% systematic 
for the flux errors, which is typical of PSF modeling uncertainties 
(e.g., \S5.3 of \citealt{yang14photoz}). 

Following the approach of \cite{yang14photoz}, we measure the photo-zs
using the SED-fitting code {\sc eazy} \citep{photozeazy} using the default galaxy 
templates and settings, and an additional obscured AGN template from 
\cite{poll07agnsed}. As described in \S5.6 of \cite{yang14photoz}, we perform 
iterative procedures to adjust the photometric zero points; the zero-point 
corrections are $\lesssim 0.1$~mag. 
For each source, {\sc eazy} calculates a parameter $Q_z$ 
(see Eq.~8 of \citealt{photozeazy}) to indicate photometric-redshift quality. 
Of the 3418 non-broad-line \hbox{X-ray} sources with forced photometry,
we consider the 2105 ($\approx 62\%$) photo-zs with $Q_z<1$ as reliable 
(see \S6.3 of \citealt{yang14photoz}). 
The fraction of sources with high-quality photo-zs becomes higher for brighter sources.
For instance, sources with VIDEO $Ks$-band magnitude in the brightest 25th, 50th, and 75th 
percentiles (corresponding to $Ks < 19.77$, 20.83, and 21.78) 
have $77\%, 76\%$, and $71\%$ high-quality photoz-s, 
because fainter sources have larger photometric uncertainties and 
fewer photometric points. 
With the deep NIR coverage from VIDEO and SERVS, 
we can detect the Balmer break even for high redshift sources,
hence the range of our $Q_z<1$ photo-zs extends to $z\approx 4$.
There are 536 sources with $Q_z<1$ and reliable spec-zs.
Of these sources, \ngpz\ of them have spectroscopic classifications 
from at least one of the public redshifts catalogs and are not classified as a broad-line AGN.
Since we excluded broad-line sources, the spectroscopic redshift range of these sources is 0.02--1.5,
with a median value of 0.79. 
We use these \ngpz\ sources to assess the quality of the 2105 photo-z measurements.
The normalized median absolute deviation (NMAD) is $\sigma_{\rm NMAD} = 0.040$, 
with an outlier fraction (defined as $\mathopen| \Delta z\mathclose|/(1+z_{\rm spec}) > 0.15$) 
of $f_{\rm outlier} = 8.7\%$, which is comparable to the photometric-redshift reliability 
reported in \cite{yang14photoz} for the CDF-N. 
A small fraction of sources are found to be outliers. This is likely caused by the photo-z code mistakenly 
identifying the location and strength of the prominent spectroscopic feature, the Balmer break 
(the lack of radiation at wavelength range $< 3646$ \AA\ ), 
due to photometric redshift uncertainties in one or more bands. 
We note that the majority of the outliers have $z_{\rm spec} < z_{\rm phot}$.
This is expected for the $z_{\rm spec} \lesssim 1$ sources (e.g., see Fig.~14 of \citealt{yang14photoz}),
because only less than three photometric bands 
cover the rest-frame wavelength range of the Balmer break and it is difficult to identify a spectral break with only three photometric bands. Therefore, it is less likely for the aforementioned misidentifications to 
cause a $z_{\rm phot}$ lower than these low-spec-z sources.
For this work, we do not include the broad-line AGN (BLAGN) templates as 
\cite{yang14photoz} did. This is primarily driven by the worse
photometric redshift qualities when including the BLAGN templates.
In addition, \cite{yang17link} estimated that the fraction of broad-line AGNs 
missed by spectroscopic campaigns in the COSMOS field is likely less than 
$\approx 18\%$. Considering the comparable surface density of the spectroscopically
confirmed BLAGNs in this work and that in the COSMOS field, only a small
fraction of sources would require an additional BLAGN template.  
In fact, the vast majority of our sources (excluding spectroscopically confirmed BLAGNs)
can be well-characterized with galaxy templates alone, and the high fraction (see below) of 
our sources with high-quality photo-zs also justifies our choice of fitting templates. 

Fig.~\ref{fig:photoz} compares the photometric and spectroscopic redshifts for 
the \ngpz\ non-broad-line sources with reliable photo-zs. 
The 4.5 deg$^2$ area covered by N18 contains 
1543 reliable photo-zs for sources that do not have spectroscopic redshift 
measurements, increasing the fraction of sources with redshifts from $\approx 32\%$ 
to $\approx 70\%$. 
We expect to expand the photometric-redshift measurements to all of our \hbox{X-ray} sources 
in the full XMM-SERVS:XMM-LSS field when the photometry catalog with data from both the
SERVS DeepDrill survey ({\it Spitzer} Program ID 11086; Lacy et al., in preparation) and the VEILS survey (see Table~\ref{tab:servsmw}) becomes available.
We have also run our photometric-redshift codes on all sources in the 
preliminary N18 catalog, and we report the photo-zs for the \nallgpz\ sources in Appendix \ref{sec:pzapp}. The full details of the Tractor catalog over the 4.5 deg$^2$ field will be presented in Nyland et al. (2018, in preparation).

\begin{figure*}
\includegraphics[width=1.0\textwidth]{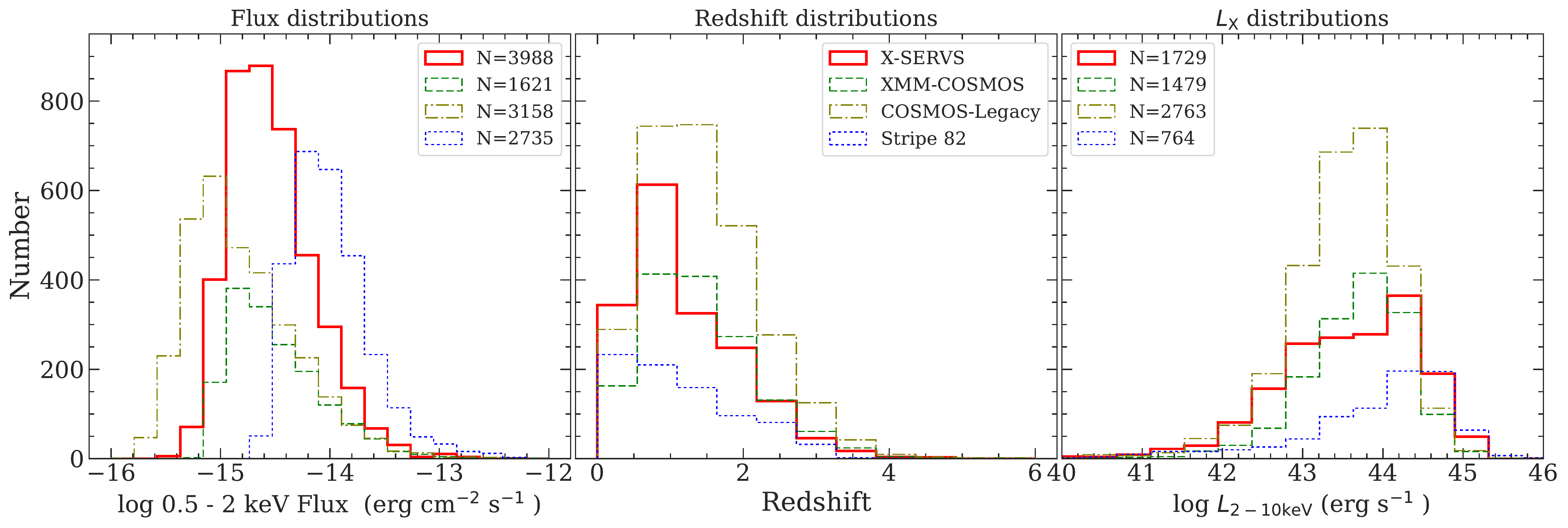}
\caption{A comparison between 
this work (solid red line), 
XMM-COSMOS (green dashed line),
COSMOS-Legacy (brown dash-dotted line),
 and Stripe 82--X (blue dotted line).
Distributions shown in panels from left to right are: 
\hbox{0.5--2}~keV flux, redshift, and $\log L_{\rm 2-10~keV}$, respectively.
The left panel shows the distribution of soft-band fluxes for the soft-band detected sources in each catalog;
no redshift information is required. 
The numbers of the soft-band sources are listed in the left panel.
For the middle and right panels, 
the histograms are for the subset of sources with redshift measurements
(regardless of the detection bands), 
with source numbers marked in the right panel. 
}
\label{fig:xcats}
\end{figure*}

\begin{figure}    
\includegraphics[width=\columnwidth]{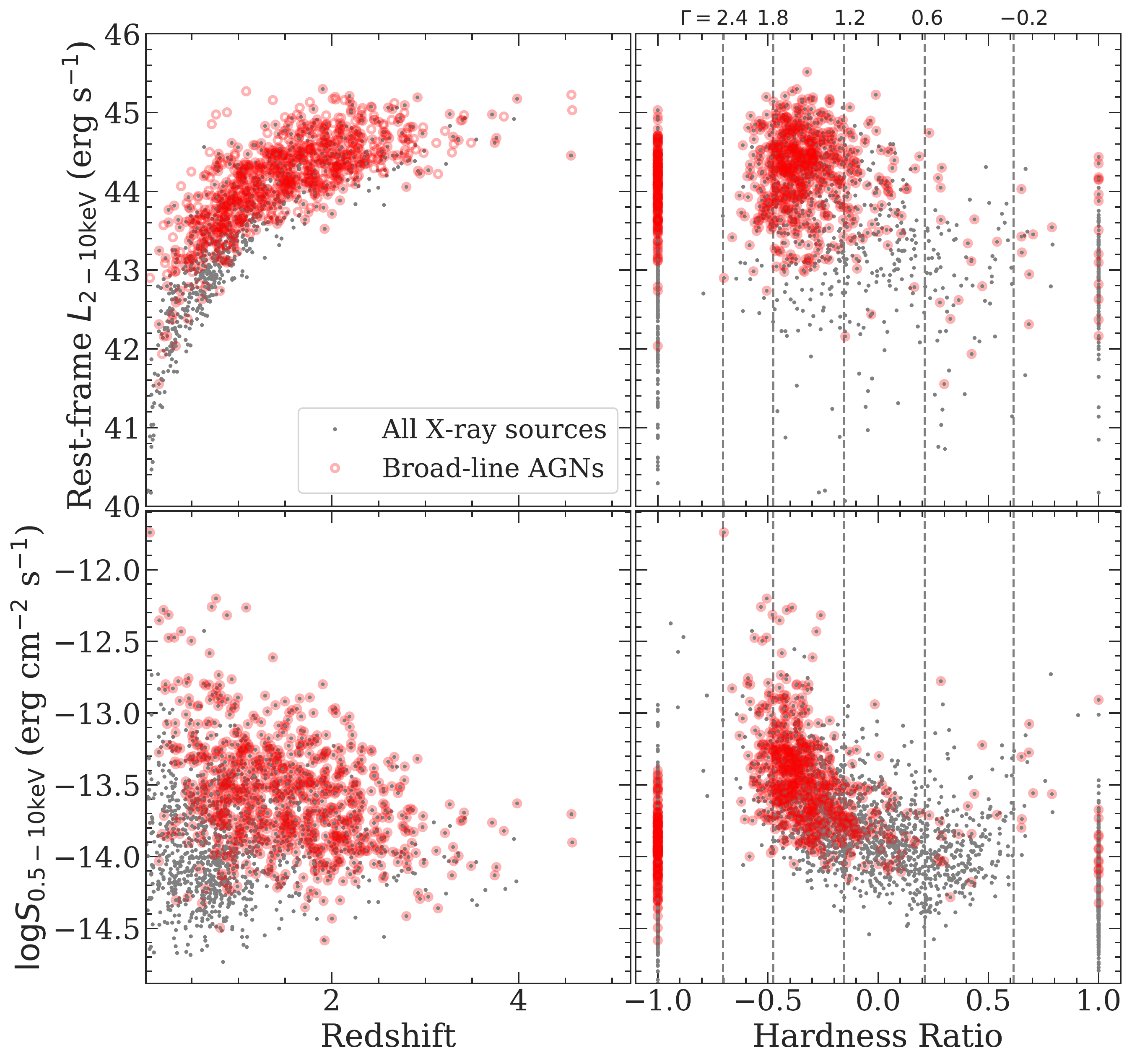}
\caption{
Properties of the 1782 \hbox{X-ray} sources with spectroscopic-redshift measurements,
including the (1) $L_{\rm 2-10~keV}$ vs. $z$ distribution (top-left),
(2) $L_{\rm 2-10~keV}$ vs. hardness ratio (top-right),
(3) \hbox{0.5--10}~keV flux vs. redshift (bottom-left),
(4) \hbox{0.5--10}~keV flux vs. hardness ratio (bottom-right).
Broad-line AGNs are marked as the red open circles.
In the right panels, the expected hardness ratios 
for power-law spectra (with Galactic column density) with different photon indices 
are plotted as the vertical dashed lines. Sources detected only in the soft or hard bands 
have their HR set at $-1$ and 1, respectively.
}
\label{fig:z_lx_hr}
\end{figure}

\section{Source properties and classification}\label{sec:class}
\begin{figure*}
\DeclareGraphicsExtensions{.png}
\includegraphics[width=0.48\textwidth]{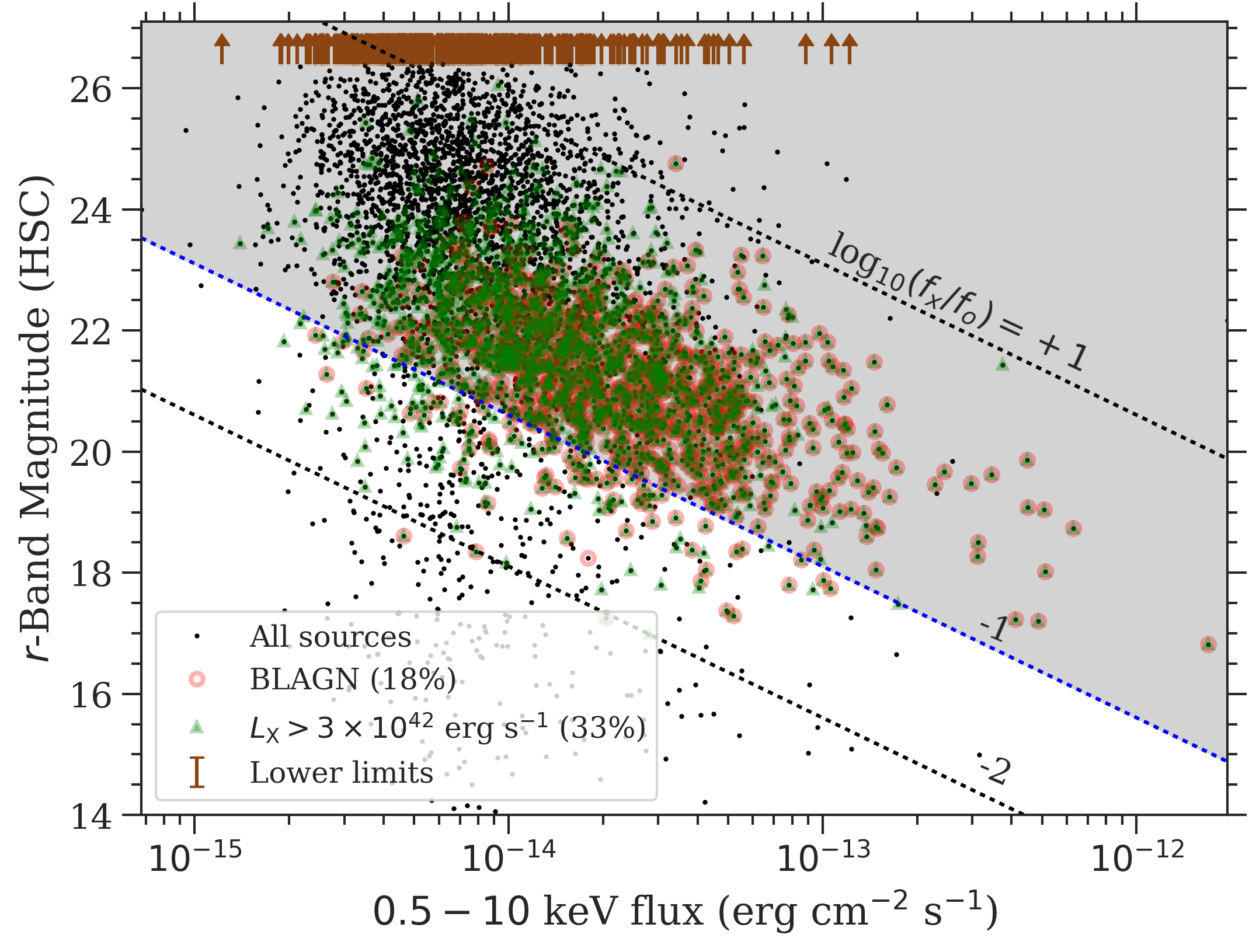}
\includegraphics[width=0.48\textwidth]{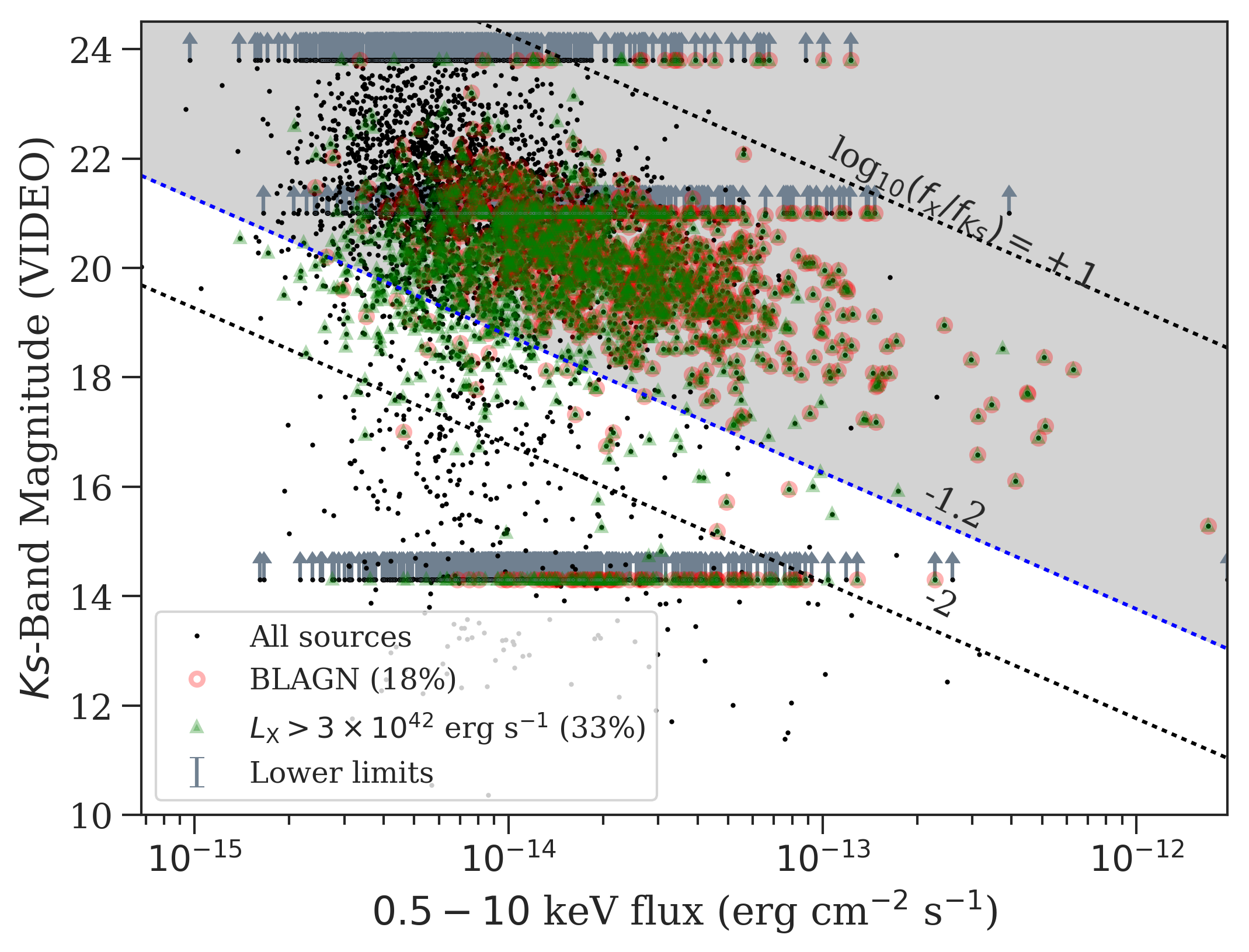}
\caption{
{\it Left} -- Distributions of the optical flux in the {\it r}-band vs. full-band (\hbox{0.5--10}~keV) \hbox{X-ray} flux. Lower limits for sources not detected in the HSC-SSP wide survey 
{\it r}-band are displayed as the brown upward arrows.  
{\it Right} -- The near-IR flux in the {\it Ks}-band 
versus  full-band \hbox{X-ray} flux. Lower limits for sources not detected in VIDEO are also show as the upward arrows. Since only $\approx 85\%$ of the \hbox{X-ray} catalog region is covered with VIDEO, 
some of the \hbox{X-ray} sources have lower limits from 2MASS ($Ks = 14.3$),
UKIDSS DXS survey ($Ks = 21$), and VIDEO ($Ks = 23.8$). 
In both plots, the shaded regions mark the ``AGN'' regime as defined by the $\log_{10} f_x/f_o > -1$ (left) or the $\log_{10} f_x/f_{\rm Ks} > -1.2$ (right) thresholds as described in \S4.5 of \protect\cite{catcdfs7ms}.
For sources with spectroscopic redshift measurements, we also mark those with $L_{\rm X} > 3\times10^{42}$~erg~s$^{-1}$ as the green triangles. The 930 sources with optical spectra consistent with broad-line AGNs are also marked as open red circles.}
\label{fig:fxfo}
\end{figure*}

In this section we briefly discuss some of the properties of the 4858 sources 
with reliable counterparts. 
For the 1782 \hbox{X-ray} sources with secure spec-zs, 
we calculate their rest-frame \hbox{2--10}~keV ``apparent'' luminosity assuming 
a $\Gamma=1.7$ power-law spectrum corrected for Galactic absorption. 
Fig.~\ref{fig:xcats} compares the flux, redshift, and luminosity distributions 
of our sample to those from archival \hbox{X-ray} surveys, including XMM-COSMOS, 
COSMOS-Legacy, and Stripe 82X. 
Fig.~\ref{fig:z_lx_hr} displays the $L_{\rm X}-z$ distribution of our sample, 
along with the $L_{\rm X}$ vs. HR, HR vs. full-band flux, and full-band flux vs. redshift distributions. 
The comparisons in the middle and right panels of 
Fig.~\ref{fig:xcats}
are limited to sources with available spec-zs in the Stripe-82 
and XMM-LSS regions. 
The left panel of Fig.~\ref{fig:xcats} demonstrates that our catalog occupies 
a valuable region of parameter space among \hbox{X-ray} surveys by more than 
doubling the source counts of the XMM-COSMOS survey, which will enable a wide 
range of science that was previously limited by either survey sensitivity 
or cosmic variance. 

For this work,
we also include the basic AGN identification results in our catalog. 
Detailed source classifications using multiwavelength SED and \hbox{X-ray} spectroscopic fitting results
will be saved for future works.
For sources with spectroscopic observations, we directly make use of the 
spectroscopic classifications when available. 
Since each spectroscopic survey has its own unique design and methodology, 
we only make use of the ``broad-line'' classifications 
provided in the SDSS, VIPERS, VVDS, and PRIMUS catalogs to identify broad-line AGNs. 
The information on spectroscopic classifications is not yet publicly available for the other
spectroscopic surveys.
For each \hbox{X-ray} source with optical spectroscopic coverage, we have included the spectroscopic flags from all available redshift catalogs (see Column 184 of the main \hbox{X-ray} catalog described in Appendix \ref{sec:catcols}).
A total of 930 sources are 
classified as AGNs based on the broad-line spectroscopic flags specified in the 
SDSS, VIPERS, VVDS, or PRIMUS catalogs. 
Since 90\% of the spec-zs for our \hbox{X-ray} sources are 
culled from one of these four catalogs, 
we expect the vast majority of the remaining \hbox{X-ray} sources with spectroscopic coverage to have galaxy-like spectra. For sources without spectroscopic observations, only a small fraction of them is expected to be broad-line AGNs 
(see \S\ref{subsec:zphot} and \citealt{yang17link} for details).
For the other sources, we use the criteria 
described in \cite{catcdfs7ms} to select AGNs: 
(1) An \hbox{X-ray} luminosity threshold where we regard sources with rest-frame 
$L_{\rm 2-10~keV} > 3\times10^{42}$~erg~s$^{-1}$ as an AGN. 
A total of 
1625 sources satisfy this criterion.
(2) \hbox{X-ray} bright sources with \hbox{X-ray}-to-optical or 
\hbox{X-ray}-to-near-IR flux ratios larger than $\log f_x/f_{r} > -1$ or  
$\log f_x/f_{Ks} > -1.2$, respectively. 
To calculate the flux ratios, 
we use the HSC-SSP {\it r}-band photometry of the primary counterpart.
For sources without a detection in the HSC-SSP {\it r}-band, 
we make use of CFHTLS or SDSS {\it r}-band photometry when available. 
For the 265 sources that are not detected in HSC-SSP, CFHTLS, or SDSS, 
we calculate their flux-ratio lower limits using the HSC-SSP wide survey upper limit, 
$r = 26.4$.
For the \hbox{X-ray}-to-near-IR flux ratios, we use the VIDEO {\it Ks}-band photometry.
For sources within the VIDEO coverage but not detected in the {\it Ks}-band,
we calculate the lower limits for $\log f_x/f_{Ks}$ 
assuming a $Ks=23.8$ upper limit. 
For sources outside the VIDEO coverage, we make use of the UKIDSS DXS survey {\it Ks}-band 
photometry when possible and assign an upper limit of $Ks=21$ for the 
non-detected sources. 
For sources outside the coverage of VIDEO and UKIDSS, the 
shallow photometric depth of 2MASS ($ Ks < 14.3$) 
cannot be used to select AGNs, since no sources this bright would have a high
$\log f_x/f_{Ks}$ ratio satisfying the AGN selection criterion. 
There are 
4998 sources with $\log f_x/f_{r} > -1$ and 
4700 sources with $\log f_x/f_{Ks} > -1.2$, 
totaling 5064 sources that can be classified as an AGN based 
on their $f_x/f_{r}$ or $f_x/f_{Ks}$ values.
The flux-ratio distributions are displayed in Fig.~\ref{fig:fxfo}. 

\begin{figure*}
\includegraphics[width=0.48\textwidth]{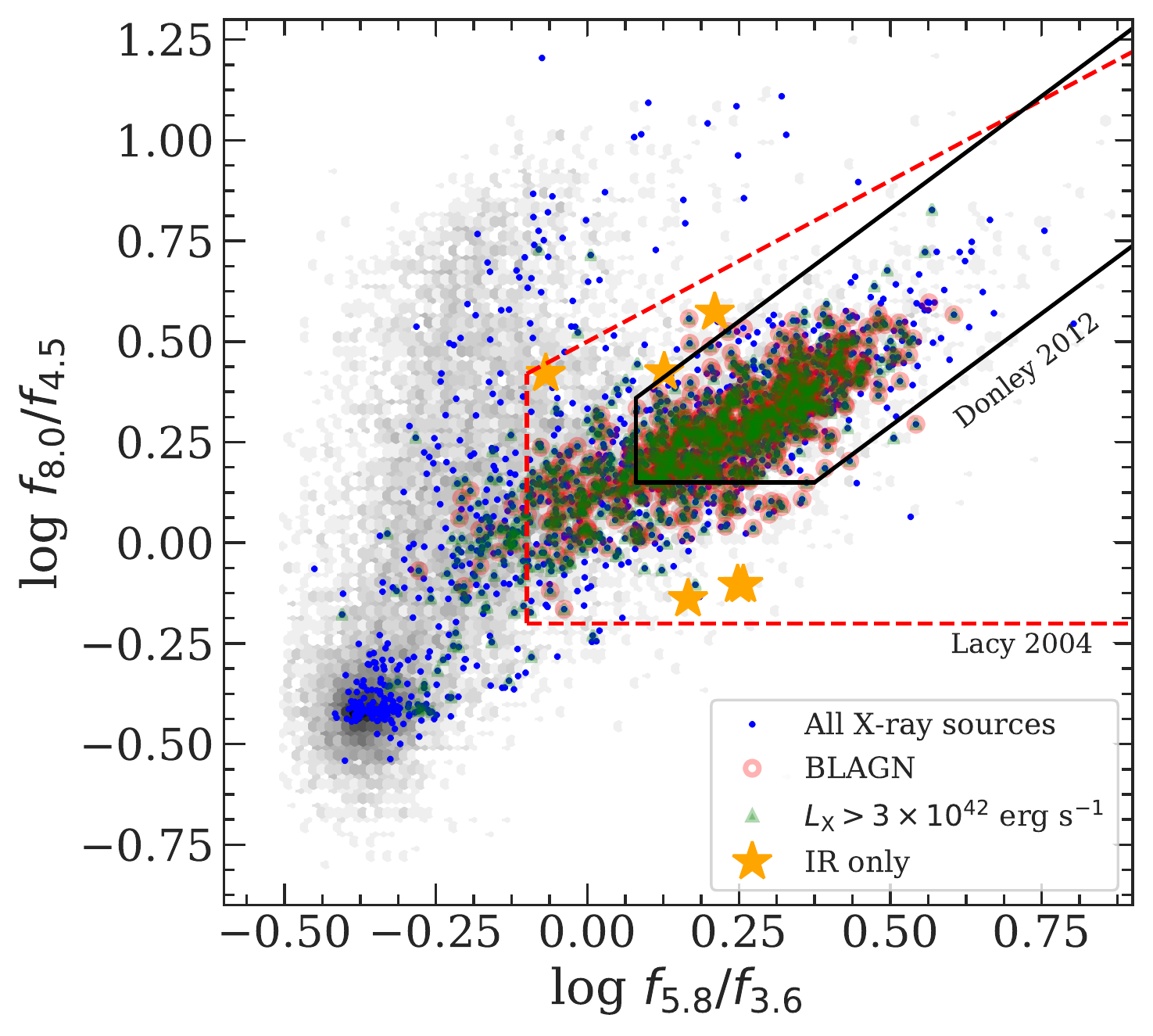}
\includegraphics[width=0.48\textwidth]{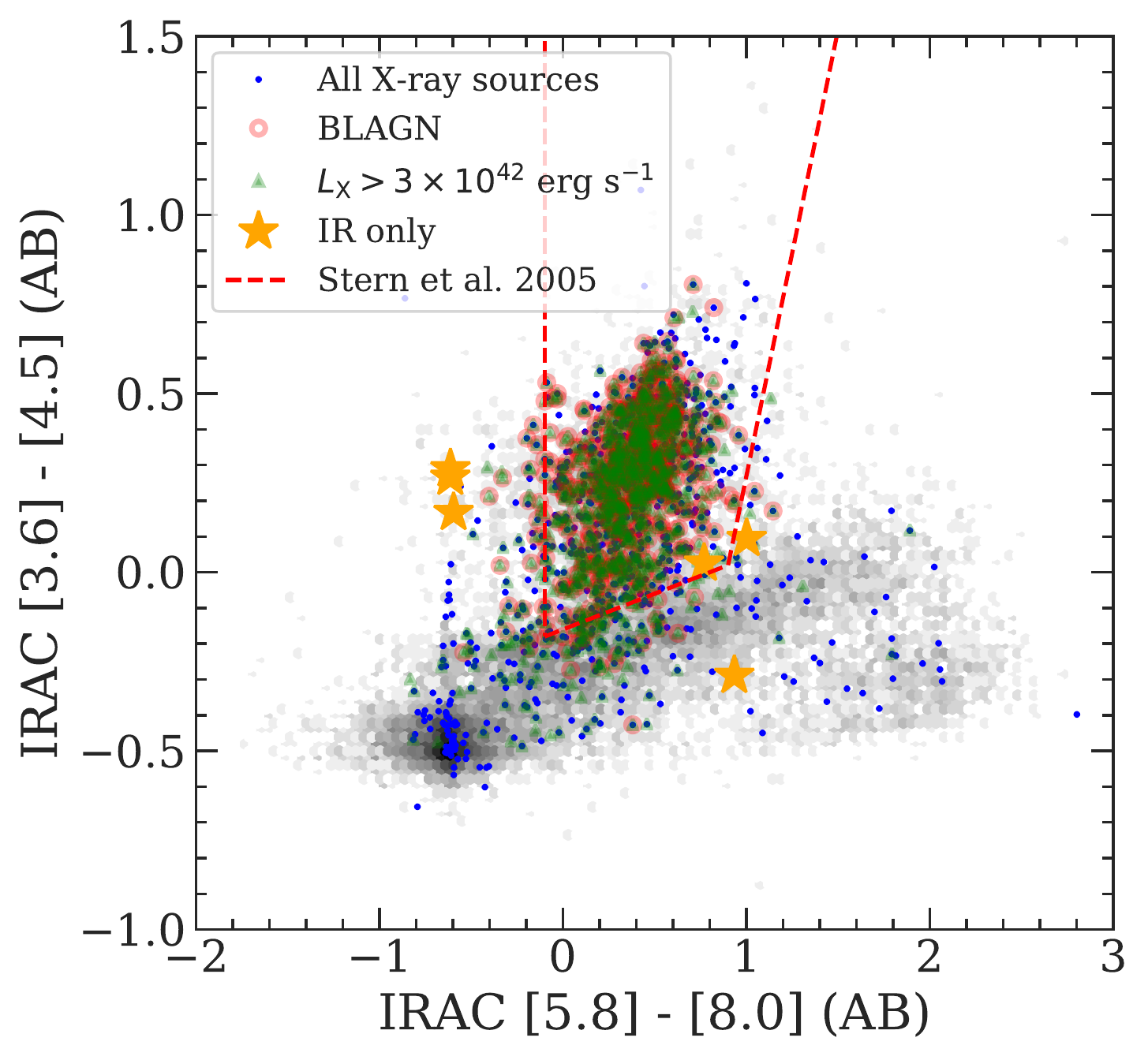}
\caption{{\it Left} -- $\log f_{5.8}/f_{3.6}$ vs, $\log f_{8.0}/f_{4.5}$
distribution. 
{\it Right} -- IRAC$[5.8] - [8.0]$ vs. $[3.6] - [4.5]$ (AB) distributions.
In both panels, the {\it Spitzer} IRAC two-color distribution 
for the 1716 \hbox{X-ray} sources with detections in all four IRAC bands 
are shown as the blue dots.
X-ray luminous AGNs and spectroscopically confirmed quasars
are also shown as the green triangles and the open red circles, respectively. 
There are a total of 1325 sources with IRAC colors satisfying one of the 
three mid-IR color AGN selection criteria (\citealt{lacy04}, \citealt{ster05}, and \citealt{donl12}). 
The six sources that are only identified as an AGN based on their IRAC colors are displayed
as the large orange stars. The color distribution for the 12990 SWIRE sources in our \hbox{X-ray} catalog region is also shown as the gray hexagonal cells, where darker color indicates 
higher source density.}
\label{fig:iraccolor}
\end{figure*}

In addition to the classification methods described above, AGNs can also be 
identified based on the distinctive red mid-IR color arising from 
hot dust heated by SMBH accretion \citep[e.g.,][]{lacy04,ster05,ster12wise,donl12,asse13wise,mate13}. 
We select these mid-IR AGNs based on three different selection criteria from 
\citet{lacy04}, \citet{ster05}, and \citet{donl12}.
The \citet{lacy04} and \citet{ster05} criteria have higher completeness while the 
\citet{donl12} criterion is more reliable (i.e., has much less star-forming 
galaxy interlopers). 
Only 1716 \hbox{X-ray} sources have 
a primary counterpart that is detected in 
all four IRAC bands,
which is a requirement of using these IRAC two-color selection criteria. 
Of these 1716 sources, 1300, 1158, and 834 satisfy the \citet{lacy04}, \citet{ster05}, and \citet{donl12} criteria, respectively,
for a total of 1325 individual \hbox{X-ray} sources.
Notably, six sources are only identified as an AGN based on their IRAC colors. 
On the other hand, of all \hbox{X-ray} sources with detections in all four IRAC bands,
257 AGNs identified using the X-ray AGN selection criteria described in the previous paragraph
do not satisfy any of the three IRAC color-color selection criteria.
The IRAC color distributions of the 1716 \hbox{X-ray} sources,
and the 12990 SWIRE sources in our \hbox{X-ray} catalog region are displayed in Fig~\ref{fig:iraccolor}. 

The total number of sources classified as AGNs is 5071, 
or $96.7\%$ of the total sample. 
For the sources not classified as AGNs,
54 of them have spectroscopic-redshift measurements, 
including 13 stars with $z \approx 0$, and 41 galaxies hosting low \hbox{X-ray} luminosity sources. They may be powered by star-formation processes in galaxies given their relatively weak \hbox{X-ray} to OIR ratios and low \hbox{X-ray} luminosities. 
The remaining 117 sources are relatively bright in the optical and NIR bands
(median {\it r}-band and {\it Ks}-band magnitudes are 15.5 and 14.1, respectively),
and thus all of them have optical-to-X-ray and NIR-to-X-ray flux ratios 
lower than the AGN selection thresholds, suggesting they are either foreground stars 
or low-redshift galaxies hosting \hbox{X-ray} sources powered by stellar processes.

\section{Summary}\label{sec:sum}
In this work, we present a new \hbox{X-ray} point-source catalog in the XMM-LSS region constructed using 
both considerable new AO-15 and archival {\it XMM-Newton} data. The main results are the following:
\begin{enumerate}[label=\arabic*.]
    \setlength{\itemindent}{-1ex}

    \item Our \hbox{X-ray} catalog is constructed based on data in a $5.3$ deg$^2$ rectangular region centered at RA$=35.580^{\circ}$, DEC=$-4.965^{\circ}$. 
    A total of 155 pointings from 149 different {\it XMM-Newton} ObsIDs are used, with a total of $2.7$~Ms background-filtered exposure time (1.1~Ms from AO-15). The median value of the cleaned PN exposure time is 46~ks for the full 5.3 deg$^2$ field (see \S\ref{sec:datasum}). 
    Our survey has a flux limit of 
	$1.7\times10^{-15}$,
	$1.3\times10^{-14}$, and
	$6.5\times10^{-15}$~erg~cm$^{-2}$~s$^{-1}$ over 90\% of its 5.3~deg$^2$ area in the soft, hard, and full bands, respectively, reaching the desired uniformity and survey depth (see \S\ref{subsec:Sens}).

    \item We use Monte Carlo simulations to estimate the fraction of spurious sources ($f_{\rm spurious}$)
    as a function of {\sc det\_ml} values for each band, and we consider sources with $f_{\rm spurious} \leq 1\%$ to be reliably detected. This corresponds to {\sc det\_ml} thresholds of 4.8, 7.8, and 6.2 in the soft, hard, and full bands, respectively (\S\ref{subsec:simulation}).

    \item The main \hbox{X-ray} source catalog is generated using {\sc ewavelet} and {\sc emldetect}. 
    All 5242 sources with {\sc emldetect} {\sc det\_ml} 
    values satisfying the $f_{\rm spurious}\leq 1\%$ criterion in the soft band (0.5--2~keV), hard band (2--10~keV), or full band (0.5--10~keV) are included. 
    Of the 5242 sources, 2861 are the same 
    \hbox{X-ray} sources identified in previous \hbox{X-ray} surveys in our survey area 
    \citep[e.g., the XMM-XXL-North survey;][]{liu16xxl}, and 2381 are newly discovered \hbox{X-ray} sources (see \S\ref{subsec:mainxcat}). There are 2967 sources with more than 100 \hbox{X-ray} counts in the full band (PN + MOS), and 126 sources with more than 1000 \hbox{X-ray} counts (see \S\ref{subsec:mainxcat}).

    \item The absolute astrometry of the {\it XMM-Newton} catalog is registered to the WCS frame of the Subaru HSC-SSP survey (\S\ref{subsec:astrometry}). The positional uncertainties for the \hbox{X-ray} sources are determined based on an empirical relation between the \hbox{X-ray}-to-optical positional offsets and the \hbox{X-ray} source counts.
    Our empirical positional uncertainties are well-characterized by the Rayleigh distribution. The median positional uncertainties in the soft, hard, and full bands are 
    1\farcs35, 1\farcs37, and 1\farcs31, respectively (see \S\ref{subsec:poserr}).

    \item We search for OIR counterparts in the SERVS, VIDEO, CFHTLS, and HSC-SSP surveys;
    $98\%$ (5147/5242) of the \hbox{X-ray} sources have at least one OIR counterpart candidate within the $99.73\%$ positional uncertainties ($r_{99\%}$). 
    A total of $\approx 93\%$ (4858/5242) of the \hbox{X-ray} sources have at least one reliable OIR counterpart (\S\ref{subsec:lrmatching}). 
    There are 1782 secure spectroscopic redshifts 
    from SDSS, VIPERS, VVDS, UDSz, PRIMUS, CSI, and 3D-HST (\S\ref{subsec:zspec}). 
    For a 4.5~deg$^2$ subfield in our survey region covered by SERVS, 
    we make use of the forced-photometry catalog from N18 to compute photometric redshifts (\S\ref{subsec:zphot}), achieving $> 70\%$ spectroscopic+photometric redshift completeness 
    for $85\%$ of our survey area. 
    We expect to expand the photometric redshift measurements to all of our \hbox{X-ray} sources 
    when SERVS DeepDrill survey ({\it Spitzer} Program ID 11086; Lacy et al., in preparation) and the VEILS survey (see Table~\ref{tab:servsmw}) are completed.

    \item We test the matching results using a subsample of 223 \hbox{X-ray} sources with a reliable \chandra\ counterpart from CSC 2.0.  Approximately $97\%$ of the matching results from {\it XMM-Newton} and \chandra\ are identical, demonstrating our multiwavelength matching results are highly reliable (see \S\ref{subsec:matchingcheck}). 

    \item We classify 5071 \hbox{X-ray} sources as AGNs based on their optical spectra from SDSS, VIPERS, or VVDS (930); 
    \hbox{X-ray} luminosity larger than $3\times10^{42}$~erg~s$^{-1}$  (1625);
    large \hbox{X-ray}-to-optical and/or \hbox{X-ray}-to-NIR flux ratios (5064);
    and {\it Spitzer} IRAC colors (1325). See \S\ref{sec:class} for details.

\end{enumerate}

The \hbox{X-ray} source catalog presented in this work is the first 
$> 2$ deg$^2$ \hbox{X-ray} survey with sensitivity comparable to that of COSMOS. 
This 5.3 deg$^2$ wide-area and 46~ks depth survey will enable 
a wide range of studies.
For instance,  
the large AGN sample and the excellent multiwavelength coverage will provide a means of 
exploring the behavior of AGNs in the multidimensional space of galaxy parameters.
The wide area of this survey will also enable studies of AGN triggering mechanisms 
as a function of environment. 
In the near future, the combination of AGN samples from this work, COSMOS, 
and the other XMM-SERVS fields
will sample the full range of cosmic large-scale structures, 
alleviating the cosmic-variance uncertainties present in previous COSMOS results  \citep[e.g.,][]{mene09,torr10,skib14} as well as advancing our understanding of the coevolution of SMBHs and their host galaxies.

\section*{Acknowledgments}
We thank the referee for suggestions that improved the manuscript. 
We thank Matthias Ehle, Norbert Schartel, and the {\it XMM-Newton} Science Operations Centre for help with scheduling the {\it XMM-Newton} AO-15 observations.
We also thank Florian Paucaud and the XMM-XXL team for the helpful discussions 
during the survey-planning stage of this work.
We thank Johannes Buchner and Stephanie LaMassa for helpful discussions,  
and Francesca Civano and Stefano Marchesi for providing comparison data. 
We acknowledge the support of NASA grant NNX17AF07G (CTJC, WNB, GY, and FV), 
National Key R\&D Program of China grant 2016YFA0400702 (BL),
and National Natural Science Foundation of China grant 11673010 (BL).
DMA and IS acknowledges support from 
Science and Technology Facilities Council (STFC) grant code ST/P000541/1, 
and IS also acknowledges support from STFC (ST/P000541/1), the ERC Advanced Investigator programme DUSTYGAL 321334 and a Royal Society/Wolfson Merit Award.
FEB acknowledges support from CONICYT-Chile 
(Basal-CATA PFB-06/2007, FONDECYT Regular 1141218),
the Ministry of Economy, Development, and Tourism's Millennium Science
Initiative through grant IC120009, awarded to The Millennium Institute
of Astrophysics, MAS. 
MJJ  was supported by the Oxford Centre for Astrophysical Surveys, 
which is funded through generous support from the Hintze Family Charitable Foundation. 
MJJ and BH acknowledge support from STFC (ST/N000919/1).
YQX was supported by NSFC-11473026, NSFC-11421303, and the CAS Frontier 
Science Key Research Program (QYZDJ-SSW-SLH006).
This work made use of data products from CFHTLS, HSC-SSP, SDSS, and VIDEO. 
The CFHTLS survey was based on observations obtained with MegaPrime/MegaCam, 
a joint project of CFHT and CEA/IRFU, at the Canada-France-Hawaii Telescope (CFHT) which is operated by the National Research Council (NRC) of Canada, 
the Institut National des Science de l'Univers of the Centre National de la Recherche Scientifique (CNRS) of France, 
and the University of Hawaii. 
This work is based in part on data products produced at Terapix available at the 
Canadian Astronomy Data Centre as part of the Canada-France-Hawaii Telescope Legacy Survey, 
a collaborative project of NRC and CNRS. 
The Hyper Suprime-Cam (HSC) collaboration includes the astronomical communities 
of Japan and Taiwan, and Princeton University. 
The HSC instrumentation and software were developed by the National Astronomical Observatory of Japan (NAOJ), 
the Kavli Institute for the Physics and Mathematics of the Universe (Kavli IPMU), 
the University of Tokyo, the High Energy Accelerator Research Organization (KEK), 
the Academia Sinica Institute for Astronomy and Astrophysics in Taiwan (ASIAA), 
and Princeton University. 
Funding was contributed by the FIRST program from Japanese Cabinet Office, 
the Ministry of Education, Culture, Sports, Science and Technology (MEXT), 
the Japan Society for the Promotion of Science (JSPS), Japan Science and Technology Agency (JST), 
the Toray Science Foundation, NAOJ, Kavli IPMU, KEK, ASIAA, and Princeton University. 
Funding for SDSS-III has been provided by the Alfred P. Sloan Foundation, the Participating Institutions, 
the National Science Foundation, and the U.S. Department of Energy Office of Science. The SDSS-III web site is http://www.sdss3.org/.
The observations for the VIDEO survey were made with ESO telescopes 
at the La Silla Paranal Observatories under ESO programme ID 179.A-2006.

\appendix

\section{Main Catalog Description}\label{sec:catcols}
Here we describe the columns of the main \hbox{X-ray} source catalog, 
Table~\ref{tab:mainxtab}.
Throughout the table,
we mark null values as $-99$. All celestial coordinates are given in equinox J2000.
\hfill\\
\noindent
\textbf{\textit{X-ray properties}}\\
Columns 1--112 give the \hbox{X-ray} properties of our sources. 
Columns for the soft-band results are marked with the ``{\sc SB\_}'' prefix. 
Columns for the hard-band and full-band results are marked with the ``{\sc HB\_}'' and ``{\sc FB\_}'' prefixes, respectively.
Note that we have calculated the upper limits on counts, count rates, and fluxes
for the non-detections (Eq.~\ref{eq:uplim}). 
For these upper limits, their corresponding uncertainty columns are set as $-99$.

\begin{enumerate}[label=(\arabic*)]
    \setlength{\itemindent}{-1ex}
    \item Column 1: The unique source ID (XID) assigned to each \hbox{X-ray} source.
    \item Columns 2--3: RA and DEC in degrees of the \hbox{X-ray} source.
    The positions are determined based on {\sc emldetect}.
    Based on availability, we use the positions from, in priority order, the full band, 
    soft band, and hard band as the primary position of the \hbox{X-ray} source. Band-specific positions are listed in Columns 8--13.
    \item Column 4: \hbox{X-ray} positional uncertainty ($\sigma_x$) in arcsec
    based on the empirical relation between source counts and positional offsets to the HSC-SSP catalog. 
    Note that this is not the $\sigma$ of a 2D-Gaussian distribution but rather 
    the scaling parameter of the univariate Rayleigh distribution (see \S\ref{subsec:poserr} 
    and \citealt{pine17match} for details). The positional uncertainties are based on 
    those of the full band. For sources without a full-band detection, 
    the soft-band or hard-band positional uncertainties are listed.
    See \S\ref{subsec:poserr} for details.
    \item Columns 5--6: 68\% and 99.73\% \hbox{X-ray} positional uncertainties in arcsec based on the Rayleigh distribution; 
    see \S\ref{subsec:poserr} for details. 
    \item Column 7: Positional uncertainties calculated by {\sc emldetect}, $\sigma_{\rm eml}$, in arcsec. Similar to $\sigma_{\rm x}$,
    we list the full-band values when possible and list soft-band or hard-band $\sigma_{\rm eml}$ for sources not detected in the full band.
    \item Columns 8--13: RA and DEC in degrees of the source in the soft, hard, and full bands, respectively.
    \item Columns 14--16: The source-detection threshold in each band, {\sc det\_ml},
    which is computed using {\sc emldetect}. 
    \item Columns 17--19: The source-detection reliability parameter in each band, defined as $1 - f_{\rm spurious}$,
    where $f_{\rm spurious}$ is the expected spurious fraction based on simulations described in \S\ref{subsec:simulation}. Due to the limited numerical precision, all sources with spurious fractions smaller than $0.01\%$ have a reliability of 1. For this work, we consider sources with $f_{\rm spurious} \leq 1\%$ to be  detected robustly.
    \item Columns 20--22: Total (PN + MOS1 + MOS2) exposure time in seconds in each band.
    \item Columns 23--31: PN, MOS1, and MOS2 exposure time in seconds in each band.
    \item Columns 32--34: Total background-map values (PN + MOS1 + MOS2) in counts per pixel in each band. 
    \item Columns 35--43: PN, MOS1, and MOS2 background-map values in counts per pixel in each band. 
    \item Columns 44--46: Total (PN + MOS1 + MOS2) net counts in each band.
    \item Columns 47--55: PN, MOS1, and MOS2 net counts in each band.
    \item Columns 56--67: Uncertainties of total, PN, MOS1, and MOS2 net counts in each band.
    \item Columns 68--79: Total, PN, MOS1, and MOS2 net count rates in each band, in count s$^{-1}$.
    \item Columns 80--91: Uncertainties of total, PN, MOS1, and MOS2 net count rates in each band, in count s$^{-1}$.
    \item Columns 92--97: Flux and flux uncertainty in each band, in erg~cm$^{-2}$~s$^{-1}$.
    The conversion factors between count rates and fluxes are derived assuming a power-law spectrum with a $\Gamma=1.7$ photon index and the Galactic absorption column density for each EPIC detector. Note that no correction is made for possible intrinsic absorption.
    See \S\ref{subsec:mainxcat} for details. 
    The fluxes and uncertainties reported here are the error-weighted average of all EPIC detectors.
    \item Columns 98--100: Hardness ratio, defined as $(H-S)/(H+S)$, 
    where $H$ is the total (PN + MOS1 + MOS2) net counts divided by the total exposure time in the hard band
    and $S$ is the total net counts divided by the total exposure time in the soft band. 
    The uncertainties on the HRs are calculated based on the count uncertainties 
    using the error-propagation method described in \S1.7.3 of Lyons (1991).
    Sources detected only in the full band are set to $-99$ in all three columns.
    The HR values for sources detected only in the soft-band are 
    calculated assuming their hard-band counts are at the upper limits calculated using Eq.~\ref{eq:invg}.
    For sources detected only in the hard-band we calculate their HR values assuming their soft-band counts 
    are the upper limits. See \S\ref{subsec:mainxcat} for details. 
    The upper and lower uncertainties for these sources with non-detections in the soft or the hard band
    are set to $-99$. 
    We note that one of the CCDs on MOS1 was affected by a micrometeorite impact,
    therefore $H$ and $S$ are sometimes calculated based on only results from the two cameras
    with non-zero exposure time.
    \item Columns 101--109: Hardness ratios $(H-S)/(H+S)$ and the 68\% lower and upper bounds 
    for each EPIC detector calculated using BEHR. 
    Sources detected only in the full band are set to $-99$ in all three columns.

    \item Column 110: Rest-frame, ``apparent'' \hbox{2--10}~keV \hbox{X-ray} luminosity (only corrected for Galactic absorption)
    computed as in \S\ref{sec:class}.
    \item Column 111: CSC 2.0 source name of the nearest \chandra\ source in the CSC within 10$^{\prime\prime}$.
    \item Column 112: XMM-XXL-North  catalog source name of the nearest \xmm\ source in \cite{liu16xxl} within 10$^{\prime\prime}$. 
    %\item Columns 110: SXDS catalog source name of the nearest \xmm\ source in \cite{catsxdf}.
\end{enumerate}

\hfill\\
\noindent
\textbf{\textit{Multiwavelength-matching results}}\\
Columns 113--122 list the multiwavelength-matching results based on the $LR$ method described in 
\S\ref{subsec:lrmatching}. 
In these columns, the 99.73\% positional-uncertainty radius represents the quadratic sum of the 
positional uncertainties of each \hbox{X-ray} source and the corresponding OIR catalog (see Table~\ref{tab:matching}).
\begin{enumerate}[label=(\arabic*)]
    \setlength{\itemindent}{-1ex}
    \item Columns 113--116: Number of counterpart candidates from each OIR catalog 
    within the 10$^{\prime\prime}$ search radius 
    of each \hbox{X-ray} source. 
    \item Columns 117--120: Number of sources from each OIR catalog 
    that satisfy $LR\geq LR_{\rm th}$. 
    \item Column 121: Flag set to 1 if a reliable counterpart has been identified for the 
    \hbox{X-ray} source. See \S\ref{subsec:lrmatching} for details.
    \item Column 122: Flag set to 1 if the primary counterpart of the \hbox{X-ray} source is from the SERVS catalog and might suffer from source blending. There are a total of 318 flagged sources. 
    See \S\ref{subsec:lrmatching} for details.
\end{enumerate}

\hfill\\
\noindent
\textbf{\textit{Multiwavelength properties}}\\
Columns 123--198 provide the multiwavelength properties from each OIR catalog 
for the primary counterparts matched to \hbox{X-ray} sources using the $LR$ method. 
Properties from SERVS, SWIRE, VIDEO, CFHTLS, and HSC-SSP 
are marked with additional prefixes ``SERVS\_'', ``SWIRE\_'', ``VIDEO\_'', ``CFHT\_'', and ``HSC\_'', respectively.

\begin{enumerate}[label=(\arabic*)]
    \setlength{\itemindent}{-1ex}
    \item Column 123: Catalog from which the primary counterpart is selected.
    The primary counterpart is chosen in priority order from SERVS, VIDEO, CFHTLS, and HSC-SSP,
    which is based on the matching reliability of each OIR catalog. See \S\ref{subsec:matchingcheck} for details.
    \item Column 124--126: RA and DEC in degrees of the primary counterpart and 
    its separation in arcsec from the \hbox{X-ray} source.
    \item Column 118: The matching likelihood ratio ($LR$) of the primary counterpart.    
    \item Columns 127--143: RA, DEC, Object ID, and the matching reliability ($MR$) of the primary counterpart culled from the original OIR catalogs.
    \item Columns 144--147: SERVS 1.9$^{\prime\prime}$ aperture photometry and the associated uncertainties
    in the $3.6\mu$m and $4.5\mu$m bands.
    \item Columns 148--155: SWIRE 1.9$^{\prime\prime}$ aperture photometry and the associated uncertainties
    in the $3.6\mu$m, $4.5\mu$m, $5.8\mu$m, and $8.0\mu$m bands.
    \item Columns 156--157: SWIRE 5.25$^{\prime\prime}$ aperture photometry and the associated uncertainty
    in the $24\mu$m band.
    \item Columns 158--167: VIDEO PSF photometry and uncertainties in AB magnitude in the {\it Z, Y, J, H}, and {\it Ks} bands.
    \item Columns 168--177: CFHTLS PSF photometry and uncertainties in AB magnitude in the {\it u, g, r, i}, and {\it z} bands.
    \item Columns 178--187: HSC CModel photometry and uncertainties in AB magnitude in the {\it g, r, i, z}, and {\it y} bands.
    \item Columns: 188--190: RA, DEC, and Object ID from the original redshift catalogs for the primary counterparts.
    \item Column 191: Spectroscopic redshift adopted for the \hbox{X-ray} source.
    The redshifts are chosen based on the spectral resolution of the observations and the redshift reliabilities. See \S\ref{subsec:zspec} for details. 
    \item Column 192: The catalog that provided the redshift.
    \item Column 193: Original redshift flag from one of the redshift catalogs. 
    For SDSS, see \url{http://www.sdss.org/dr14/algorithms/bitmasks/#ZWARNING} for the definition of flags. 
    For VVDS, see \S3.4 of \cite{catspeczvvds} for the definition of flags.
    For VIPERS, see \S4.3 of \cite{catspeczvipers} for the definition of flags.
    For PRIMUS, see \url{http://primus.ucsd.edu/version1.html#ztags}  for the definition of flags.
    For CSI, see \S4.6 of \cite{catspeczcsi}  for the definition of flags.
    For UDSz, see \cite{catspeczudsz2} for the definition of flags.
    For the 3D-HST catalog, we only select redshifts with $\sigma_z/(1+z) \leq 0.003$ and thus no redshift flags are included. 
    \item Column 194--197:
    Photometric redshift, the associated upper and lower uncertainties, and
    the photometric-redshift quality parameter ($Q_z$). See \S\ref{subsec:zphot}.
    The photometric-redshift measurements are limited to the 4.5 deg$^2$ area with forced-photometry
    from N18. See \S\ref{subsec:zspec} for details.
    \item Column 198: 
    A five-digit AGN classification flag, each digit represents the flag for an AGN classification criterion described in \S\ref{sec:class}.
    From left to right: spectroscopic classification, \hbox{X-ray} luminosity classification, 
    \hbox{X-ray} to optical flux ratio classification, \hbox{X-ray} to near-IR flux ratio classification, 
    and IRAC color classification. 
    For each digit, the number ``1'' means the source is not classified as an AGN using the corresponding criterion. 
    The number ``2'' means the source is classified as an AGN.
    If the given criterion cannot be used to classify the \hbox{X-ray} source (e.g., there is no spectroscopic coverage), 
    the numeric expression is ``3''. 
    For instance, if an \hbox{X-ray} source does not have optical spectral coverage, 
    has $L_{\rm X}>3\times10^{42}$ erg~s$^{-1}$ and high X-ray-to-optical as well as X-ray-to-NIR flux ratios, 
    but is not an mid-IR AGN, the source is flagged as ``32221''.

\end{enumerate}

\hfill\\
\noindent
\textbf{\textit{Multiwavelength properties for additional counterparts}}\\
In our source catalog, there are 1034 \hbox{X-ray} sources with 
two $LR \geq LR_{\rm th}$ counterparts where the 
second-highest $LR$ counterpart also satisfies $LR \geq 0.5$ $LR_{\rm primary}$ (see \S\ref{subsec:lrmatching} for details). 
The highest $LR$ counterparts are considered as ``primary'' with properties reported in 
Columns 123--193. Here we report the multiwavelength properties of the 
``secondary'' counterparts in Columns \hbox{199--269}, which are identical as Columns \hbox{123--193}
except for the ``SECONDARY\_'' prefixes. 
There are also 29 \hbox{X-ray} sources with three
$LR \geq LR_{\rm th}$ counterparts, where the secondary and the tertiary counterparts both satisfy
the $LR \geq 0.5$ $LR_{\rm primary}$ criterion.
The multiwavelength properties of the secondary counterparts for these 29 sources are also reported in 
Columns \hbox{199--269}. 
The properties for the tertiary counterparts are reported in 
Columns \hbox{270--340}, which are identical as Columns \hbox{123--193}
except for the additional ``TERTIARY\_'' prefixes.

\hfill\\
\noindent
\textbf{\textit{Supplementary multiwavelength properties for primary counterparts}}\\
In our catalog, a small number of primary counterparts do not have reliable photometry 
from VIDEO, CFHTLS, and HSC-SSP due to the lack of areal coverage or various
instrumental artifacts (see \S{\ref{subsec:lrmatching}). 
Columns 341--369 report supplementary properties for sources in 
SDSS DR12, 2MASS, and UKIDSS-DXS that are matched within 1$^{\prime\prime}$ of the primary counterparts. 
These columns are marked with ``SUPPLEMENTARY\_'' prefixes.

\begin{enumerate}[label=(\arabic*)]
    \setlength{\itemindent}{-1ex}
    \item Columns 341--353: Source ID, RA, DEC (J2000, in degrees), and photometry and the associated uncertainties in the SDSS {\it u, g, r, i}, and {\it z} bands (CModel magnitudes).
    \item Columns 354--362: Source ID, RA, DEC (J2000, in degrees), and photometry and the associated uncertainties in the 2MASS {\it J, H}, and {\it Ks} bands (in AB magnitudes). 
    \item Columns 363--369: Source ID, RA, DEC (J2000, in degrees), and photometry and the associated uncertainties in the UKIDSS-DXS {\it J} and {\it Ks} bands (in AB magnitudes). 
\end{enumerate}

\section{Supplementary catalog from {\sc NWAY}}\label{sec:nwaycols}

Here we describe the columns of the supplementary multiwavelength matching results table obtained with {\sc NWAY}
(see \S\ref{subsec:nwaymatching}).
Only the counterparts with {\sc match\_flag$\geq 1$} are included.
Similar to the $LR$ matching results, 
some of the \hbox{X-ray} sources have multiple probable counterparts. 
In this table, 
the same \hbox{X-ray} source can have multiple counterparts
and the information for each counterpart is given in an independent row.
Similarly to columns 114--184 of Table A, properties from SERVS, VIDEO, CFHTLS, and HSC-SSP are 
marked with the prefixes ``SERVS\_'', ``VIDEO\_'', ``CFHT\_'', and ``HSC\_'', respectively.
Null values are marked as $-99$ throughout the table.
\begin{enumerate}[label=(\arabic*)]
    \setlength{\itemindent}{-1ex}
    \item Column 1: The unique source ID (XID) assigned to the \hbox{X-ray} source.
    \item Column 2: The posterior probability of the \hbox{X-ray} source having any correct counterparts,
    $p\_any$, for each \hbox{X-ray} source. 
    \item Column 3: The relative probability of a counterpart to be the correct match, 
    $p\_i$. 
    \item Columns 4--11: RA and DEC of the counterpart in each OIR catalog in degrees.
    \item Columns 12--15: The original Object ID of the counterpart from each OIR catalog.
    \item Columns 16--19: Separation of the \hbox{X-ray} position from the counterpart in each OIR catalog in arcseconds.
    \item Columns 20--23: SERVS 1\farcs9 aperture photometry and the associated uncertainties
    in the $3.6\mu$m and $4.5\mu$m bands.
    \item Columns 24--31: VIDEO PSF photometry and uncertainties in AB magnitude in the {\it Y, J, H}, and {\it Ks} bands.
    \item Columns 32--41: CFHTLS PSF photometry and uncertainties in AB magnitude in the {\it u, g, r, i}, and {\it z} bands.
    \item Columns 42--51: HSC CModel photometry and uncertainties in AB magnitude in the {\it g, r, i, z}, and {\it y} bands.
    \item Columns 52: Matching flag, {\sc match\_flag}. For the most-probable counterparts the flag is set to 1.
    For other counterparts that are almost as likely as the most-probable counterpart 
    (i.e., with $p\_i \geq p\_i_{\rm Best}$), the flag is set to 2.
\end{enumerate}

\section{Photometric redshifts for galaxies in the 4.5 deg$^2$ SERVS region}\label{sec:pzapp}
Since one of the major scientific goals of the XMM-SERVS survey is to study the interactions between AGN activity and large-scale structures, it is important to simultaneously consider the \hbox{X-ray} AGNs and the galaxies in the same survey region. This requires photometric-redshift measurements for the full galaxy population. 
To this end, we have also computed photo-zs for the \nallpz\ sources from the N18 catalog
that have reliable (SNR $> 5$) detections in at least 5 bands. 
For these sources, the 25th, 50th, and 75th percentiles of the number of bands 
with reliable detections are 8, 11, and 12, respectively.
The methodology and the multiwavelength data used are identical to those 
described in \S\ref{subsec:zphot}, except we do not include the Seyfert 2 
template in our fitting for these sources. 
These photo-zs were calculated using all 13-band OIR photometry when available. 
Here we report the \nallgpz\ high-quality photo-zs (with $Q_z \leq 1.0$, see \S\ref{subsec:zphot} and \citealt{yang14photoz}), 
which accounts for $\approx 74\%$ of the sources in the {\it Ks}-band selected VIDEO catalog (see \S\ref{sec:mw}).
To assess the quality of these photo-zs, we make use of the \nallpzsz\ spec-zs culled from the same redshift catalogs reported in \S\ref{subsec:zspec}. 
The normalized median absolute deviation (NMAD) is $\sigma_{\rm NMAD} = 0.035$, 
with an outlier fraction (defined as $\mathopen| \Delta z\mathclose|/(1+z_{\rm spec}) > 0.15$) 
of $f_{\rm outlier} = 5.4\%$. 
The median value of $ \Delta z/(1+z_{\rm spec})$ is $-0.018$,
which is a typical systematic offset for photo-z catalogs
(e.g., see Fig. 5 of \citealt{salv11} or \S6.4 of \citealt{yang14photoz}).
Note that %$\sigma_{\rm NMAD}$ and $f_{\rm outlier}$ are not affected by this systematic offset, and 
this offset is nearly negligible compared to the upper and lower 68\% limits reported in our catalog. 
For the 106 sources with $z_{\rm spec} > 2.0$, their NMAD is $\sigma_{\rm NMAD} = 0.078$, and the outlier fraction is $f_{\rm outlier} = 20.8\%$. 
The vast majority of the 20.8\% of outliers have $z_{\rm spec} = 2-2.5$,
and at higher redshifts there are fewer outliers owing to the strength of the Lyman break signature. 
For these high-z sources, the median value of $ \Delta z/(1+z_{\rm spec})$ is $0.02$, 
which is also negligible compared to their photo-z uncertainties.
Fig.~\ref{fig:appphotoz} compares the photometric and spectroscopic redshifts for 
the \nallpzsz\ sources with reliable photometric and spectroscopic redshifts. 

For comparison, photo-zs for HSC-detected sources in our survey region were also reported in \citet{tana17photoz}. As part of HSC-SSP PDR1, these photo-zs were derived using the {\it g, r, i, z}, and {\it y} band photometry and a number of different photometric-redshift algorithms \citep{tana17photoz}, yielding $\sigma_{\rm NMAD}\approx 0.05$ and an outlier fraction of $f_{\rm outlier}\approx 15\%$ for $i < 25$ galaxies with $0.2 \lesssim z_{\rm phot} \lesssim 1.5$. For the full HSC photometric redshift sample, the mean NMAD is 
$\sigma_{\rm NMAD} = 0.08$, and the outlier fraction is $22.7\%$. 
Thanks to the infrared photometric data from VIDEO and SERVS, 
our photo-zs are not restricted by limits set by requiring the Balmer break stay within the wavelength range of the HSC bands, therefore covering a wider redshift range while reducing the NMAD and outlier fractions compared to the HSC-SSP redshifts. In Table C, we report our photo-zs as well as basic redshift flags reported in the publicly available catalogs. The descriptions of the columns are included in the table caption. The high quality and wide range of these redshifts will enable a wide array of science. 

\begin{figure}
\includegraphics[width=\columnwidth]{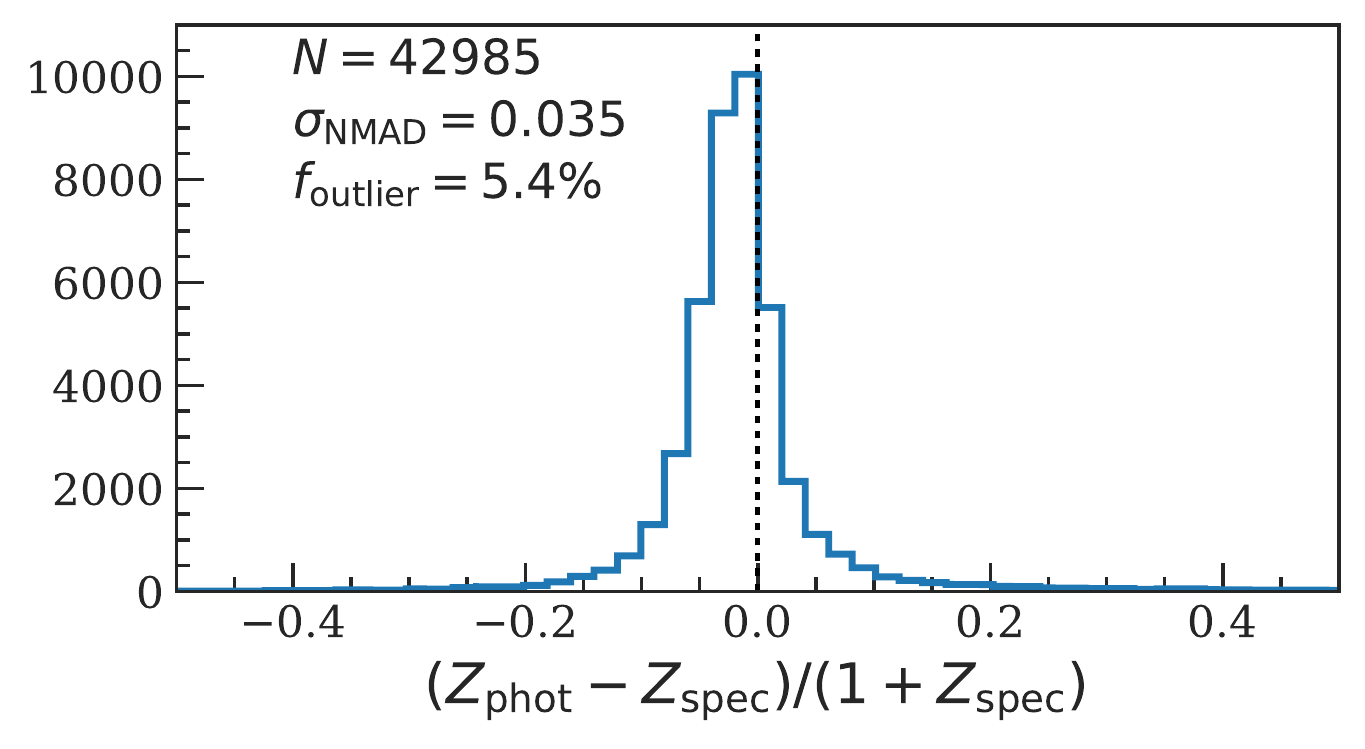}
\DeclareGraphicsExtensions{.png} 
\includegraphics[width=\columnwidth]{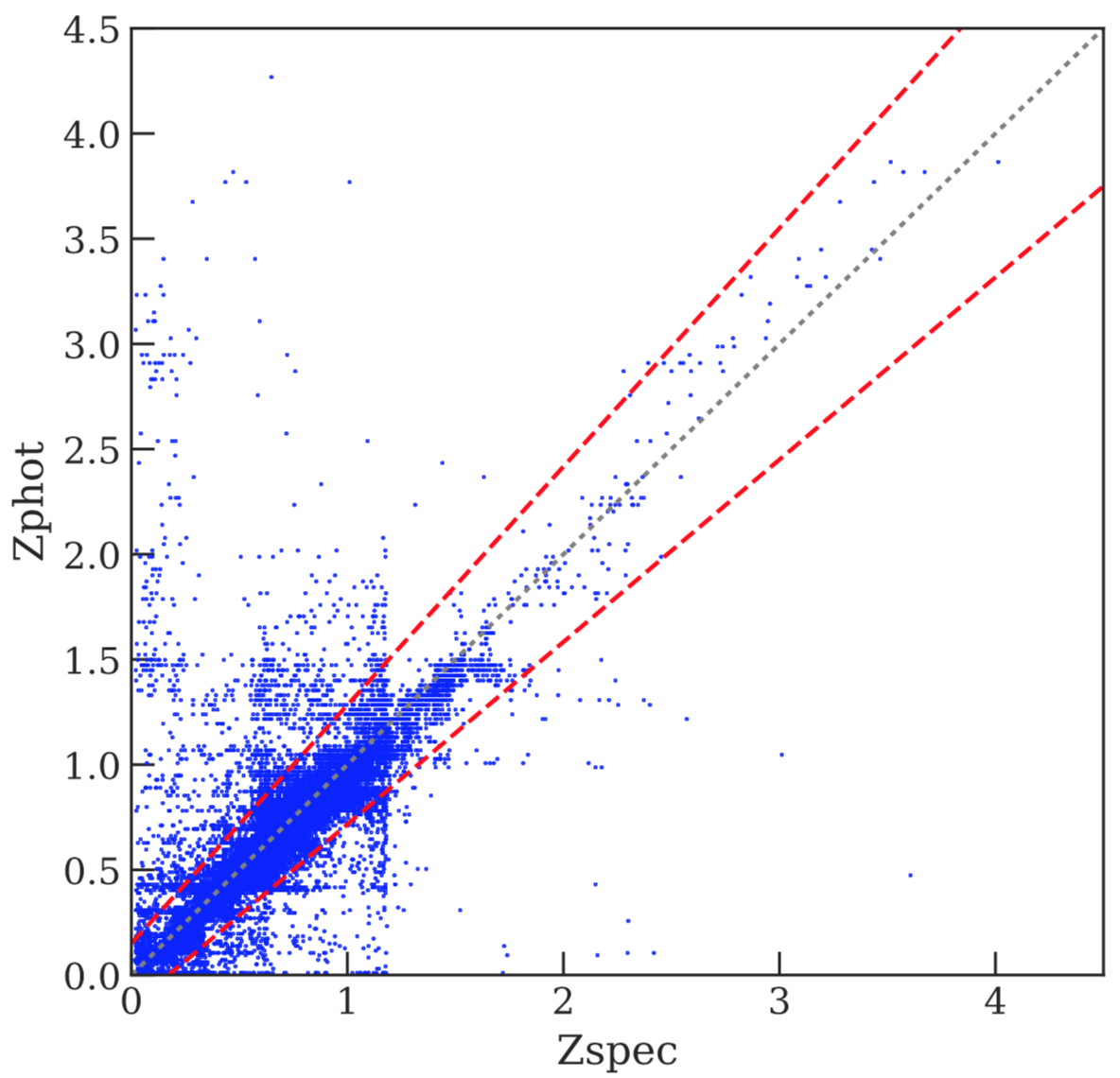}
\caption{
Spectroscopic and photometric redshifts for the \nallpzsz\ sources with high-quality photo-zs and spec-zs in the 4.5~deg$^2$ region
covered by the N18 forced-photometry catalog (sources identified as a broad-line AGN were excluded, see \S\ref{subsec:zspec}).
The top panel shows the histogram of the fractional difference between the spec-zs and the photo-zs. 
The bottom panel shows the direct comparison between the spec-zs and the photo-zs. 
The black dotted lines in both panels mark the $z_{\rm spec} = z_{\rm phot}$ relation.
In the bottom panel, the red dashed lines mark the $\mathopen|\Delta z\mathclose|/(1+z_{\rm spec}) = 0.15$ 
thresholds for outliers. 
}
\label{fig:appphotoz}
\end{figure}

\begin{landscape}
\renewcommand{\thetable}{\Alph{table}}

\begin{table}
\scriptsize
\caption{\label{tab:mainxtab}
The main \hbox{X-ray} source catalog with a selection of columns. Empty or null values are marked as $-$99.
The numbers listed in the second row of this table is the column numbers of the full \hbox{X-ray} catalog.
See Appendix~\ref{sec:catcols} for a detailed description of each column. This table is available in its entirety in machine-readable form online.
}
\begin{tabular}{cccccccccccccccc}
    \hline
	XID &  RA &  DEC &  XPOSERR &  FB\_DET\_ML &  FB\_EXP &  FB\_BKG &  FB\_SCTS &  FB\_FLUX &  HR 
	&  LX &  FLAG\_RELIABLE &  OIR\_CATALOG &  ZBEST &  ZSOURCE &  CLASS \\
    (1) & (2) & (3) & (4)    &  (16)       &  (19)    &  (31)    &  (43)    &  (93)    & (95)   
    &  (98) &  (109)      &   (111)      & (179)  & (180)        & (186)
   \\
    \hline
XMM00000 & 34.200220 & $-$4.035250 & 1.44 & 19.0  & 59076.2  & 1.74 & 83.04  & 8.92$\times 10^{-15}$  & $-$99.0 & $-$99		        	& 1 & HSC 	& $-$99.0 & $-$99 & 33213 \\
XMM00001 & 34.200710 & $-$4.933730 & 1.45 & 63.0  & 61051.8  & 1.00 & 82.03  & 8.53$\times 10^{-15}$  & $-$1.0  & 9.66$\times 10^{43}$  & 1 & SERVS & 1.82 & UDSz & 32212 \\
XMM00002 & 34.201450 & $-$5.556720 & 1.96 & 16.4  & 29731.6  & 0.80 & 29.64  & 5.71$\times 10^{-15}$  & 1.0 	& 2.64$\times 10^{42}$ 	& 1 & SERVS & 0.459 & VIPERS & 32213 \\
XMM00003 & 34.201470 & $-$4.499310 & 1.50 & 23.3  & 72553.8  & 1.76 & 72.37  & 5.71$\times 10^{-15}$  & $-$1.0 	& 1.47$\times 10^{43}$  & 1 & SERVS & 0.959 & PRIMUS & 32212 \\
XMM00004 & 34.201950 & $-$4.555520 & 0.93 & 316.8 & 87846.9  & 1.81 & 351.91 & 2.87$\times 10^{-14}$  & $-$0.43 & 1.03$\times 10^{43}$ 	& 1 & SERVS & 0.41 & SDSS & 32213 \\
XMM00005 & 34.202640 & $-$5.690720 & 1.66 & 16.5  & 26430.1  & 1.01 & 52.23  & 1.40$\times 10^{-14}$  & $-$1.0 	& 1.81$\times 10^{44}$  & 1 & CFHTLS & 1.932 & VIPERS & 32213 \\
XMM00006 & 34.203280 & $-$4.315290 & 1.55 & 29.2  & 107957.7 & 1.79 & 65.42  & 3.16$\times 10^{-15}$  & $-$99.0 & $-$99					& 1 & SERVS & $-$99.0 & $-$99 & 33213 \\
XMM00007 & 34.203750 & $-$5.433790 & 1.77 & 11.3  & 78270.8  & 1.54 & 41.87  & 5.06$\times 10^{-15}$  & $-$99.0 & $-$99					& 1 & VIDEO & $-$99.0 & $-$99 & 33213 \\
XMM00008 & 34.203820 & $-$4.595270 & 1.17 & 114.8 & 83485.0  & 1.49 & 168.25 & 1.31$\times 10^{-14}$  & $-$0.48 & 2.21$\times 10^{42}$ 	& 1 & VIDEO & 0.294 & SDSS & 33213 \\
XMM00009 & 34.204670 & $-$5.378240 & 1.35 & 57.1  & 93769.9  & 1.40 & 101.92 & 7.31$\times 10^{-15}$  & $-$1.0 	& $-$99					& 1 & SERVS & $-$99.0 & $-$99 & 33213 \\
 ... & ... & ... &  ... & ... & ... &  ... & ... & ... &  ... & ... & ... & ... & ... & ... & ... \\

\hline
\end{tabular} 
\end{table}

\begin{table}
\scriptsize
\caption{\label{tab:nwaytab}
The {\sc NWAY} matching results with a selection of columns. Empty or null values are marked as $-$99.
See Appendix~B for a detailed description of each column. This table is available in its entirety in machine-readable form online.
}
\begin{tabular}{cccccccccccc}
    \hline
XID & P\_ANY & P\_I & SERVS\_ID & SERVS\_MAG1 & VIDEO\_ID & VIDEO\_KSMAG & CFHT\_ID & CFHT\_IMAG & HSC\_ID & HSC\_IMAG & MATCH\_FLAG\\
(1) & (2) & (3) & (12) & (20) & (13) & (30) & (14) & (38) & (15) & (46) & (52)\\
\hline

XMM00000 & 0.00281 & 0.922088 & 701845.0 & 19.32 & -99 & -99.0 & -99.0 & -99 & -99 & -99.0 & 1 \\
XMM00001 & 0.99867 & 0.990685 & 408032.0 & 19.36 & 644246149826 & 20.17 & 20.17 & 1114\_171196 & 37485121644815869 & -99.0 & 1 \\
XMM00002 & 0.99234 & 0.979547 & 162933.0 & 19.72 & 644245967165 & 19.36 & 19.36 & 1114\_017717 & 37485108759910516 & -99.0 & 1 \\
XMM00003 & 0.99192 & 0.987072 & 595262.0 & 18.96 & 644246286360 & 19.57 & 19.57 & 1105\_044095 & 37485134529712709 & -99.0 & 1 \\
XMM00004 & 0.99369 & 0.987703 & 571059.0 & 18.82 & 644246268652 & 18.37 & 18.37 & 1105\_032861 & 37485130234751017 & -99.0 & 1 \\
XMM00005 & 0.64455 & 0.993912 & -99.0 & -99.0 & -99 & -99.0 & -99.0 & 1123\_209193 & 37485104464927930 & -99.0 & 1 \\
XMM00006 & 0.03568 & 0.313800 & 647512.0 & 22.23 & 644246338512 & 23.19 & 23.19 & 1105\_084654 & 38549431720610501 & -99.0 & 1 \\
XMM00007 & 0.01958 & 0.192855 & -99.0 & -99.0 & 644246003165 & 22.05 & 22.05 & -99 & -99 & -99.0 & 2 \\
XMM00007 & 0.01958 & 0.333900 & -99.0 & -99.0 & 644246382177 & 21.59 & 21.59 & -99 & -99 & -99.0 & 1 \\
XMM00007 & 0.01958 & 0.174318 & -99.0 & -99.0 & 644246413618 & 22.54 & 22.54 & -99 & -99 & -99.0 & 2 \\
XMM00008 & 0.99216 & 0.992165 & -99.0 & -99.0 & 644246255264 & 18.86 & 18.86 & 1105\_022859 & 37485130234749091 & -99.0 & 1 \\
... & ... &  ... & ... & ... &  ... & ... & ... &  ... & ... & ... & ... \\

\hline
  \end{tabular}
 \end{table}

\begin{table}
\scriptsize
\caption{\label{tab:allphotoz}
The photo-zs for galaxies detected in the N18 forced-photometry catalog; see Appendix~C for details. 
Columns 1--3: VIDEO object ID, RA, and DEC (J2000).
Columns 4--6: SERVS object ID, RA, and DEC (J2000).
Column 7: photometric redshift.
Column 8--9: Upper and lower 68\% limits of the photometric redshift based on the probability distribution 
($p(z)$) of the photo-zs. Note that our photo-zs 
correspond to the peak value of $p(z)$, and thus for a small fraction of sources, the most probable redshifts are 
not in the range covered by the upper and lower 68\% limits. 
See Eq. (12) and \S7.6 of \protect\cite{yang14photoz} for details.
Column 10: photometric redshift quality parameter; see \S\ref{subsec:zphot} for details.
Column 11: Spectroscopic redshift. Null values are filled with -99.
Columns 12--13: Similar to Columns 183--184 of Table A. 
This table is available in its entirety in machine-readable form online.}
\begin{tabular}{ccccccccccccc}
\hline
VIDEO\_ID & VIDEO\_RA & VIDEO\_DEC & SERVS\_ID & SERVS\_RA & SERVS\_DEC & 
ZPHOT & PZ\_ULIM & PZ\_LLIM & $Q_z$ & ZSPEC & ZSOURCE & ZFLAG \\
(1) & (2) & (3) & (4) & (5) & (6) & (7) & (8) & (9) & (10) & (11) & (12) & (13)\\
\hline
644246417039 & 34.559784 & -4.966628 & 419320 & 34.552429 & 
-4.971455 & 0.890 & 0.810 & 0.978 & 0.166874 & -99 & ... & ... \\
644246232645 & 34.826369 & -4.664200 & 463179 & 34.826399 & 
-4.664191 & 0.729 & 0.628 & 0.805 & 0.025925 & 0.7615 & VIPERS & 2.5 \\
644246414844 & 34.865246 & -5.255360 & 217192 & 34.866427 & 
-5.256932 & 1.239 & 0.986 & 1.942 & 0.767938 & -99 & ... & ... \\
644245112476 & 35.980791 & -5.384739 & 79929 & 35.980755 & 
-5.384734 & 0.799 & 0.644 & 0.982 & 0.351430 & 1.009558 & PRIMUS & 3 \\
644245112310 & 35.990999 & -5.385472 & 79380 & 35.991092 & 
-5.385497 & 0.431 & 0.193 & 0.492 & 0.224409 & 0.513350 & PRIMUS & 4 \\
644245264327 & 36.514834 & -4.820542 & 242344 & 36.513482 & 
-4.818775 & 0.010 & 0.023 & 0.132 & 0.189810 & 0.2036 & VVDS & 3 \\
 ... & ... & ... &  ... & ... & ... &  ... & ... & ... &  ... & ... & ... & ... \\
\hline
\end{tabular}
\end{table}

\end{landscape}

\bibliographystyle{mnras}
\bibliography{XMM-XSERV} % if your bibtex file is called example.bib

% Don't change these lines
\bsp    % typesetting comment
\label{lastpage}
\end{document}